\documentclass[twocolumn]{aastex631}

\usepackage{amsmath}
\usepackage{multirow}

\definecolor{DarkOrange}{RGB}{204, 85, 0}
\definecolor{LincolnGreen}{RGB}{17, 102, 0}
\def\ion#1#2{#1$\;${\footnotesize\rm{#2}}\relax}

\usepackage{xspace}

\newcommand\srg{\textit{SRG}\xspace}

\newcommand\galex{\textit{GALEX}\xspace}

\newcommand\gaia{\textit{Gaia}\xspace}
\newcommand\chandra{\textit{Chandra}\xspace}

\shorttitle{TDE Demographics}
\shortauthors{Yao et al.}

\def \caltech {{Cahill Center for Astrophysics, California Institute of Technology, 
MC 249-17, 1200 E California Boulevard, Pasadena, CA 91125, USA}}

\def \coo {{Caltech Optical Observatories, California Institute of Technology, Pasadena, CA 91125, USA}}

\def \gsfc {{Astrophysics Science Division, NASA Goddard Space Flight Center, Greenbelt, MD 20771, USA}}

\begin{document}

\title{Tidal Disruption Event Demographics with the Zwicky Transient Facility: \\ 
Volumetric Rates, Luminosity Function, and Implications for the Local Black Hole Mass Function}

\correspondingauthor{Yuhan Yao}
\email{yuhanyao@berkeley.edu}

\author[0000-0001-6747-8509]{Yuhan Yao}
\affiliation{\caltech}
\affiliation{Miller Institute for Basic Research in Science, 468 Donner Lab, Berkeley, CA 94720, USA}
\affiliation{Department of Astronomy, University of California, Berkeley, CA 94720, USA}

\author[0000-0002-7252-5485]{Vikram Ravi}
\affiliation{\caltech}

\author[0000-0003-3703-5154]{Suvi Gezari}
\affiliation{Space Telescope Science Institute, 3700 San Martin Drive, Baltimore, MD 21218, USA}
\affiliation{Department of Physics and Astronomy, Johns Hopkins University, Baltimore, MD 21218, USA}

\author[0000-0002-3859-8074]{Sjoert van Velzen}
\affiliation{Leiden Observatory, Leiden University, Postbus 9513, 2300 RA, Leiden, The Netherlands}

\author[0000-0002-1568-7461]{Wenbin Lu}
\affiliation{Department of Astronomy, University of California, Berkeley, CA 94720, USA}

\author[0000-0001-6797-1889]{Steve Schulze}
\affiliation{Department of Physics, The Oskar Klein Centre, Stockholm University, AlbaNova, SE-10691 Stockholm, Sweden}

\author[0000-0001-8426-5732]{Jean J.~Somalwar}
\affiliation{\caltech}

\author[0000-0001-5390-8563]{S.~R.~Kulkarni}
\affiliation{\caltech}

\author[0000-0002-5698-8703]{Erica Hammerstein}
\affiliation{Department of Astronomy, University of Maryland, College Park, MD 20742, USA}

\author[0000-0002-2555-3192]{Matt Nicholl}
\affiliation{Birmingham Institute for Gravitational Wave Astronomy and School of Physics and Astronomy, University of Birmingham, Birmingham B15 2TT, UK}
\affiliation{Astrophysics Research Centre, School of Mathematics and Physics, Queens University Belfast, Belfast BT7 1NN, UK}

\author[0000-0002-3168-0139]{Matthew J. Graham}
\affiliation{\caltech}

\author[0000-0001-8472-1996]{Daniel A.~Perley}
\affiliation{Astrophysics Research Institute, Liverpool John Moores University, IC2, Liverpool Science Park, 146 Brownlow Hill, Liverpool L3 5RF, UK}

\author[0000-0003-1673-970X]{S.~Bradley Cenko}
\affiliation{\gsfc}
\affiliation{Joint Space-Science Institute, University of Maryland, College Park, MD 20742, USA}

\author[0000-0003-2434-0387]{Robert Stein}
\affiliation{\caltech}

\author[0000-0001-5287-0452]{Angelo Ricarte}
\affiliation{Center for Astrophysics $\vert$ Harvard \& Smithsonian, 60 Garden Street, Cambridge, MA 02138, USA}
\affiliation{Black Hole Initiative at Harvard University, 20 Garden Street, Cambridge, MA 02138, USA}

\author{Urmila Chadayammuri}
\affiliation{Center for Astrophysics $\vert$ Harvard \& Smithsonian, 60 Garden Street, Cambridge, MA 02138, USA}

\author{Eliot Quataert}
\affiliation{Department of Astrophysical Sciences, Princeton University, Princeton, NJ 08544, USA}


\author[0000-0001-8018-5348]{Eric C. Bellm}
\affiliation{DIRAC Institute, Department of Astronomy, University of Washington, 3910 15th Avenue NE, Seattle, WA 98195, USA}
 
\author[0000-0002-7777-216X]{Joshua S. Bloom}
\affiliation{Department of Astronomy, University of California, Berkeley, CA 94720, USA}

\author{Richard Dekany}
\affiliation{\coo}

\author{Andrew J. Drake}
\affiliation{\caltech}

\author[0000-0001-5668-3507]{Steven L. Groom}
\affiliation{IPAC, California Institute of Technology, 1200 E. California Blvd, Pasadena, CA 91125, USA}

\author[0000-0003-2242-0244]{Ashish~A.~Mahabal}
\affiliation{Division of Physics, Mathematics and Astronomy, California Institute of Technology, Pasadena, CA 91125, USA}
\affiliation{Center for Data Driven Discovery, California Institute of Technology, Pasadena, CA 91125, USA}

\author[0000-0002-8850-3627]{Thomas A. Prince}
\affiliation{\caltech}

\author[0000-0002-0387-370X]{Reed Riddle}
\affiliation{\coo}

\author[0000-0001-7648-4142]{Ben Rusholme}
\affiliation{IPAC, California Institute of Technology, 1200 E. California Blvd, Pasadena, CA 91125, USA}

\author[0000-0003-4531-1745]{Yashvi Sharma}
\affiliation{\caltech}

\author[0000-0003-1546-6615]{Jesper Sollerman}
\affiliation{Department of Astronomy, The Oskar Klein Centre, Stockholm University, AlbaNova, SE-10691 Stockholm, Sweden}

\author[0000-0003-1710-9339]{Lin Yan}
\affiliation{\caltech}

\begin{abstract}
We conduct a systematic tidal disruption event (TDE) demographics analysis using the largest sample of optically selected TDEs.
A flux-limited, spectroscopically complete sample of 33 TDEs is constructed using the Zwicky Transient Facility over 3\,yr (from October 2018 to September 2021). 
We infer the black hole (BH) mass ($M_{\rm BH}$) with host galaxy scaling relations, showing that the sample $M_{\rm BH}$ ranges from $10^{5.1}\,M_\odot$ to $10^{8.2}\,M_\odot$.
We developed a survey efficiency corrected maximum volume method to infer the rates.
The rest-frame $g$-band luminosity function can be well described by a broken power law of $\phi (L_g)\propto [(L_g / L_{\rm bk})^{0.3} + (L_g / L_{\rm bk})^{2.6}]^{-1}$, with $L_{\rm bk}=10^{43.1}\,{\rm erg\,s^{-1}}$.
In the BH mass regime of $10^{5.3}\lesssim (M_{\rm BH}/M_\odot) \lesssim 10^{7.3}$, the TDE mass function follows $\phi(M_{\rm BH})\propto M_{\rm BH}^{-0.25}$, which favors a flat local BH mass function ($dn_{\rm BH}/d{\rm log}M_{\rm BH}\approx{\rm constant}$). 
We confirm the significant rate suppression at the high-mass end ($M_{\rm BH}\gtrsim  10^{7.5}\,M_\odot$), which is consistent with theoretical predictions considering direct capture of hydrogen-burning stars by the event horizon. 
At a host galaxy mass of $M_{\rm gal}\sim 10^{10}\,M_\odot$, the average optical TDE rate is $\approx 3.2\times 10^{-5}\,{\rm galaxy^{-1}\,yr^{-1}}$. 
We constrain the optical TDE rate to be [3.7, 7.4, and 1.6$]\times 10^{-5}\,{\rm galaxy^{-1}\,yr^{-1}}$ in galaxies with red, green, and blue colors.
\end{abstract}

\vspace{1em}

\section{Introduction} \label{sec:intro}
In the local universe, a small fraction ($\sim 10$\%) of galaxies host active massive black holes (BHs) in their nuclei \citep{Kewley2006, Aird2012}. 
The remaining massive BHs are quiescent, but can be temporarily awakened when a star comes too close to it and becomes disrupted by tidal forces. 
The stellar debris evolves into an elongated stream, approximately half of which comes back to get accreted \citep{Rees1988}. This produces an electromagnetic flare if the tidal radius $R_{\rm T}$ (where the self gravity of the star balances the tidal forces) is greater than the size of the BH event horizon. 
Since $R_{\rm T}\propto M_{\rm BH}^{1/3}$ and the size of the event horizon $\propto M_{\rm BH}$, there exists a maximum BH mass for an observable TDE --- the so-called Hills mass. For Sun-like stars, $M_{\rm Hills}\sim$10$^8\,M_\odot$ (\citealt{Hills1975}). 

The first tidal disruption event (TDE) was identified with the \textit{ROSAT} all-sky X-ray survey, where the soft X-rays are thought to come from a newly formed accretion disk \citep{Bade1996, Grupe1999, Saxton2020}. Recently, the eROSITA telescope \citep{Predehl2021} on board the Spektrum-Roentgen-Gamma (\srg) X-ray mission \citep{Sunyaev2021} reported 13 TDEs selected from the second eROSITA all-sky survey \citep{Sazonov2021}. 
Low-temperature (${\rm few}\times 10^4$\,K) thermal emission from TDEs has been discovered with UV and optical sky surveys
\citep{
Gezari2006, 
vanVelzen2011, 
Gezari2012,
Arcavi2014,
Holoien2014,
Hung2017}, 
which has been postulated to arise from either energy dissipation within a stream-stream collision shock \citep{piran15_shock_model, Jiang2016_self_crossing_shock} or reprocessing of high-energy photons \citep{metzger16_reprocessing, roth16_reprocessing}. 
In the latter scenario, the physical origin of the ``reprocessing layer'' may be the optically thick gas from the self-collision shock \citep{Lu2020}, a radiation-driven outflow formed under super-Eddington accretion \citep{Miller2015_disk_wind, Dai2018, Thomsen2022_disk_wind}, or a quasi-static weakly bound envelope \citep{loeb97_spherical_envelope, coughlin14_ZEBRA, metzger22_colling_envelope}.

Theoretically, the TDE rate is determined by processes that govern stellar diffusion into the ``loss cone'', which defines a phase-space volume of orbits with angular momentum $J \leq J_{\rm lc} \equiv \sqrt{2GM_{\rm BH}R_{\rm T}}$ \citep{Alexander2017, Stone2020}.
Observational constraints on TDE demography can help address various open questions in astrophysics. First, the TDE luminosity function (LF) provides clues to how the emission mechanism is tied to the loss cone filling \citep{Kochanek2016, Stone2016, Stone2020} and provides an essential input to predict TDE rates in future sky surveys.  

Moreover, measuring the volumetric rate of TDEs as a function of $M_{\rm BH}$ 
offers a unique approach to trace the local BH population.
At the low-mass end ($M_{\rm BH}\lesssim 10^6\,M_\odot$), the TDE mass function depends on the unknown bottom end of the massive black hole mass function (BHMF). 
The space density of such intermediate-mass black holes (IMBHs) encodes formation mechanisms of primordial BHs in the early Universe at redshifts of $z>10$ \citep{Ricarte2018b, Woods2019, Greene2020, Chadayammuri2023}. 
The mergers of IMBHs and extreme mass-ratio inspirals are prime targets for the upcoming space-based gravitational-wave detector Laser Interferometer Space Antenna \citep{Amaro-Seoane2017, Jani2020, Amaro-Seoane2022}. 

At the high-mass end, the location of the TDE mass function's cutoff is set by the size of the event horizon, which probes the spin distribution of BHs in the mass range of $10^{7.5}\,M_\odot \lesssim M_{\rm BH}\lesssim 10^{8.5}\,M_\odot$ \citep{Kesden2012, Stone2019, Du2022, Huang2023}. 
The spin of such quiescent BHs cannot be measured via the traditional method of X-ray reflection spectroscopy \citep{Reynolds2021} developed for X-ray binaries and active galactic nulei (AGN). 

\cite{vanVelzen2018} made the first attempt to construct the TDE LF and mass function. 
Using a sample of 13 objects selected from five different UV and optical sky surveys, the authors inferred a rest-frame $g$-band LF of $dN/dL_g \propto L_g^{-5/2}$ for $L_g \in (10^{42.3}, 10^{44.8})\,{\rm erg\,s^{-1}}$ and a nearly constant TDE mass function for $M_{\rm BH} \in (10^{5.8}, 10^{7.3})\,M_\odot$. While these early results have demonstrated the important role that TDEs play in understanding BH demographics, they are susceptible to small number statistics and the heterogeneous nature of the sample.

Over the past few years, time domain sky surveys have led to a surge of TDE discoveries. The Zwicky Transient Facility (ZTF; \citealt{Bellm2019b, Graham2019}) is one of the most prolific optical discovery engines. Previous ZTF TDE sample studies have made significant progress on characterizing the photometric and spectroscopic properties of TDEs \citep{vanVelzen2021, Hammerstein2023}. However, since the classification completeness of photometric candidates was not assessed, recent studies that attempt to constrain the TDE optical LF using previously published ZTF TDE samples (e.g., \citealt{Lin2022, Charalampopoulos2023}) had to rely on false assumptions regarding the spectroscopic completeness. In this work, we aim to put new observational constraints on TDE demography. To this end, we constructed a flux-limited, spectroscopically complete sample of 33 TDEs selected from 3\,yr of the ZTF operation. 

This paper is organized as follows. The procedures of the TDE sample selection, observation, and classification are outlined in \S\ref{sec:sample}. UV and optical light curve fitting is described in \S\ref{sec:lc_fitting}. Host galaxy observation and analysis (including measurements of the $M_{\rm BH}$) are presented in \S\ref{sec:host}. The survey efficiency is assessed in \S\ref{sec:method}. We compute and discuss the volumetric rate of optical TDEs as a function of $M_{\rm BH}$, $L_g$, as well as other host galaxy and transient properties in \S\ref{sec:rate}. We summarize our conclusions in \S\ref{sec:conclusion}.

UT time is used throughout the paper. We assume a basic cosmology of $\Omega_{\rm M} = 0.3$, $\Omega_{\Lambda}=0.7$, and $h=0.7$. Optical magnitudes are reported in the AB system. Assuming $R_V = 3.1$, we correct the observed photometry for Galactic extinction using the \citet{Cardelli1989} extinction law and the \citet{Schlafly2011} extinction map. The coordinates are given in J2000. We use $t$ to denote rest-frame time relative to the maximum-light epoch. 

\section{Sample Construction} \label{sec:sample}

\subsection{The ZTF TDE Experiment}

\begin{deluxetable*}{c|c|c}[htp]
\tabletypesize{\scriptsize}
\tablecaption{Steps for Selecting TDE Candidates. \label{tab:selection}}
\tablehead{
\colhead{Step}
& \colhead{Criteria}
&\colhead{\# TDE Candidates} 
} 
\startdata
    1 & Initial cuts to select nuclear transients   & 890,266  \\
    2 & More detailed cuts to select nuclear transients & 143,731\\
    3 & Cuts on peak magnitude, transient duration, and number of detections  & 9426 \\
    4 & Cuts on the peak color, PS1 machine-learning classification, and IR variability; remove known quasars & 1390 \\
    5 & Alert photometry: cuts on color, cooling rate, and rise and decline timescales  & 174 \\
    6 & Forced photometry: cuts on color, cooling rate, and rise and decline timescales & 90 \\
    7 & Cuts on peak magnitude (of forced photometry) & 55 \\
    8 & Spectroscopic classification for 50 objects; photometric and contextual classification for 5 objects & 33\\
\enddata
\end{deluxetable*}

ZTF is an optical time domain sky survey operated by the Palomar Observatory. It uses the Palomar Oschin Schmidt 48 inch telescope (P48) equipped with a 47\,$\rm deg^2$ camera \citep{Dekany2020} to scan the entire northern visible sky at ${\rm decl.}>-35.2^{\circ}$. The three ZTF filters ($g$, $r$, and $i$) were designed to maximize throughput by avoiding major Palomar sky lines. The typical survey depth is $\sim 20.5$\,mag \citep{Graham2019}. 

Image processing and reference subtraction are performed by the ZTF Science Data System \citep{Masci2019}. Every 5$\sigma$ point-source detection is saved as an ``alert'' in the Avro format and distributed to community brokers via the ZTF Alert Distribution System  \citep{Patterson2019}. The alerts are enhanced with additional contextual information such as the machine-learning real-bogus score \citep{Duev2019, Mahabal2019}, the proximity to the nearest object in archival catalogs \citep{Soumagnac2018}, and the star--galaxy classifier \citep{Tachibana2018}.

ZTF phase I (hereafter ZTF-I) ran from March 2018 to September 2020, during which 40\% of the total time was dedicated to two public sky surveys, including a Northern sky survey (1 $g$ + 1 $r$ every 3 days) and a Galactic Plane survey \citep{Bellm2019a}. On 2020 October 1, ZTF increased the MSIP/NSF-funded public program to 50\% of the total time, and the Northern sky survey cadence was shortened from 3 to 2 days. Therefore, in this paper, we use 2020 October 1 as the start of ZTF phase II (hereafter ZTF-II).\footnote{Note that some other publications from the ZTF collaboration (such as \citealt{Hammerstein2023}) consider December 2020 as the start of ZTF-II, as the Phase II Partnership surveys did not begin until that time.}

The ZTF team selects nuclear transients in real-time by filtering public alerts with the \texttt{AMPEL} broker \citep{Nordin2019}. Details of our filtering techniques are described in \citet{vanVelzen2019, vanVelzen2021}.
AT2018zr is the first TDE selected by the ZTF nuclear transient filter \citep{vanVelzen2019}. 
Afterwards, \citet{vanVelzen2021} presented 17 TDEs selected within the first 1.5\,yr of ZTF-I operation, and introduced three distinct spectroscopic subclasses of optically selected TDEs (TDE-H, TDE-H+He, and TDE-He) based on the existence of a combination of broad emission lines around H$\alpha$, H$\beta$, and \ion{He}{II} $\lambda 4686$. 
Recently, \citet{Hammerstein2023} presented a sample of 30 spectroscopically classified TDEs from the entirety of ZTF-I, and reported a new spectroscopic subclass called ``TDE-featureless'', which is characterized by a lack of broad emission lines in optical spectra. 

Entering into ZTF-II, the TDE experiment was carried out with more spectroscopic follow-up resources allocated from the Keck and Palomar Observatories, which allowed us to classify a larger number of fainter TDE candidates. 

The follow-up campaign in ZTF was conducted on a best effort basis. We tried to classify as many TDE candidates as possible, with higher priorities of spectroscopic observations given to objects with brighter peak magnitudes. 
Unlike previous ZTF work, we here seek to construct a flux-limited sample of TDEs, enabling a systematic study of optical TDE demographics.
Therefore, we performed a retrospective search of nuclear transients using historical ZTF alerts, and applied a set of well-defined criteria to select TDE candidates (see \S\ref{subsec:filtering}). We then find the peak magnitude limits (in ZTF-I and ZTF-II separately) below which our spectroscopic classification is almost ($\gtrsim90$\%) complete (see step (7) in \S\ref{subsec:filtering}). And for the few candidates with no (or ambiguous) spectroscopic classification, we determine the transient type using the photometric properties and other information (see details in \S\ref{subsec:classification}).

\subsection{Retrospective Candidate Filtering} \label{subsec:filtering}

\begin{figure}[thbp]
    \centering
    \includegraphics[width=\columnwidth]{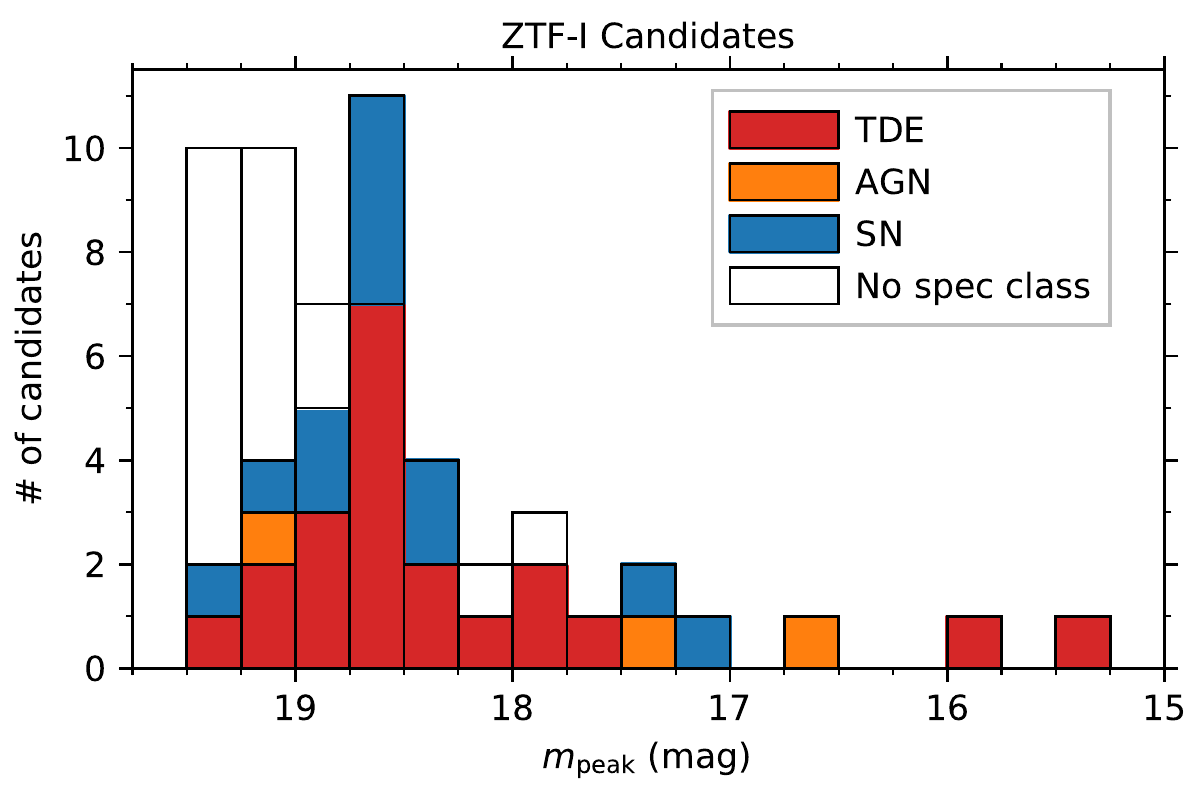}
    \includegraphics[width=\columnwidth]{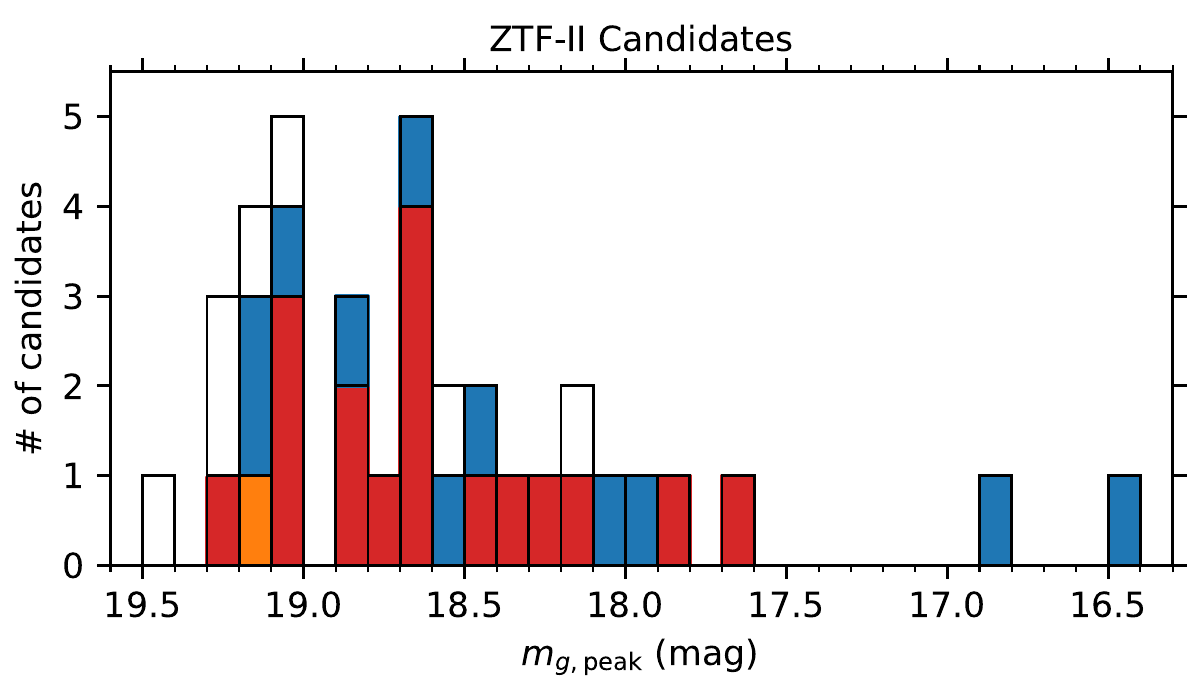}
    \caption{Histograms of the photometric TDE candidates that passed the filtering step (6) (see Table~\ref{tab:selection}), color-coded by their spectroscopic classifications. For ZTF-I candidates, the spectroscopic classification is $\sim93$\% complete at $m_{\rm peak}<18.75$. For ZTF-II candidates, the spectroscopic classification is $\sim89$\% complete at $m_{g, \rm peak}<19.1$.
    \label{fig:cand}}
\end{figure}

Table~\ref{tab:selection} presents a summary of the candidate filtering steps. 
\begin{enumerate}
    \item We applied basic cuts to select nuclear transients. 
    We kept alerts with a real-bogus score \texttt{rb>0.5} \citep{Mahabal2019} or a deep learning score \texttt{drb>0.65} \citep{Duev2019}\footnote{The deep learning score was not included in the alert packets until 2019 June 19. Therefore, we used \texttt{rb} and \texttt{drb} for alerts released before and after that date, respectively.}, a position within $0.6^{\prime\prime}$ to the location of the nearest object in the Panoramic Survey Telescope and Rapid Response System Data Release 1 (PS1; \citealt{Chambers2016a}) catalog (\texttt{distpsnr1<0.6}) or hostless (\texttt{distpsnr1==-999}). 
    We removed alerts in negative subtractions. 
    We kept alerts in coincidence with objects with galaxy-like morphologies, selected using a cut on the star--galaxy score \citep{Tachibana2018} of \texttt{sgscore1<0.8}.
    This step left 890,266 unique sources.
    \item We kept objects first detected between 2018 October 1 and 2021 September 30, i.e., the last 2\,yr of ZTF-I\footnote{Due to a likely low recovery efficiency for TDEs detected in the reference images, we do not consider events first detected before 2018 October 1, when ZTF reference images for most fields were still being constructed.} and the first year of ZTF-II. 
    We require that, in either $g$ or $r$ band, the transient is within $0.6^{\prime\prime}$ to the location of the nearest object in the ZTF reference image (\texttt{distnr<0.6}). 
    If the nearest reference object is brighter than 15\,mag (\texttt{magnr<=15}), we require \texttt{sgscore1<0.2}; similarly, we require  \texttt{sgscore1<=0.5} for \texttt{15<magnr<=18} and \texttt{sgscore1<0.8} for \texttt{magnr>18}. 
    This left 143,731 sources.
    \item We define $n_g$ ($n_r$) as the number of detections in $g$ band ($r$ band), and $t_{\rm dur}$ as the duration of all detections. The peak magnitudes in the $g$ and $r$ bands are $m_{g, {\rm peak}}$ and $m_{r, {\rm peak}}$, respectively. 
    We required $m_{g, \rm peak}< 19.5$\,mag, $m_{r, \rm peak}< 19.5$\,mag, $t_{\rm dur} > 30$\,d, $n_g>10$ and $n_r>10$. 
    This left 9426 sources.
    \item We applied a few cuts to remove stellar and AGN variability. 
    We required $m_{g, {\rm peak}} - m_{r, {\rm peak}}<1$, and that the closest object in the ``Pan-STARRS1 Source Types and Redshifts with Machine learning'' catalog \citep{Beck2021} is not classified as ``QSO'' or ``STAR''. 
    We removed objects with a counterpart in the Million Quasars catalog (Milliquas v6.3, \citealt{Flesch2019}). 
    We constructed a $W1$-band light curve from the NeoWISE \citep{Mainzer2011} photometry prior to the first ZTF detection, and rejected any galaxies with significant variability in the $W1$ band ($\chi^2/ $degrees of freedom$ > 10$).
    This left 1390 sources.
    \item We selected candidates based on the alert photometry. 
    We kept objects with at least 5 nights of post-peak multi-band photometry. 
    We required the rate of post-peak $g-r$ color change to be $<0.02\,{\rm mag\, day^{-1}}$, and the mean $g-r$ color to be $<0.2$\,mag. 
    We calculated the rise and decay e-folding times in the alert photometry light curve (smoothed with a Gaussian process). We required the rise e-folding time to be $ 2< t_{e \rm , rise}<300$\,d, and the decline e-folding time to be $ 2<t_{e \rm , decline}<300$\,d.
    This step left 174 sources, including 104 sources first detected during ZTF-I, and 70 sources first detected during the first year of ZTF-II. 
    \item We ran forced point spread function (PSF) photometry, which provide more accurate light curves. 
    We also visually examined the light curves and excluded 8 objects\footnote{ZTF18accdkxa, ZTF18acenyfr, ZTF18acpjddi, ZTF19acblzqb, ZTF19abkftuu, ZTF19abukbuc, ZTF20absxaaj, and ZTF20abzpysa show stochastic variability.} that are reminiscent of AGN and one object\footnote{ZTF21abiplqz has a fast rise, a rapid decline followed by a sudden flux frop, and a blue optical conterpart.} with a typical dwarf nova light curve.
    We applied the criteria outlined in step (4) to the ZTF forced photometry.
    This left 90 sources, including 54 in ZTF-I and 36 in ZTF-II. 
    \item We found that for candidates in ZTF-I, our spectroscopic classification completeness was $\sim93$\% at $m_{\rm peak}< 18.75$; for candidates in ZTF-II, our spectroscopic classification completeness was $\sim 89$\% complete at $m_{g, \rm peak}< 19.1$ (see Figure~\ref{fig:cand}).
    Therefore, we kept ZTF-I sources with $m_{\rm peak}< 18.75$, and ZTF-II sources with $m_{g, \rm peak}< 19.1$. 
    This left 55 sources, including 27 in ZTF-I and 28 in ZTF-II.
\end{enumerate}

A few notes are worth mentioning. 
First, as pointed out in \citet{vanVelzen2021}, by applying step (4), our search is biased against TDEs hosted by AGN, such as PS1-16dtm \citep{Blanchard2017} and ZTF20abisysx/AT2020nov \citep{Dahiwale2020_ZTF20abisysx}. The local AGN fraction for galaxies throughout the stellar mass range of $9.5< {\rm log}(M_{\rm gal}/M_\odot) <12$ is $\lesssim 10$\% \citep{Kewley2006, Aird2012}, and the fraction is even lower in dwarf galaxies \citep{Latimer2021_AGN_fraction}. Therefore, the majority of TDEs should be hosted by quiescent galaxies without strong AGN activity, unless the rate is enhanced by a factor $\sim10$ in AGN. Second, unlike previous ZTF TDE sample studies, we do not reject candidates based on the mean $W1-W2$ color of their host galaxies, since recent studies have found that some star-forming dwarf galaxies also exhibit red neoWISE colors \citep{Latimer2021_color}.
Third, in steps (5) and (6), the cuts on color and cooling rate are defined such that all TDEs presented in \citet{vanVelzen2021}, \citet{Angus2022}, and \citet{Hammerstein2023} satisfy the selection criteria. Finally, we show in Appendix~\ref{sec:details_selection} that our cuts on \texttt{sgscore1}, $t_{e, {\rm rise}}$ and $t_{e, {\rm decline}}$ do not hit the boundary of the selection.

\subsection{Observations} 

\subsubsection{UV and Optical Photometry}
For all TDE candidates, we constructed the optical and UV light curves using data from ZTF, the Asteroid Terrestrial-impact Last Alert System (ATLAS; \citealt{Tonry2018, Smith2020, Shingles2021}), and the Ultra-Violet/Optical Telescope (UVOT; \citealt{Roming2005}) on board the \textit{Neil Gehrels Swift Observatory} \citep{Gehrels2004}. Data reduction procedures follow those outlined in \citet{vanVelzen2021} and \citet{Hammerstein2023}. We show the Galactic extinction-corrected $g-r$ evolution in ZTF forced photometry in Figure~\ref{fig:color_evol}. The photometry of the final sample of 33 TDEs is presented in Appendix~\ref{sec:obs_log}.

\begin{figure}[htbp!]
    \centering
    \includegraphics[width=\columnwidth]{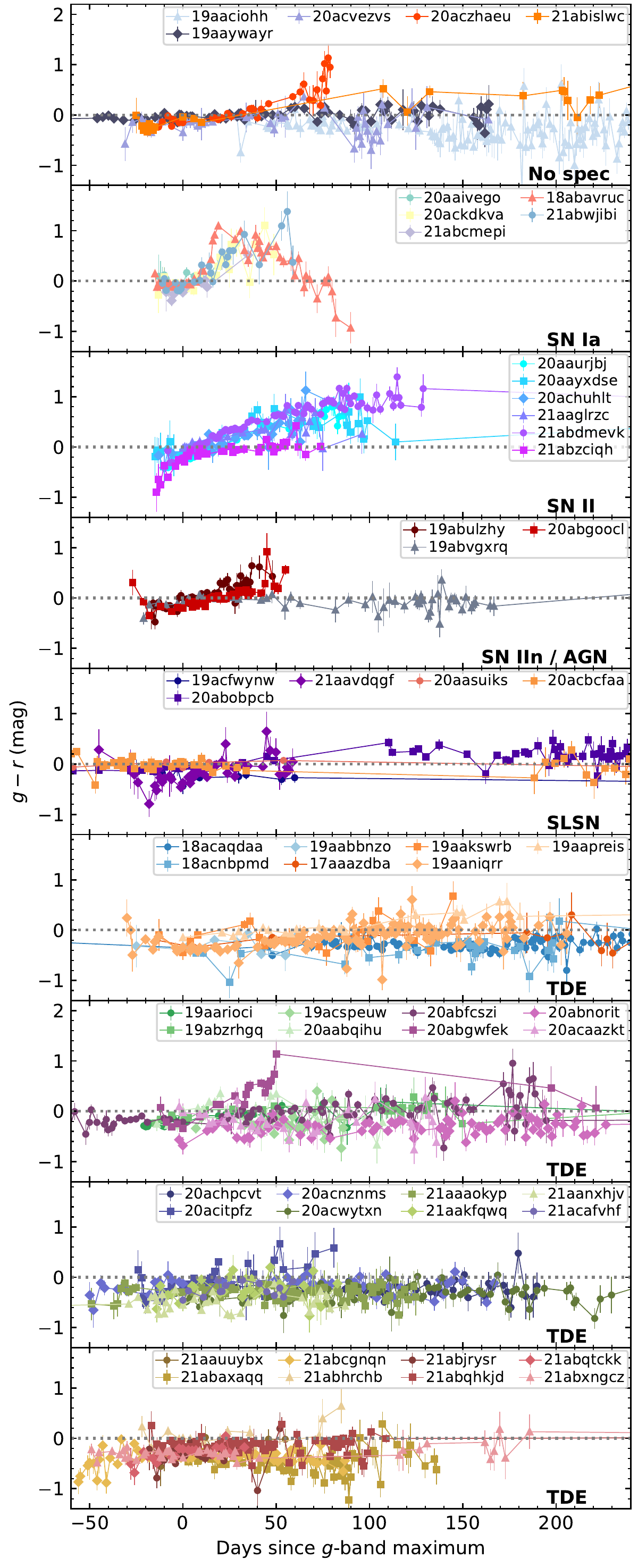}
    \caption{$g-r$ evolution of the 55 TDE candidates. The top panel shows 5 objects without spectroscopic classifications, and the other 8 panels show 50 spectroscopically classified objects. 
    Color has been corrected for Galactic extinction. \label{fig:color_evol}}
\end{figure}

\subsubsection{Optical Spectroscopy}
To spectroscopically classify the TDE candidates, we obtained low-resolution optical spectra with the Spectral Energy Distribution Machine (SEDM, \citealt{Blagorodnova2018}, \citealt{Rigault2019}, \citealt{Kim2022}) on the robotic Palomar 60 inch telescope (P60; \citealt{Cenko2006}), the Low Resolution Imaging Spectrograph (LRIS; \citealt{Oke1995}) on the Keck I telescope, the Double Spectrograph (DBSP; \citealt{Oke1982}) on the 200 inch Hale telescope, and the De Veny Spectrograph on the Lowell Discovery Telescope (LDT). Note that all DBSP observations are affected by a CCD malfunction, which results in a wavelength gap between 5750 and 6200\,\AA. The instrument configurations and data reduction procedures follow those described in Appendix~B of \citet{Yao2022}. 

We also made use of spectra uploaded to the transient name server (TNS) by other groups.
For each TDE that was not previously reported in the literature, we release at least one optical spectrum in this paper. 
An observing log of the released data is provided in Appendix~\ref{sec:obs_log} (Table~\ref{tab:spec_lowres}).\footnote{Upon publication, all spectra in Table~\ref{tab:spec_lowres} will be available in electronic format on the Weizmann Interactive Supernova Data Repository \citep{Yaron2012}.}

\subsection{Classification} \label{subsec:classification}
As mentioned in \S\ref{subsec:filtering}, five of the 55 photometrically selected TDE candidates do not have spectroscopic classifications. Using light curves, host galaxy spectroscopy, and multi-wavelength information (see details below), we classify ZTF19aaciohh and ZTF20acvezvs as TDE?, ZTF19aaywayr as AGN?, and ZTF20aczhaeu and ZTF21abislwc as SN?. Table~\ref{tab:false_positive} and Table~\ref{tab:tab_info} summarize 22 false positives and 33 TDEs. Below, we comment on the individual events.

\subsubsection{False Positives}

Among the list of 22 false positives, spectroscopic classifications are available for 19 objects: five were classified as Type Ia supernovae (SNe Ia); six were classified as Type II SNe (SNe II); two were classified as Type IIn SNe (SNe IIn); three were classified as hydrogen-poor superluminous SNe (SLSNe-I); two were classified as hydrogen-rich SLSNe (SLSNe-II); one was classified as an AGN.

ZTF20aczhaeu and ZTF21abislwc are probably SNe since their post-peak color reddened significantly, which is different from known TDEs (see Figure~\ref{fig:color_evol}).

ZTF19aaywayr is probably a slow AGN flare. In the forced photometry light curve, it has two peaks: the first at $m_r = 19.9$\,mag in 2019 June, and the second at $m_r = 18.1$\,mag in 2020 September. The rise time of the second peak is $\approx 400$\,days, which is a factor of $\sim10$ longer than the typical rise time of the spectroscopically classified TDE sample. Therefore, we think it is more likely to be an AGN.

\begin{deluxetable}{ccc}[htbp!]
    \tabletypesize{\scriptsize}
\tablecaption{Spectroscopic Classifications of 22 False Positives. \label{tab:false_positive}}
\tablehead{
	\colhead{ZTF name}
    & \colhead{Class}  
	& \colhead{Reference}
	}
\startdata
ZTF18abavruc & SN Ia & \citet{Angus2021_ZTF18abavruc} \\
ZTF20aaivego & SN Ia & \citet{Dahiwale2020_ZTF20aaivego} \\
ZTF20ackdkva & SN Ia & \citet{Dahiwale2020_ZTF20ackdkva} \\
ZTF21abcmepi & SN Ia & \citet{SNIascore2021_ZTF21abcmepi} \\
ZTF21abwjibi & SN Ia & \citet{Yao2022_ZTF21abwjibi} \\
ZTF20aaurjbj & SN II & \citet{Siebert2020_ZTF20aaurjbj} \\
ZTF20aayxdse & SN II & \citet{Dahiwale2020_ZTF20aayxdse} \\
ZTF20achuhlt & SN II & \citet{Yan2020_ZTF20achuhlt} \\
ZTF21aaglrzc & SN II & \citet{Dahiwale2021_ZTF21aaglrzc} \\
ZTF21abdmevk & SN II & \citet{Bruch2021_ZTF21abdmevk} \\
ZTF21abzciqh & SN II & \citet{Chu2021_ZTF21abzciqh} \\
ZTF19abulzhy & SN IIn & \citet{Dahiwale2020_ZTF19abulzhy} \\
ZTF20abgoocl & SN IIn & \citet{Perley2020_ZTF20abgoocl}\\
ZTF19acfwynw & SLSN-I & \cite{Nicholl2019_ZTF19acfwynw}\\
ZTF20abobpcb & SLSN-I & \citet{Perez-Fournon2020_ZTF20abobpcb}\\
ZTF21aavdqgf & SLSN-I & \citet{Yao2021_ZTF21aavdqgf} \\
ZTF20aasuiks & SLSN-II & \citet{Tucker2021_ZTF20aasuiks}\\
ZTF20acbcfaa & SLSN-II & \citet{Pessi2020_ZTF20acbcfaa}\\
ZTF19abvgxrq & AGN & \citet{Frederick2021, Yu2022}\\
\hline 
ZTF21abislwc & SN? & This work\\
ZTF20aczhaeu & SN? & This work\\
ZTF19aaywayr & AGN? & This work\\
\enddata 
\end{deluxetable}

\begin{deluxetable*}{rllrrlcl}[htbp!]
\tabletypesize{\scriptsize}
    \tablecaption{Basic Information of 33 TDEs in Our Sample.\label{tab:tab_info}}
	\tablehead{
	\colhead{ID}
	& \colhead{ZTF Name} 
	& \colhead{IAU Name}
    & \colhead{R.A. (deg)}
    & \colhead{Decl. (deg)}
	& \colhead{Redshift}
	& \colhead{TDE Report}
	& \colhead{Spectral Subtype}
    }
\startdata
1 & ZTF18acaqdaa & AT2018iih & 262.0163662 & 30.6920758 & 0.212 & \citet{vanVelzen2021} & TDE-He \\
2 & ZTF18acnbpmd & AT2018jbv & 197.6898587 & 8.5678292 & 0.340 & \citet{Hammerstein2023} & TDE-featureless \\
3 & ZTF19aabbnzo & AT2018lna & 105.8276892 & 23.0290953 & 0.0914 & \citet{vanVelzen2021} & TDE-H+He \\
4 & ZTF19aaciohh & AT2019baf & 268.0005082 & 65.6266546 & 0.0890 & This paper; J. Somalwar et al. (in preparation) & Unknown \\
5 & ZTF17aaazdba & AT2019azh & 123.3206388 & 22.6483180 & 0.0222 & \citet{Hinkle2021} & TDE-H+He \\
6 & ZTF19aakswrb & AT2019bhf & 227.3165243 & 16.2395720 & 0.121 & \citet{vanVelzen2021} & TDE-H \\
7 & ZTF19aaniqrr & AT2019cmw & 282.1644974 & 51.0135422 & 0.519 & This paper; J. Wise et al. (in preparation) & TDE-featureless \\
8 & ZTF19aapreis & AT2019dsg & 314.2623552 & 14.2044787 & 0.0512 & \citet{Stein2021} & TDE-H+He \\
9 & ZTF19aarioci & AT2019ehz & 212.4245268 & 55.4911223 & 0.0740 & \citet{vanVelzen2021} & TDE-H \\
10 & ZTF19abzrhgq & AT2019qiz & 71.6578313 & $-$10.2263602 & 0.0151 & \citet{Nicholl2020} & TDE-H+He \\
11 & ZTF19acspeuw & AT2019vcb & 189.7348778 & 33.1658869 & 0.0890 & \citet{Hammerstein2023} & TDE-H+He \\
12 & ZTF20aabqihu & AT2020pj & 232.8956925 & 33.0948917 & 0.0680 & \citet{Hammerstein2023} & TDE-H+He \\
13 & ZTF20abfcszi & AT2020mot & 7.8063109 & 85.0088329 & 0.0690 & \citet{Hammerstein2023} & TDE-H+He \\
14 & ZTF20abgwfek & AT2020neh & 230.3336852 & 14.0696032 & 0.0620 & \citet{Angus2022} & TDE-H+He \\
15 & ZTF20abnorit & AT2020ysg & 171.3584535 & 27.4406021 & 0.277 & \citet{Hammerstein2023} & TDE-featureless \\
16 & ZTF20acaazkt & AT2020vdq & 152.2227354 & 42.7167535 & 0.0450 & This paper; J. Somalwar et al. (in preparation) & Unknown \\
\hline
17 & ZTF20achpcvt & AT2020vwl & 232.6575481 & 26.9824432 & 0.0325 & \citet{Hammerstein2021_20vwl_CR} & TDE-H+He \\
18 & ZTF20acitpfz & AT2020wey & 136.3578499 & 61.8025699 & 0.0274 & \citet{Arcavi2020} & TDE-H+He \\
19 & ZTF20acnznms & AT2020yue & 165.0013942 & 21.1127532 & 0.204 & This paper & TDE-H? \\
20 & ZTF20acvezvs & AT2020abri & 202.3219785 & 19.6710235 & 0.178 & This paper & Unknown \\
21 & ZTF20acwytxn & AT2020acka & 238.7581288 & 16.3045292 & 0.338 & \citet{Hammerstein2021_20acka_CR} & TDE-featureless \\
22 & ZTF21aaaokyp & AT2021axu & 176.6514953 & 30.0854257 & 0.192 & \citet{Hammerstein2021_21axu_CR} & TDE-H+He \\
23 & ZTF21aakfqwq & AT2021crk & 176.2789219 & 18.5403839 & 0.155 & This paper & TDE-H+He? \\
24 & ZTF21aanxhjv & AT2021ehb & 46.9492531 & 40.3113468 & 0.0180 & \citet{Yao2022} & TDE-featureless \\
25 & ZTF21aauuybx & AT2021jjm & 219.8777384 & $-$27.8584845 & 0.153 & \citet{Yao2021_21jjm_CR} & TDE-H \\
26 & ZTF21abaxaqq & AT2021mhg & 4.9287185 & 29.3168745 & 0.0730 & \citet{Chu2021_21mhg_CR} & TDE-H+He \\
27 & ZTF21abcgnqn & AT2021nwa & 238.4636684 & 55.5887978 & 0.0470 & \citet{Yao2021_21nwa_CR} & TDE-H+He \\
28 & ZTF21abhrchb & AT2021qth & 302.9121723 & $-$21.1602187 & 0.0805 & This paper & TDE-coronal \\
29 & ZTF21abjrysr & AT2021sdu & 17.8496154 & 50.5749060 & 0.0590 & \citet{Chu2021_21sdu_CR} & TDE-H+He \\
30 & ZTF21abqhkjd & AT2021uqv & 8.1661654 & 22.5489257 & 0.106 & \citet{Yao2021_21uqv_CR} & TDE-H+He \\
31 & ZTF21abqtckk & AT2021utq & 229.6212498 & 73.3587323 & 0.127 & This paper & TDE-H \\
32 & ZTF21abxngcz & AT2021yzv & 105.2774821 & 40.8251799 & 0.286 & \citet{Chu2022_21yzv_CR} & TDE-featureless \\
33 & ZTF21acafvhf & AT2021yte & 103.7697396 & 12.6341503 & 0.0530 & \citet{Yao2021_21yte_CR} & TDE-H+He \\
\enddata
\tablecomments{The first 16 objects were selected from ZTF-I (from 2018 October 1 to 2020 September 30) with $m_{\rm peak}< 18.75$. 
The last 17 objects were selected from the first year of ZTF-II (from 2020 October 1 to 2021 September 30) with $m_{g, \rm peak}< 19.1$.
In the ``TDE report'' column, we include a refereed paper if existent. }
\end{deluxetable*}

\subsubsection{True Positives}

The TDE classifications of 15 objects (IDs 1--3, 5--6, 8--15, 18, 24) have been previously reported in refereed papers \citep{Arcavi2020, Nicholl2020, Hinkle2021, Stein2021, vanVelzen2021, Yao2022, Angus2022, Hammerstein2023}.

Two objects were detected in the radio band with the Very Large Array Sky Survey (VLASS; \citealt{Lacy2020}).
In short, \textbf{ZTF19aaciohh/AT2019baf (ID 4)} is hosted by a galaxy with Seyfert-like emission line ratios. Multi-wavelength properties suggest that it is likely a TDE associated with a jet.
\textbf{ZTF20acaazkt/AT2020vdq (ID 16)} can be spectroscopically classified as a TDE based on the existence of intermediate-width ($\sim700\,{\rm km\,s^{-1}}$) transient Balmer lines, \ion{He}{II}, and \ion{Fe}{X} emission lines. 
Detailed properties of these two events will be presented as part of a sample of VLASS-selected TDE (candidates) with optical flares (see J. Somalwar et al. in preparation). 

\begin{figure}[htbp!]
    \centering
    \includegraphics[width=\columnwidth]{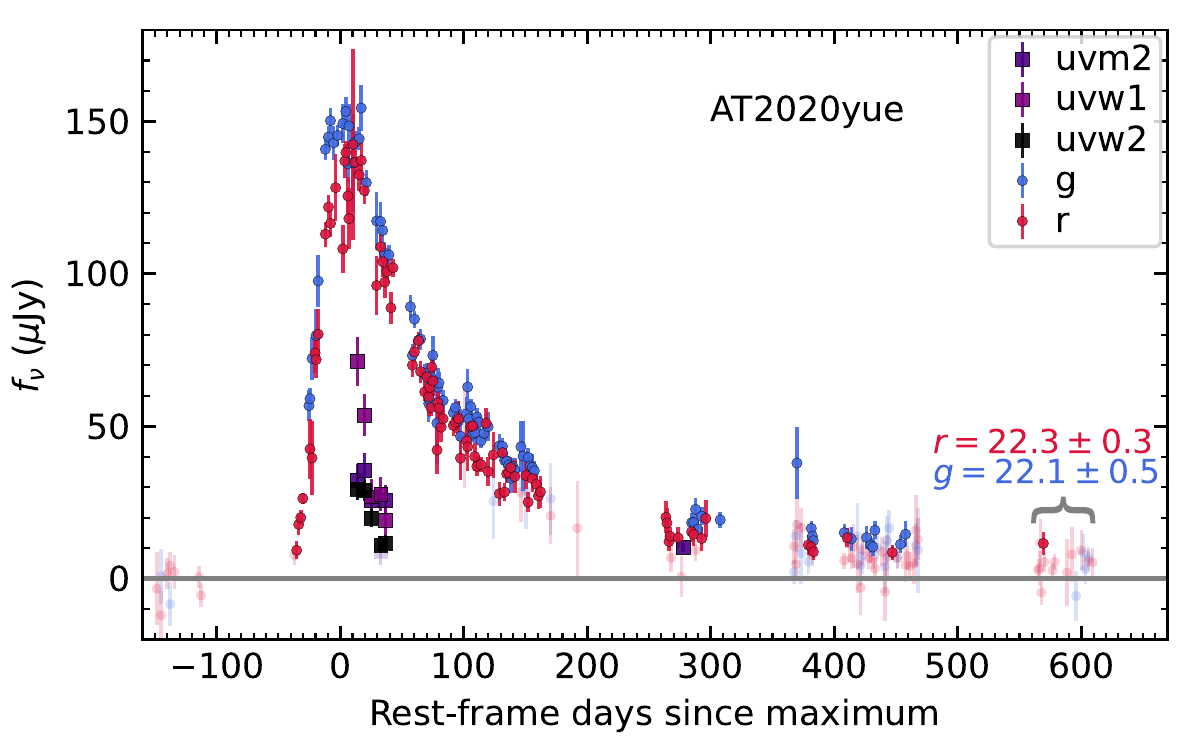}
    \includegraphics[width=\columnwidth]{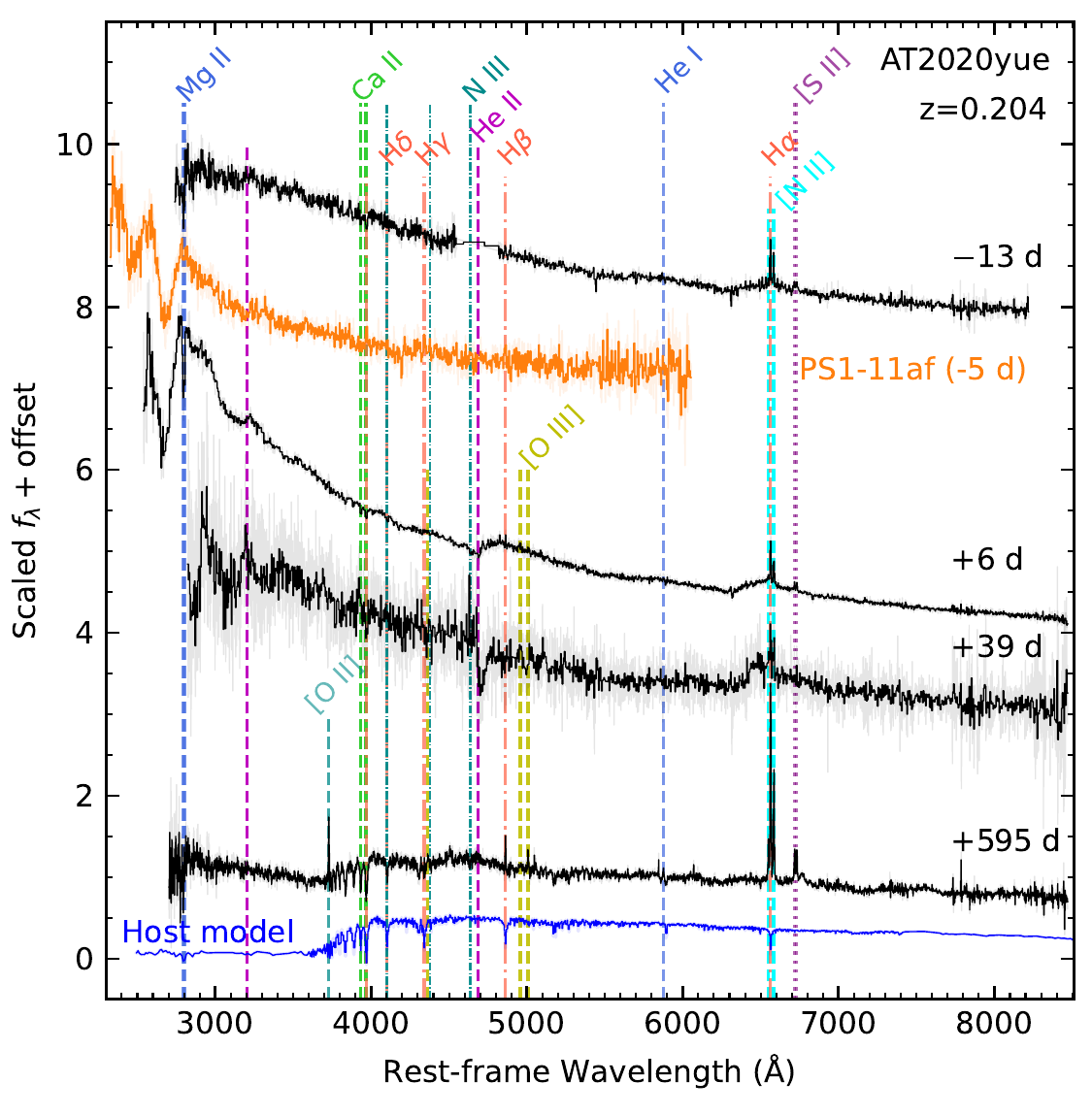}
    \includegraphics[width=\columnwidth]{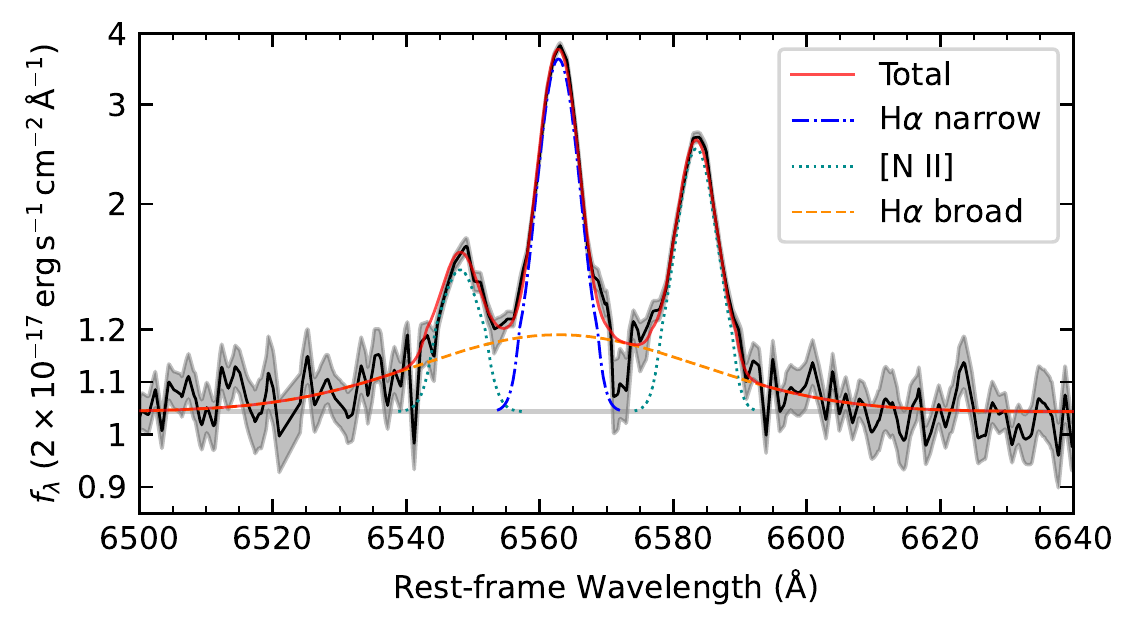}
    \caption{
    UV and optical properties of AT2020yue.
    \textit{Upper}: ZTF and UV light curves of AT2020yue. Detections at $>3\sigma$ are shown with high opacity; other observations are shown in semitransparent.
    \textit{Middle}: optical spectra of AT2020yue. For comparison, we also show the host-subtracted optical spectrum of PS1-11af \citep{Chornock2014}, and the host galaxy model derived in \S\ref{subsubsec:host_sed_fit}.
    \textit{Bottom}: the $+595$\,days spectrum zoomed around H$\alpha$. To highlight the broad H$\alpha$ component, the $y$-axis is shown in linear-scale below 1.2, and in log scale above 1.2.  
    \label{fig:opt_spec_20yue}}
\end{figure}

\textbf{ZTF19aaniqrr/AT2019cmw (ID 7)} was first reported by \citet{Perley2020} as a peculiar transient discovered in the ZTF Bright Transient Survey (BTS; \citealt{Fremling2020, Perley2020}). With an absolute magnitude of $M<-23$\,mag, it was the most luminous event in the BTS sample. Its high luminosity and featureless optical spectra make it similar to events previously classified as TDE-featureless by \citet{Hammerstein2023}. Detailed analysis and modeling of this object will be presented by J. Wise et al. (in preparation).

\textbf{ZTF20acnznms/AT2020yue (ID 19)} was previously classified as a SLSN-II by \citet{Kangas2022}. However, some observed properties of this object favor a TDE interpretation. The upper panel of Figure~\ref{fig:opt_spec_20yue} shows the UV and optical light curves. The color $uvm2-r$ is $1.56\pm 0.19$, $1.47\pm0.22$, and $0.37\pm 0.19$\,mag at $t\approx 14$, 37, and 278\,days, respectively. This indicates a significant increase of temperature from 37 to 278\,days post peak, which is not uncommon in TDEs \citep{Hammerstein2023}, but not observed in SLSNe. 

The middle panel of Figure~\ref{fig:opt_spec_20yue} shows the three optical spectra published in \citet{Kangas2022}, as well as a deep late-time optical spectrum obtained by us in November 2022 using 85\,min of LRIS on-source time (see details in Table~\ref{tab:spec_lowres}).
Broad H$\alpha$ emission is seen in the $-13$, $+6$, and $+39$\,days spectra.
In the +6\,days LRIS spectrum, we clearly identified narrow absorption lines of the \ion{Mg}{II} $\lambda2800$ doublet as well as a broad absorption trough around rest-frame 2660\,\AA, which can be attributed to blueshifted \ion{Mg}{II} absorption. 
Such near-UV features have been observed in both SLSNe \citep{Quimby2011, Chomiuk2011} and the TDE PS1-11af \citep{Chornock2014}.

At $\approx 595$\,d, the transient flux is still detected at $r=22.3\pm0.3$ in the ZTF forced photometry. No broad lines characteristic of SLSN nebular emission (such as [\ion{O}{I}] $\lambda6300$ and [\ion{Ca}{II}] $\lambda 7300$; \citealt{Nicholl2019_SLSN}) are observed. 
The 6500--6640\,\AA\ spectrum can be decomposed into three narrow components (from the host galaxy) and a broader component that originates from the transient (see the bottom panel of Figure~\ref{fig:opt_spec_20yue}). 
The late-time luminosity of the broad H$\alpha$ component is $1.8\times 10^{40}\,{\rm erg\,s^{-1}}$, which is a factor of 5--10 times brighter than that observed in the optically selected TDEs ASASSN-14li and ASASSN-14ae \citep{Brown2017_14li} but similar to the radio-selected TDE VLASS J1008 (J. Somalwar et al. in preparation).
The full-width half-maximum of the transient H$\alpha$ line decreased from $\approx 14000\,{\rm km\,s^{-1}}$ at early time to $\approx 2250 \,{\rm km\,s^{-1}}$ at $\approx 595$\,days. Such a narrowing phenomenon has been observed in a few known TDEs \citep{Brown2017_14li, Onori2019, Nicholl2020} and can be explained by a decrease in the optical depth of the line-emitting region \citep{Roth2018}.

\begin{figure}[htbp!]
    \centering
    \includegraphics[width=\columnwidth]{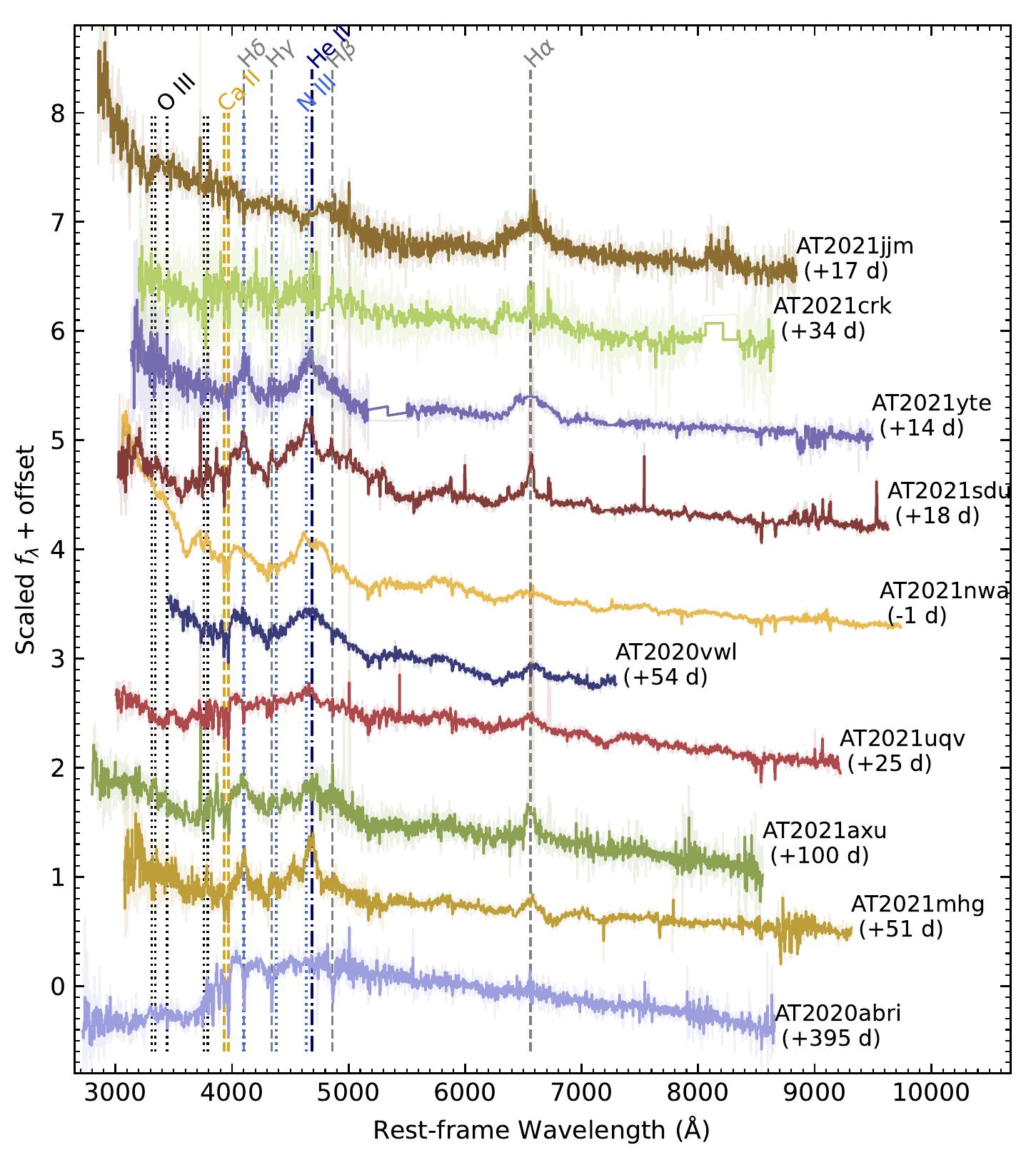}
    \caption{Optical spectra of 10 objects. Strong atmospheric telluric features have been masked. 
    The top 9 objects show broad emission lines characteristic of spectral classes of TDE-H, TDE-H+He, and TDE-He. 
    In a few objects, we have subtracted the blue blackbody continua and masked strong host galaxy narrow emission lines. 
    The bottom spectrum was obtained for the host galaxy of AT2020abri. 
    \label{fig:spec_gallery_H_He}}
\end{figure}

\textbf{ZTF20acvezvs/AT2020abri (ID 20)} has no optical spectrum obtained during the optical flare. 
A post-flare spectrum clearly shows host galaxy absorption lines at $z=0.178$ (see Figure~\ref{fig:spec_gallery_H_He}). 
Following the procedures adopted by \citet{Sazonov2021}, we measure the equivalent width (EW) of the H$\alpha$ emission line and the Lick H$\delta_{\rm A}$ index, resulting in EW(H$\alpha_{\rm em})=3.22$\,\AA, and Lick H$\delta_{\rm A, abs} = 5.52$\,\AA. 
We consider this object to be a probable TDE since (i) its color remains blue ($g-r\approx -0.2$\,mag) for $\sim$200\,days (see Figure~\ref{fig:color_evol}), and the lack of cooling makes it different from most SNe; (ii) the relatively strong H$\delta$ absorption and weak H$\alpha$ emission suggest that the host is a post-starburst galaxy, which is overrepresented in previous samples of TDE host galaxies \citep{French2016, Law-Smith2017, French2020, Hammerstein2021}. 

\textbf{ZTF21aakfqwq/AT2021crk (ID 23)} has a DBSP spectrum obtained during the optical flare, which is not of high signal-to-noise ratio (S/N; see Figure~\ref{fig:spec_gallery_H_He}). A broad emission line at H$\alpha$ is clearly present (with the red wing slightly affected by telluric absorptions), while the \ion{He}{II} wavelength region is affected by the DBSP CCD malfunction. Therefore, we tentatively assign a spectral subtype of TDE-H+He? for this object.

\begin{figure}[htbp!]
\centering
    \includegraphics[width=\columnwidth]{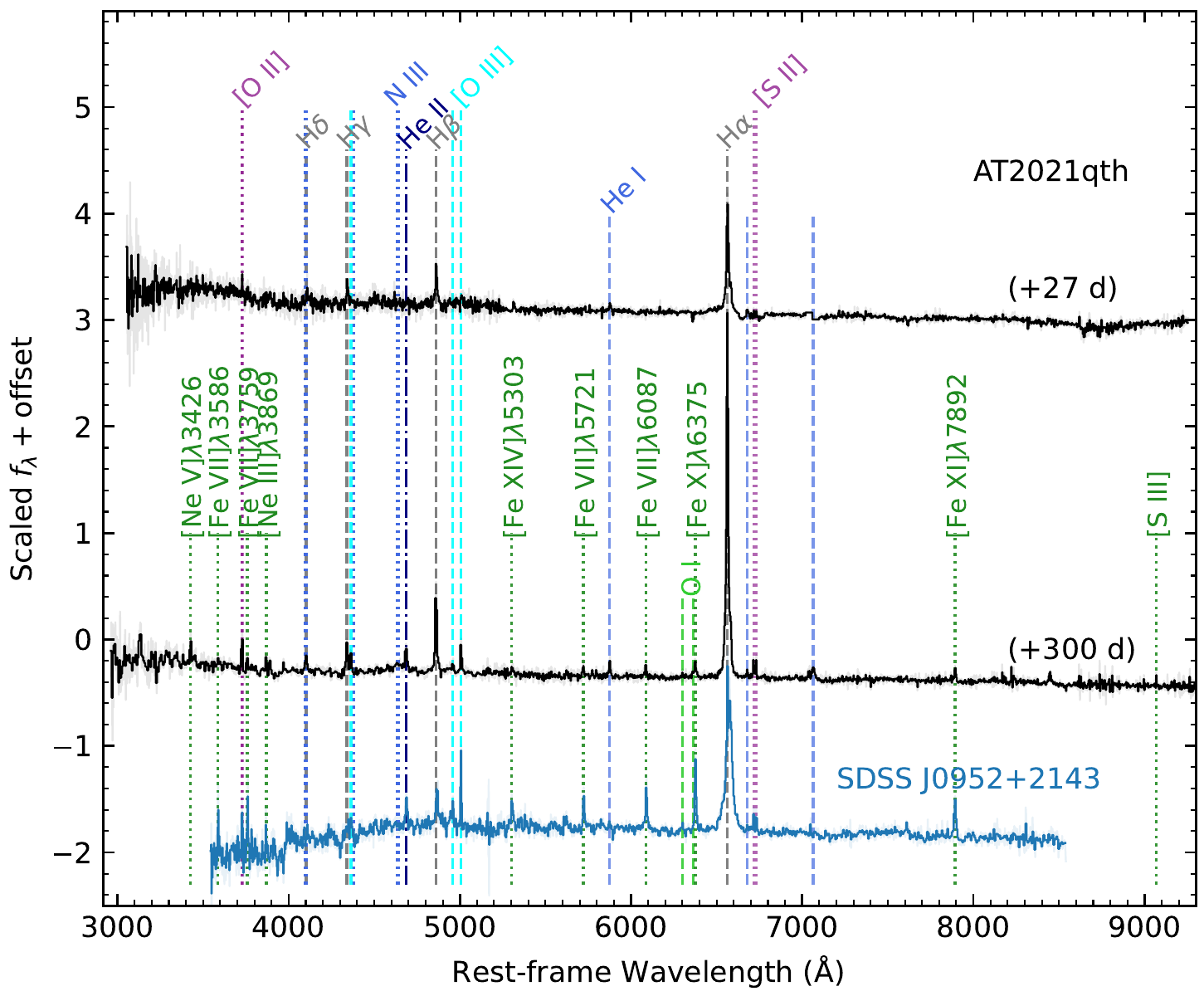}
    \caption{Optical spectra of AT2021qth, compared with the SDSS spectrum of the prototype extreme coronal line emitter SDSS\,J0952+2143 \citep{Komossa2008, Palaversa2016}.
    \label{fig:spec_21qth}}
\end{figure}

\textbf{ZTF21abhrchb/AT2021qth (ID 28)} was missed by real-time selection with optical surveys, but was later revealed to be a TDE based on an X-ray detection at $L_{\rm X}\sim 6\times 10^{42}\,{\rm erg\,s^{-1}}$ from \srg/eROSITA (private communication). X-ray data of this object will be presented as part of a sample of \srg-selected TDEs with strong optical flares by M. Gilfanov et al. (in preparation). Such a high X-ray luminosity is not theoretically expected in interaction-powered SNe (see Fig.~3 of \citealt{Margalit2022}), and $>\times 10$ brighter than the peak of the most X-ray luminous known SN IIn (see, e.g., Fig.~7 of \citealt{Yao2022_20mrf}). 
Figure~\ref{fig:spec_21qth} shows that its late-time optical spectrum exhibits highly ionized narrow emission lines of [\ion{Ne}{III}], [\ion{Ne}{V}], [\ion{Fe}{VII}], [\ion{Fe}{X}], [\ion{Fe}{XI}], and [\ion{Fe}{XIV}] --- reminiscent of the known class of extreme coronal line emitters \citep{Komossa2008, Somalwar2022}. 

\begin{figure}[htbp!]
    \centering
    \includegraphics[width=\columnwidth]{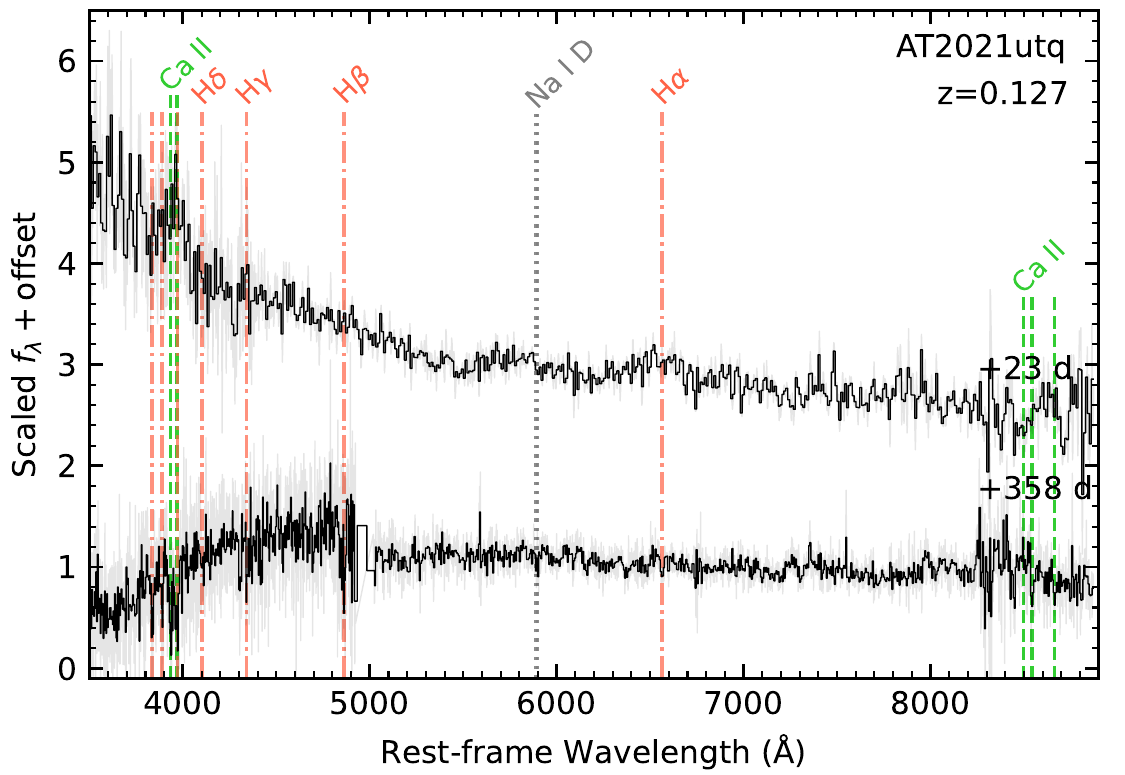}
    \caption{Optical spectra of AT2021utq. The late-time DBSP spectrum shows host galaxy absorption lines (\ion{Na}{I}, \ion{Ca}{II}, Balmer series) at $z=0.127$. At this redshift, the early-time spectrum reported by \citet{Burke2021} exihibits a broad emission line around H$\alpha$, making it consistent with the TDE-H spectral class. \label{fig:opt_spec_21utq}}
\end{figure}

\textbf{ZTF21abqtckk/AT2021utq (ID 31)} was previously classified as a variable star on TNS based on the fact that its parallax was reported by \gaia Data Release 2 (DR2) and that the distance was estimated by \citet{Bailer-Jones2018} to be $\sim1$\,kpc \citep{Burke2021}. 
However, both the \gaia parallax ($\varpi = -0.91\pm1.51$\,mas) and the distance estimate ($1.16_{-0.50}^{+0.81}$\,kpc) have large uncertainties.
Moreover, a post-flare optical spectrum reveals host galaxy absorption lines at $z=0.127$ (see Figure~\ref{fig:opt_spec_21utq}). At this redshift, the TNS spectrum exhibits a board emission line at H$\alpha$, suggesting a spectral class of TDE-H.

\begin{figure}[htbp!]
    \centering
    \includegraphics[width=\columnwidth]{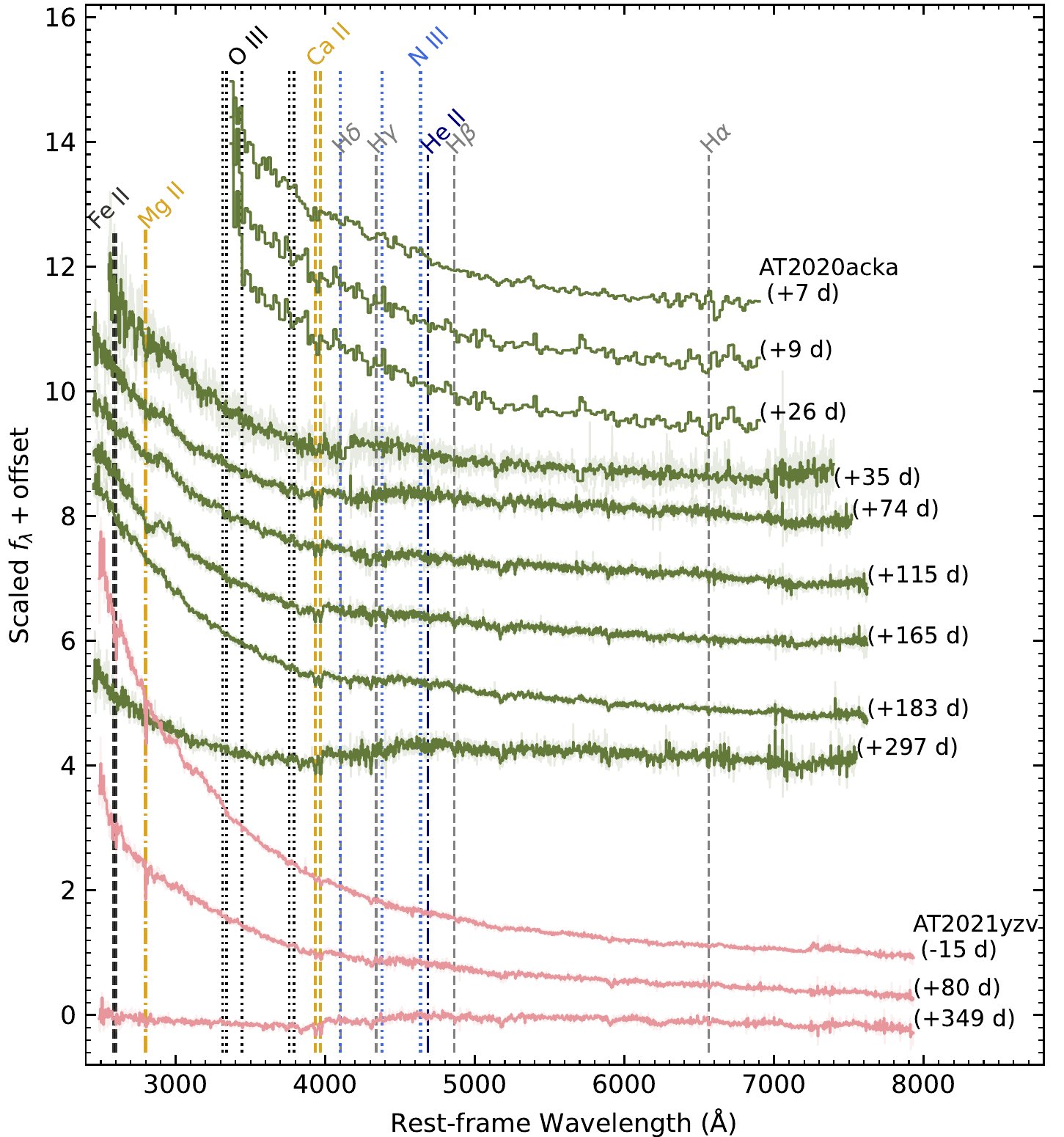}
    \caption{Optical spectra of two TDEs that belong to the TDE-featureless subclass. 
    We highlight the fact that the rest-frame H$\alpha$ region is covered by our spectra, and no discernible emission lines are present throughout the spectral evolution.
    \label{fig:spec_gallery_featureless}}
\end{figure}

The TDE classifications of the remaining 10 objects have been previously reported to TNS by the ZTF group. Their optical spectra are shown in Figure~\ref{fig:spec_gallery_H_He} for objects with broad emission lines, and in  Figure~\ref{fig:spec_gallery_featureless} for two objects in the TDE-featureless spectral class.  

We note that, although TDEs can evolve and change spectroscopic subtypes \citep{Nicholl2019, vanVelzen2020, Charalampopoulos2022_spec}, a precise labeling of the subtype is not important for this work.

\section{Light Curve Characterization} \label{sec:lc_fitting}

In this section, we aim to systematically estimate the peak-light properties and light curve evolution timescales of the 33 TDEs. 
We outline the procedures of the fitting routine in \S\ref{subsec:fitting_routine}, 
describe the choice of the light curve model in \S\ref{subsec:choose_lcmod}, 
and summarize the results in \S\ref{subsec:lc_result}.

\subsection{The Fitting Routine} \label{subsec:fitting_routine}
Model fitting was performed using the Markov Chain Monte Carlo (MCMC) approach with the \texttt{emcee} sampler \citep{Foreman-Mackey2013}. For each TDE (at redshift $z$) and each observation $i$, the input data are $t_i$ (rest-frame days relative to the visually determined light curve maximum), $L_i$, $\sigma_{i}$ (Galactic extinction corrected luminosity and its uncertainty in the observed band), and $\nu_i$ (rest-frame effective frequency of the observed band). 
We assume negligible host galaxy extinction. 

Following \citet{Yao2019}, we add a constant additional variance $\sigma_0^2$ to each of the measurement variance $\sigma_i^2$ to account for systematic uncertainties.
We use 100 walkers and $N$ steps, where $N$ is typically 1000--3000. We visually inspect the walker values as a function of step to ensure convergence. The posterior distribution is obtained after discarding the first $N-500$ steps.

\subsection{The Light Curve Model} \label{subsec:choose_lcmod}

\subsubsection{The SED Shape}
It has been shown that the UV and optical emission of TDEs can be described with a thermal blackbody \citep{Gezari2021}. 
Therefore, we assume that the UV and optical spectrum follows a blackbody $B_\nu(T_{\rm bb})$.
Our goal is to determine the blackbody parameters (temperature $T_{\rm bb}$, radius $R_{\rm bb}$, and luminosity $L_{\rm bb}$) at maximum light. 

Since the majority of known TDEs show little temperature evolution \citep{vanVelzen2020}, we assume the temperature is fixed to that near peak.
However, this assumption is not appropriate for a few TDEs in our sample 
(IDs 
2,
5,
7,
8,
10,
13,
14,
18,
21,
24,
27) 
with significant post-peak $uvw2-r$ color change. 
Since our goal is to constrain the peak-light blackbody parameters, we excluded late-time UVOT data for these objects.\footnote{We removed UVOT data at $t\gtrsim t_{\rm c/h}$, where $t_{\rm c/h}$ is the time when clear evidence of post-peak cooling or heating is observed. We chose $t_{\rm c/h} \in (5, 100)$\,days for each of the 10 objects by visually inspecting their multi-band light curves.} 

\subsubsection{The Rise Function} \label{subsubsec:choose_rise}

Following \citet{vanVelzen2021}, we first model the light curve at $t<100$\,days with a Gaussian rise and an exponential decay: 
\begin{subequations}
\begin{align}
    L_\nu(t) &= A_\nu \times
    \begin{cases}
    e^{(t-t_{\rm peak})^2 / (2 \sigma_{\rm rise}^2)} & t\leq t_{\rm peak}\\
    e^{-(t - t_{\rm peak})/ \tau_{\rm decay}}& t> t_{\rm peak}
    \end{cases} \label{eq:model1_decay} \\
     A_\nu  & = L_{\nu_0\,{\rm peak}} \frac{B_\nu(T_0)}{B_{\nu_0} (T_0)} 
\end{align} \label{eq:model1}
\end{subequations}
Here, $L_{\nu_0\,{\rm peak}}$ is the rest-frame $g$-band ($\nu_0 = 6.3\times 10^{14}$\,Hz) peak luminosity, and $t_{\rm peak}$ is the epoch of rest-frame $g$-band maximum. 

\begin{figure}[htbp!]
    \centering
    \includegraphics[width = \columnwidth]{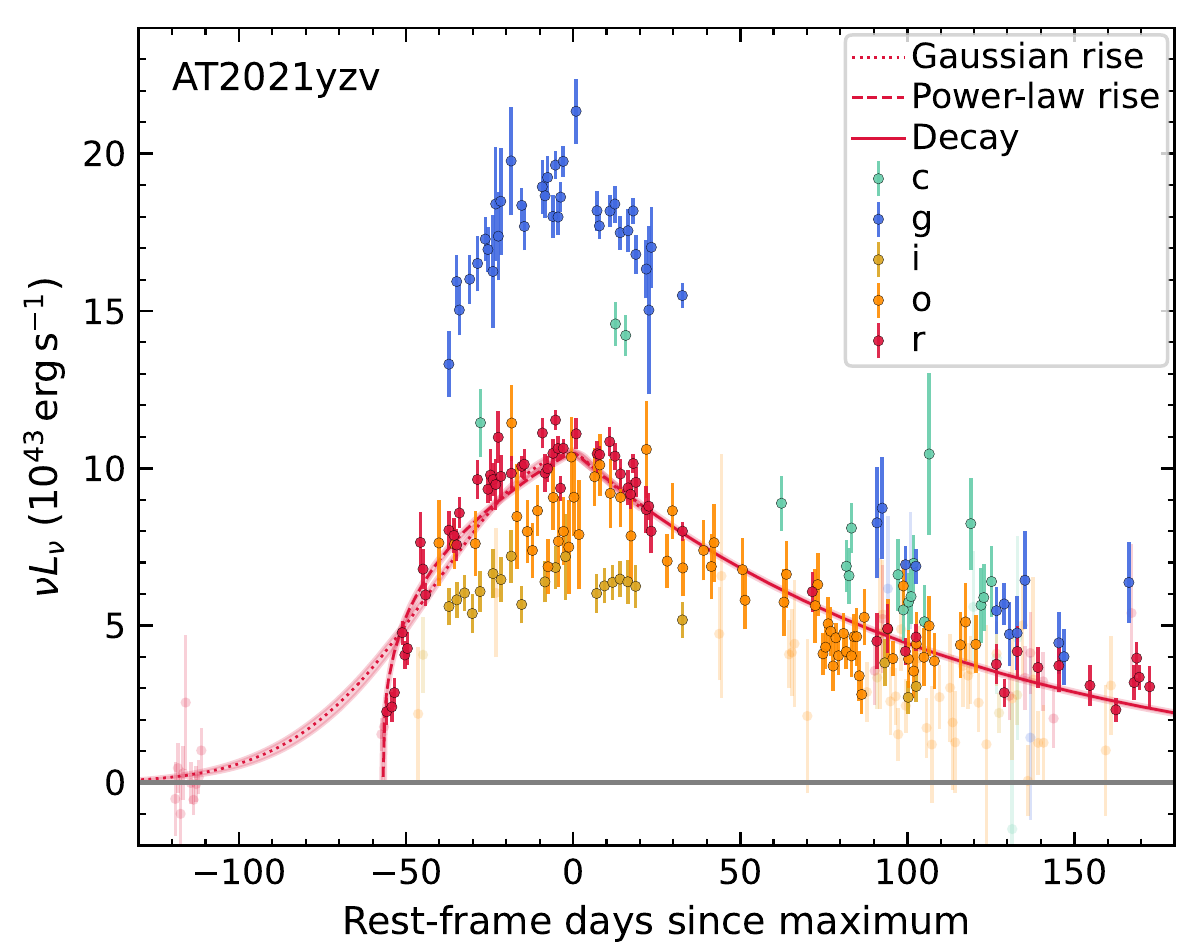}
    \caption{ZTF and ATLAS light curves of AT2021yzv, overplotted with the best-fit models in the ZTF $r$ band. Detections at $>4\sigma$ are shown with high opacity. 
    A power-law function provides a better description for the rise profile. 
    \label{fig:why_plrise}}
\end{figure}

\begin{deluxetable*}{rlccccccrrrlc}[htbp!]
\tabletypesize{\scriptsize}
    \tablecaption{Light Curve Properties and Survey Efficiencies.\label{tab:lc_pars}}
	\tablehead{
	\colhead{ID}
	& \colhead{IAU Name} 
    & \colhead{Model} 
	& \colhead{$t_{\rm peak}$}
	& \colhead{log$T_{\rm bb}$} 
	& \colhead{log$L_g$} 
	& \colhead{log$L_{\rm bb}$} 
	& \colhead{log$R_{\rm bb}$} 
	& \colhead{$t_{1/2, {\rm rise}}$}
	& \colhead{$t_{1/2, {\rm decline}}$}
	& \colhead{$D_{\rm max, t}$} 
	& \colhead{$z_{\rm max, t}$} 
	& \colhead{$f_{\rm loss}$} \\
	\colhead{}
	& \colhead{}
    & \colhead{}
	& \colhead{(MJD)}
	& \colhead{(K)}
	& \colhead{($\rm erg\,s^{-1}$)}
	& \colhead{($\rm erg\,s^{-1}$)}
	& \colhead{($\rm cm$)}
	& \colhead{(days)}
	& \colhead{(days)}
	& \colhead{(Mpc)}
	& \colhead{}
	& \colhead{}
    }
\startdata
1 & AT2018iih & \texttt{r2+d2} & $58451.13_{-2.20}^{+2.78}$ & 4.22 & 44.11 & 44.59 & 15.43 & $31.0_{-1.5}^{+2.5}$ & $86.5_{-5.0}^{+3.3}$ & 1501 & 0.291 & 0.525 \\
2 & AT2018jbv & \texttt{r1+d2} & $58470.36_{-0.00}^{+0.00}$ & 4.50 & 44.23 & 45.33 & 15.24 & $34.4_{-1.4}^{+2.1}$ & $65.9_{-1.7}^{+2.3}$ & 2052 & 0.381 & 0.328 \\
3 & AT2018lna & \texttt{r1+d1} & $58507.31_{-0.95}^{+1.20}$ & 4.49 & 43.21 & 44.27 & 14.73 & $15.5_{-1.0}^{+1.3}$ & $30.2_{-1.1}^{+1.3}$ & 488 & 0.106 & 0.241 \\
4 & AT2019baf & \texttt{r2+d6} & $58514.16_{-0.78}^{+0.82}$ & 4.10 & 43.52 & 43.81 & 15.28 & $23.2_{-1.0}^{+0.9}$ & $27.6_{-0.9}^{+0.6}$ & 668 & 0.141 & 0.475 \\
5 & AT2019azh & \texttt{r2+d2} & $58561.39_{-0.77}^{+1.05}$ & 4.46 & 43.30 & 44.31 & 14.80 & $24.7_{-1.0}^{+1.3}$ & $44.1_{-0.9}^{+1.1}$ & 547 & 0.118 & 0.652 \\
6 & AT2019bhf & \texttt{r1+d2} & $58544.78_{-1.34}^{+1.10}$ & 4.14 & 43.46 & 43.81 & 15.20 & $9.9_{-0.9}^{+0.7}$ & $29.1_{-1.4}^{+1.9}$ & 630 & 0.134 & 0.207 \\
7 & AT2019cmw & \texttt{r2+d2} & $58588.82_{-0.00}^{+0.00}$ & 4.34 & 44.68 & 45.41 & 15.60 & $14.0_{-0.3}^{+0.3}$ & $28.9_{-0.5}^{+0.7}$ & 3714 & 0.626 & 0.288 \\
8 & AT2019dsg & \texttt{r1+d1} & $58606.97_{-3.22}^{+3.51}$ & 4.41 & 43.18 & 44.05 & 14.79 & $19.7_{-2.0}^{+2.3}$ & $43.1_{-1.1}^{+1.0}$ & 465 & 0.101 & 0.526 \\
9 & AT2019ehz & \texttt{r2+d6} & $58618.69_{-0.51}^{+0.70}$ & 4.29 & 43.28 & 43.90 & 14.94 & $15.7_{-0.8}^{+0.7}$ & $28.0_{-1.0}^{+0.0}$ & 521 & 0.112 & 0.380 \\
10 & AT2019qiz & \texttt{r1+d4} & $58766.50_{-0.26}^{+0.25}$ & 4.23 & 42.90 & 43.40 & 14.81 & $11.6_{-0.3}^{+0.3}$ & $17.9_{-0.8}^{+0.7}$ & 322 & 0.0714 & 0.545 \\
11 & AT2019vcb & \texttt{r1+d1} & $58819.83_{-0.89}^{+1.08}$ & 4.11 & 43.35 & 43.65 & 15.19 & $13.6_{-0.8}^{+1.1}$ & $24.6_{-0.4}^{+0.4}$ & 546 & 0.117 & 0.309 \\
12 & AT2020pj & \texttt{r1+d2} & $58866.42_{-0.55}^{+0.58}$ & 4.10 & 42.95 & 43.24 & 14.99 & $12.4_{-0.5}^{+0.7}$ & $17.2_{-1.1}^{+1.3}$ & 335 & 0.0742 & 0.158 \\
13 & AT2020mot & \texttt{r1+d4} & $59082.04_{-1.30}^{+1.24}$ & 4.29 & 43.22 & 43.84 & 14.92 & $42.6_{-1.6}^{+1.3}$ & $46.1_{-2.1}^{+1.9}$ & 485 & 0.105 & 0.515 \\
14 & AT2020neh & \texttt{r1+d1} & $59030.93_{-0.39}^{+0.53}$ & 4.19 & 43.26 & 43.70 & 15.04 & $6.4_{-0.4}^{+0.4}$ & $16.4_{-0.6}^{+0.6}$ & 501 & 0.108 & 0.269 \\
15 & AT2020ysg & \texttt{r1+d2} & $59094.32_{-3.03}^{+3.30}$ & 4.37 & 44.24 & 45.04 & 15.35 & $24.0_{-1.5}^{+2.1}$ & $72.5_{-3.3}^{+2.1}$ & 1963 & 0.367 & 0.463 \\
16 & AT2020vdq & \texttt{r1+d2} & $59113.09_{-0.93}^{+1.00}$ & 4.16 & 42.62 & 42.99 & 14.76 & $11.9_{-1.3}^{+1.7}$ & $23.3_{-1.7}^{+1.5}$ & 227 & 0.0511 & 0.210 \\
\hline
17 & AT2020vwl & \texttt{r1+d4} & $59166.88_{-1.14}^{+1.17}$ & 4.30 & 43.13 & 43.77 & 14.86 & $22.2_{-0.7}^{+0.8}$ & $27.4_{-1.7}^{+1.9}$ & 515 & 0.111 & 0.623 \\
18 & AT2020wey & \texttt{r1+d5} & $59155.84_{-0.20}^{+0.19}$ & 4.32 & 42.47 & 43.15 & 14.51 & $13.9_{-0.4}^{+0.4}$ & $5.2_{-0.2}^{+0.2}$ & 228 & 0.0514 & 0.302 \\
19 & AT2020yue & \texttt{r1+d4} & $59179.44_{-1.12}^{+1.25}$ & 4.06 & 44.00 & 44.24 & 15.57 & $19.5_{-0.9}^{+1.0}$ & $62.8_{-1.9}^{+2.0}$ & 1399 & 0.274 & 0.465 \\
20 & AT2020abri & \texttt{r2+d3} & $59208.56_{-0.80}^{+0.83}$ & 4.10 & 43.66 & 43.95 & 15.35 & $16.7_{-0.9}^{+1.2}$ & $31.7_{-0.8}^{+0.7}$ & 948 & 0.194 & 0.261 \\
21 & AT2020acka & \texttt{r1+d5} & $59217.15_{-1.14}^{+1.38}$ & 4.45 & 44.47 & 45.44 & 15.39 & $26.9_{-1.8}^{+1.6}$ & $28.8_{-0.5}^{+0.7}$ & 3629 & 0.614 & 0.514 \\
22 & AT2021axu & \texttt{r1+d2} & $59252.50_{-0.50}^{+0.55}$ & 4.58 & 43.75 & 45.05 & 14.93 & $23.9_{-0.6}^{+0.5}$ & $33.4_{-1.0}^{+0.9}$ & 1253 & 0.249 & 0.368 \\
23 & AT2021crk & \texttt{r1+d2} & $59273.90_{-0.52}^{+0.53}$ & 4.30 & 43.50 & 44.14 & 15.05 & $10.2_{-0.4}^{+0.7}$ & $20.9_{-1.1}^{+1.1}$ & 831 & 0.173 & 0.216 \\
24 & AT2021ehb & \texttt{r1+d3} & $59314.51_{-1.90}^{+2.78}$ & 4.44 & 42.58 & 43.54 & 14.46 & $23.7_{-1.4}^{+1.9}$ & $50.5_{-3.8}^{+3.6}$ & 265 & 0.0593 & 0.661 \\
25 & AT2021jjm & \texttt{r1+d1} & $59327.68_{-0.93}^{+0.99}$ & 4.17 & 43.59 & 43.99 & 15.23 & $9.1_{-0.7}^{+0.7}$ & $29.1_{-1.7}^{+2.6}$ & 893 & 0.184 & 0.304 \\
26 & AT2021mhg & \texttt{r1+d4} & $59370.28_{-0.85}^{+0.89}$ & 4.49 & 43.22 & 44.28 & 14.74 & $17.2_{-0.7}^{+0.7}$ & $14.7_{-1.0}^{+1.1}$ & 595 & 0.127 & 0.399 \\
27 & AT2021nwa & \texttt{r1+d3} & $59402.51_{-0.68}^{+0.64}$ & 4.51 & 42.68 & 43.81 & 14.45 & $27.1_{-0.8}^{+0.6}$ & $76.2_{-1.6}^{+1.9}$ & 301 & 0.0669 & 0.483 \\
28 & AT2021qth & \texttt{r2+d4} & $59401.88_{-1.26}^{+1.26}$ & 3.96 & 43.14 & 43.30 & 15.30 & $15.8_{-1.3}^{+1.2}$ & $39.1_{-2.0}^{+1.3}$ & 481 & 0.104 & 0.374 \\
29 & AT2021sdu & \texttt{r1+d3} & $59419.36_{-0.36}^{+0.33}$ & 4.30 & 43.09 & 43.73 & 14.84 & $12.2_{-0.4}^{+0.4}$ & $11.0_{-0.4}^{+0.3}$ & 488 & 0.106 & 0.340 \\
30 & AT2021uqv & \texttt{r1+d5} & $59446.39_{-0.63}^{+0.66}$ & 4.29 & 43.15 & 43.77 & 14.87 & $14.9_{-0.7}^{+0.7}$ & $36.0_{-2.0}^{+2.2}$ & 525 & 0.113 & 0.251 \\
31 & AT2021utq & \texttt{r1+d6} & $59457.51_{-0.85}^{+0.83}$ & 4.39 & 43.39 & 44.22 & 14.91 & $14.6_{-0.6}^{+0.6}$ & $43.4_{-4.3}^{+5.8}$ & 736 & 0.155 & 0.390 \\
32 & AT2021yzv & \texttt{r2+d2} & $59511.50_{-1.38}^{+1.35}$ & 4.43 & 44.07 & 45.01 & 15.21 & $51.8_{-1.2}^{+1.4}$ & $69.9_{-2.6}^{+2.6}$ & 1920 & 0.360 & 0.456 \\
33 & AT2021yte & \texttt{r1+d3} & $59484.99_{-0.60}^{+0.59}$ & 4.29 & 42.90 & 43.52 & 14.75 & $18.4_{-0.6}^{+0.5}$ & $23.7_{-0.7}^{+0.7}$ & 385 & 0.0847 & 0.413 \\
\enddata
\tablecomments{Column 3 indicates the light curve rise and decline functional forms of the adopted model. 
\texttt{r1}: Gaussian rise, 
\texttt{r2}: power-law rise. See \S\ref{subsubsec:choose_decay} for the meaning of the six decline models. 
Columns 4--10 are light curve properties (see \S\ref{subsec:lc_result} for definitions). 
Columns 11--13 are parameters relevant to the survey efficiencies (see \S\ref{sec:method} for definitions).
}
\end{deluxetable*}

\begin{figure*}
    \centering
    \includegraphics[width=\textwidth]{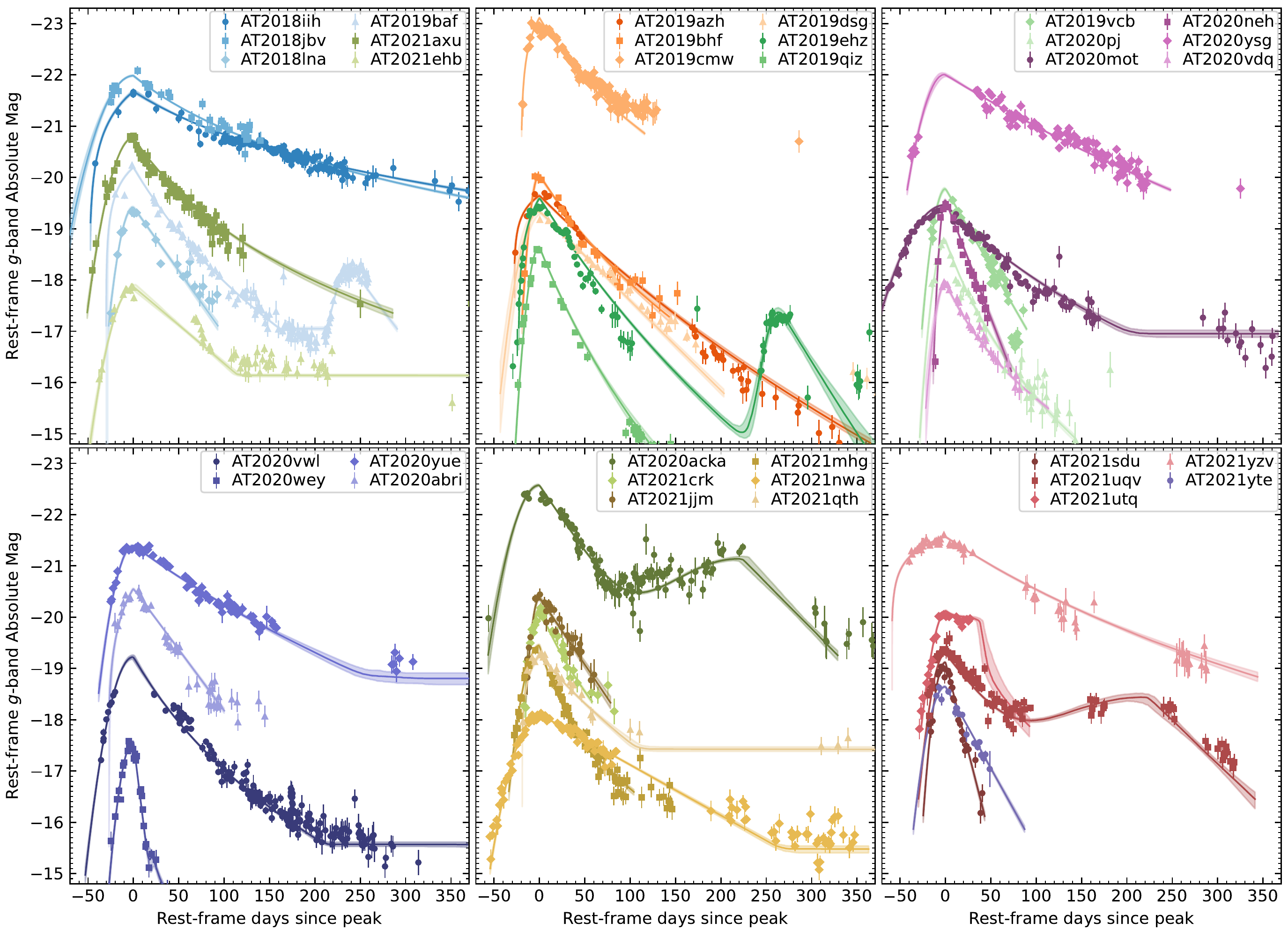}
    \caption{Rest-frame $g$-band light curves of the 33 TDEs in our sample. 
    The solid lines show the best-fit models.
    \label{fig:sample_lcs}}
\end{figure*}

A Gaussian function is generally a good model when the data sampling is sparse on the rise, since it reduces the model complexity by imposing strong assumptions on the shape of the light curve profile. However, it cannot describe a rise where the flux increase rate decreases as a function of time (e.g., see Figure~\ref{fig:why_plrise}). Therefore, for objects with good sampling on the rise\footnote{Here, good sampling is defined as follows. For each object, we select data within [$t_{\rm peak}-2\sigma_{\rm rise}$, $t_{\rm peak}+\sigma_{\rm rise}$], where $t_{\rm peak}$ and $\sigma_{\rm rise}$ are best-fit model parameters from Eq.~(\ref{eq:model1}). We require that the maximum time separation in consecutive pairs of observations is less than $\sigma_{\rm rise}$. }, we also fit the rise with a power-law function:
\begin{align}
    L_\nu(t) = A_\nu\times
    \begin{cases}
       0 & t \leq t_{\rm fl}\\
      \frac{(t - t_{\rm fl})^{n} }{ (t_{\rm peak} - t_{\rm fl})^n} & t_{\rm fl} < t\leq t_{\rm peak}\\
    \end{cases}
\end{align}
where $t_{\rm fl}$ is the first-light epoch, and $n$ is the rise power-law index. We consider the power-law rise model to be superior to the Gaussian rise model if the best-fit $\sigma_0$ is smaller, and the 68\% confidence region of $n$ is $<0.5$. 
The adopted rise function for each TDE is given in the ``Model'' column of Table~\ref{tab:lc_pars}.

\subsubsection{The Decline Function} \label{subsubsec:choose_decay}
Having decided on the rise function, we fit the light curve within $t<365$\,d with six types of decline functions:
\begin{enumerate}
    \item an exponential decline (model \texttt{d1}; Eq.~\ref{eq:model1_decay}),
    \item a power-law decline (model \texttt{d2}):
    \begin{align}
        L_\nu(t) & = A_\nu  \left( \frac{t - t_{\rm peak} + t_0}{t_0}\right)^p\;\;\; t> t_{\rm peak},
    \end{align}
    \item an exponential decline followed by a late-time plateau (model \texttt{d3}),
    \item a power-law decline followed by a late-time plateau (model \texttt{d4}),
    \item an exponential decline with a secondary peak on top of that (model \texttt{d5}),
    \item a power-law decline with a secondary peak on top of that (model \texttt{d6}).
\end{enumerate}
In functions \texttt{d5} and \texttt{d6}, we assume that the secondary peak has a Gaussian rise and an exponential decline. 
We compare the Bayesian information criterion (BIC) of the six model fits and choose the one with the smallest value of BIC.
The adopted decline function for each TDE is given in the `Model' column of Table~\ref{tab:lc_pars}.

\subsection{The Fitting Results} \label{subsec:lc_result}

Figure~\ref{fig:sample_lcs} shows the fitting results. 
The light curve properties obtained with the best-fit models are provided in Table~\ref{tab:lc_pars}, where
$t_{\rm peak}$ is the peak-light epoch,
$T_{\rm bb}$, $L_{\rm bb}$, and $R_{\rm bb}$ are the blackbody parameters at peak;
$L_{g}$ is the rest-frame $g$-band luminosity at peak (corrected for Galactic extinction).
Following conventions of transient studies \citep{Yao2022_20mrf, Ho2023_fbot_sample}, we characterize the light curve evolution speed by calculating the rest-frame duration it takes for a TDE to rise from half-max to max ($t_{1/2, {\rm rise}}$) and to decline from max to half-max ($t_{1/2, {\rm decline}}$).
The rest-frame duration above half-max light is $t_{1/2}\equiv t_{1/2, {\rm rise}} + t_{1/2, {\rm decline}}$.

\section{Host Galaxy Analysis} \label{sec:host}
\subsection{Observation}
\subsubsection{Photometry}

\begin{deluxetable*}{rlrrrrrccccc}[htbp!]
\tabletypesize{\scriptsize}
\tablecaption{Host Galaxy Properties. \label{tab:host_pars}}
\tablehead{
	\colhead{ID}
    & \colhead{IAU name}
    & \colhead{log$M_{\rm gal}$}  
    & \colhead{$^{0,0}u-r$}  
    & \colhead{$\tau_{\rm SFH}$}
    & \colhead{$t_{\rm age}$}
    & \colhead{log$Z$}
    & \colhead{$E(B-V)_{\rm h}$} 
    & \colhead{log$M_{\rm BH}$} 
    & \colhead{$\sigma_\ast$}
    & \colhead{$r_{1/2}$} 
    & \colhead{$z_{\rm max, h}$} \\
    \colhead{}
    & \colhead{}
	& \colhead{($M_\odot$)}
    & \colhead{(mag)}
    & \colhead{(Gyr)}
    & \colhead{(Gyr)}
    & \colhead{($Z_\odot$)}
	& \colhead{(mag)}
	& \colhead{($M_\odot$)}
	& \colhead{(km\,s$^{-1}$)}
    & \colhead{($^{\prime\prime}$)}
    & \colhead{}
	}
\startdata
1 & AT2018iih &
$10.69_{-0.16}^{+0.12}$ &
$2.17_{-0.13}^{+0.09}$ &
$0.33_{-0.19}^{+0.54}$ &
$8.59_{-3.63}^{+2.81}$ &
$-1.02_{-0.65}^{+0.43}$ &
$0.13_{-0.09}^{+0.10}$ &
$7.93\pm 0.35$ & $148.64 \pm 14.42$ & 1.5 & 0.60 \\
2 & AT2018jbv &
$10.20_{-0.19}^{+0.17}$ &
$1.98_{-0.19}^{+0.18}$ &
$0.71_{-0.50}^{+1.29}$ &
$7.87_{-3.73}^{+3.38}$ &
$-1.27_{-0.51}^{+0.61}$ &
$0.15_{-0.09}^{+0.08}$ &
$6.77\pm 0.40$ & --- & 1.0 & 0.52 \\
3 & AT2018lna &
$9.50_{-0.17}^{+0.12}$ &
$1.84_{-0.19}^{+0.11}$ &
$0.37_{-0.22}^{+0.60}$ &
$8.33_{-3.29}^{+2.66}$ &
$-1.43_{-0.39}^{+0.43}$ &
$0.06_{-0.04}^{+0.04}$ &
$5.56\pm 0.51$ & --- & 1.4 & 0.26 \\
4 & AT2019baf &
$10.27_{-0.05}^{+0.04}$ &
$1.75_{-0.04}^{+0.05}$ &
$3.23_{-0.95}^{+0.76}$ &
$10.57_{-2.40}^{+1.35}$ &
$-0.54_{-0.39}^{+0.27}$ &
$0.17_{-0.04}^{+0.04}$ &
$6.89\pm 0.24$ & --- & 1.8 & 0.43 \\
5 & AT2019azh &
$9.88_{-0.03}^{+0.03}$ &
$1.76_{-0.01}^{+0.01}$ &
$0.29_{-0.04}^{+0.05}$ &
$2.26_{-0.24}^{+0.28}$ &
$-0.63_{-0.10}^{+0.10}$ &
$0.06_{-0.01}^{+0.01}$ &
$6.44\pm 0.33$ & $67.99 \pm 2.03$ & 4.0 & 0.41 \\
6 & AT2019bhf &
$10.39_{-0.06}^{+0.05}$ &
$1.96_{-0.04}^{+0.04}$ &
$1.74_{-0.47}^{+0.35}$ &
$10.45_{-2.29}^{+1.49}$ &
$-0.95_{-0.44}^{+0.42}$ &
$0.12_{-0.05}^{+0.05}$ &
$7.10\pm 0.24$ & --- & 1.7 & 0.45 \\
7 & AT2019cmw &
$10.88_{-0.20}^{+0.17}$ &
$2.22_{-0.24}^{+0.12}$ &
$0.40_{-0.23}^{+1.00}$ &
$7.40_{-3.61}^{+3.39}$ &
$-0.74_{-0.85}^{+0.60}$ &
$0.16_{-0.10}^{+0.09}$ &
$7.94\pm 0.42$ & --- & 1.0 & 0.63 \\
8 & AT2019dsg &
$10.34_{-0.05}^{+0.06}$ &
$2.12_{-0.04}^{+0.04}$ &
$0.49_{-0.09}^{+0.13}$ &
$4.30_{-0.69}^{+0.96}$ &
$0.11_{-0.07}^{+0.05}$ &
$0.01_{-0.01}^{+0.02}$ &
$6.90\pm 0.32$ & $86.89 \pm 3.92$ & 2.5 & 0.42 \\
9 & AT2019ehz &
$9.65_{-0.16}^{+0.13}$ &
$1.93_{-0.04}^{+0.05}$ &
$0.76_{-0.58}^{+0.67}$ &
$6.08_{-3.05}^{+4.18}$ &
$-1.36_{-0.46}^{+0.53}$ &
$0.13_{-0.06}^{+0.04}$ &
$5.81\pm 0.46$ & --- & 1.7 & 0.32 \\
10 & AT2019qiz &
$10.28_{-0.06}^{+0.04}$ &
$2.36_{-0.06}^{+0.04}$ &
$0.26_{-0.13}^{+0.34}$ &
$10.95_{-1.88}^{+1.16}$ &
$-0.41_{-0.18}^{+0.14}$ &
$0.03_{-0.02}^{+0.03}$ &
$6.48\pm 0.33$ & $69.70 \pm 2.30$ & 9.9 & 0.27 \\
11 & AT2019vcb &
$9.77_{-0.07}^{+0.03}$ &
$1.54_{-0.03}^{+0.02}$ &
$3.00_{-0.84}^{+0.57}$ &
$10.46_{-2.48}^{+1.50}$ &
$-0.95_{-0.22}^{+0.23}$ &
$0.10_{-0.02}^{+0.02}$ &
$6.03\pm 0.36$ & --- & 1.2 & 0.44 \\
12 & AT2020pj &
$10.01_{-0.08}^{+0.07}$ &
$2.01_{-0.05}^{+0.07}$ &
$1.43_{-0.88}^{+0.47}$ &
$9.28_{-3.84}^{+2.32}$ &
$-1.35_{-0.34}^{+0.53}$ &
$0.17_{-0.05}^{+0.03}$ &
$6.44\pm 0.31$ & --- & 1.7 & 0.35 \\
13 & AT2020mot &
$10.40_{-0.08}^{+0.06}$ &
$2.20_{-0.05}^{+0.05}$ &
$1.18_{-0.50}^{+0.35}$ &
$9.52_{-2.65}^{+2.09}$ &
$-0.73_{-0.38}^{+0.32}$ &
$0.12_{-0.05}^{+0.05}$ &
$6.66\pm 0.34$ & $76.61 \pm 5.33$ & 1.4 & 0.49 \\
14 & AT2020neh &
$9.80_{-0.06}^{+0.05}$ &
$1.49_{-0.03}^{+0.03}$ &
$3.25_{-0.94}^{+0.71}$ &
$10.41_{-2.36}^{+1.46}$ &
$-1.19_{-0.24}^{+0.26}$ &
$0.12_{-0.02}^{+0.02}$ &
$5.43\pm 0.46$ & $40.00 \pm 6.00$ & 1.7 & 0.38 \\
15 & AT2020ysg &
$10.70_{-0.07}^{+0.06}$ &
$2.09_{-0.12}^{+0.17}$ &
$1.63_{-0.71}^{+0.43}$ &
$10.24_{-2.79}^{+1.65}$ &
$-0.12_{-0.37}^{+0.20}$ &
$0.07_{-0.05}^{+0.06}$ &
$8.04\pm 0.33$ & $157.78 \pm 13.03$ & 1.2 & 0.56 \\
16 & AT2020vdq &
$9.25_{-0.11}^{+0.07}$ &
$1.69_{-0.07}^{+0.09}$ &
$1.34_{-1.08}^{+0.81}$ &
$8.18_{-3.71}^{+2.95}$ &
$-1.10_{-0.53}^{+0.30}$ &
$0.06_{-0.04}^{+0.04}$ &
$5.59\pm 0.37$ & $43.56 \pm 3.07$ & 1.3 & 0.27 \\
17 & AT2020vwl &
$9.89_{-0.08}^{+0.08}$ &
$2.08_{-0.04}^{+0.03}$ &
$0.36_{-0.21}^{+0.42}$ &
$8.81_{-2.16}^{+2.18}$ &
$-0.84_{-0.28}^{+0.17}$ &
$0.05_{-0.03}^{+0.04}$ &
$5.79\pm 0.35$ & $48.49 \pm 2.00$ & 2.4 & 0.27 \\
18 & AT2020wey &
$9.67_{-0.12}^{+0.09}$ &
$2.05_{-0.03}^{+0.04}$ &
$0.61_{-0.39}^{+0.40}$ &
$7.92_{-1.85}^{+2.39}$ &
$-1.18_{-0.56}^{+0.59}$ &
$0.11_{-0.08}^{+0.04}$ &
$5.40\pm 0.38$ & $39.36 \pm 2.79$ & 2.1 & 0.24 \\
19 & AT2020yue &
$10.19_{-0.14}^{+0.10}$ &
$1.48_{-0.07}^{+0.10}$ &
$4.18_{-2.02}^{+2.94}$ &
$7.68_{-2.93}^{+3.07}$ &
$-0.51_{-0.34}^{+0.25}$ &
$0.16_{-0.04}^{+0.04}$ &
$6.75\pm 0.32$ & --- & 1.5 & 0.59 \\
20 & AT2020abri &
$9.54_{-0.17}^{+0.14}$ &
$1.85_{-0.08}^{+0.07}$ &
$0.29_{-0.15}^{+0.46}$ &
$6.74_{-3.04}^{+3.73}$ &
$-1.29_{-0.48}^{+0.49}$ &
$0.05_{-0.04}^{+0.05}$ &
$5.62\pm 0.51$ & --- & 0.9 & 0.36 \\
21 & AT2020acka &
$11.03_{-0.19}^{+0.15}$ &
$2.21_{-0.09}^{+0.08}$ &
$0.56_{-0.40}^{+0.98}$ &
$7.21_{-3.71}^{+3.58}$ &
$-1.20_{-0.50}^{+0.83}$ &
$0.21_{-0.09}^{+0.07}$ &
$8.23\pm 0.40$ & $174.47 \pm 25.30$ & 1.1 & 0.70 \\
22 & AT2021axu &
$10.20_{-0.13}^{+0.11}$ &
$1.78_{-0.05}^{+0.05}$ &
$0.42_{-0.26}^{+0.74}$ &
$7.82_{-3.24}^{+3.16}$ &
$-1.57_{-0.29}^{+0.33}$ &
$0.06_{-0.03}^{+0.04}$ &
$6.59\pm 0.55$ & $73.50 \pm 17.26$ & 1.2 & 0.51 \\
23 & AT2021crk &
$9.89_{-0.10}^{+0.11}$ &
$1.28_{-0.06}^{+0.11}$ &
$2.90_{-1.57}^{+2.62}$ &
$8.59_{-3.79}^{+2.90}$ &
$-1.09_{-0.53}^{+0.40}$ &
$0.06_{-0.04}^{+0.04}$ &
$6.12\pm 0.39$ & $57.62 \pm 6.29$ & 1.6 & 0.48 \\
24 & AT2021ehb &
$10.23_{-0.02}^{+0.01}$ &
$2.34_{-0.02}^{+0.01}$ &
$0.20_{-0.08}^{+0.21}$ &
$11.96_{-0.72}^{+0.41}$ &
$-0.43_{-0.04}^{+0.04}$ &
$0.01_{-0.00}^{+0.01}$ &
$7.16\pm 0.32$ & $99.58 \pm 3.83$ & 3.3 & 0.27 \\
25 & AT2021jjm &
$9.47_{-0.14}^{+0.13}$ &
$1.13_{-0.08}^{+0.08}$ &
$4.53_{-2.85}^{+3.34}$ &
$6.38_{-2.76}^{+3.41}$ &
$-1.23_{-0.52}^{+0.54}$ &
$0.11_{-0.05}^{+0.03}$ &
$5.51\pm 0.51$ & --- & 0.7 & 0.52 \\
26 & AT2021mhg &
$9.65_{-0.14}^{+0.12}$ &
$2.05_{-0.07}^{+0.07}$ &
$0.26_{-0.12}^{+0.45}$ &
$7.71_{-2.99}^{+3.14}$ &
$-1.27_{-0.55}^{+0.57}$ &
$0.12_{-0.07}^{+0.05}$ &
$6.13\pm 0.37$ & $57.78 \pm 5.25$ & 1.0 & 0.31 \\
27 & AT2021nwa &
$10.13_{-0.05}^{+0.03}$ &
$2.24_{-0.02}^{+0.02}$ &
$1.09_{-0.16}^{+0.12}$ &
$10.94_{-1.55}^{+1.06}$ &
$-0.58_{-0.12}^{+0.12}$ &
$0.06_{-0.02}^{+0.02}$ &
$7.22\pm 0.32$ & $102.44 \pm 5.37$ & 1.7 & 0.36 \\
28 & AT2021qth &
$9.73_{-0.21}^{+0.14}$ &
$1.91_{-0.17}^{+0.24}$ &
$2.65_{-1.82}^{+3.63}$ &
$5.17_{-3.60}^{+4.93}$ &
$-0.94_{-0.70}^{+0.67}$ &
$0.40_{-0.17}^{+0.15}$ &
$5.95\pm 0.48$ & --- & 1.2 & 0.31 \\
29 & AT2021sdu &
$10.15_{-0.09}^{+0.07}$ &
$1.45_{-0.06}^{+0.07}$ &
$2.22_{-1.28}^{+2.47}$ &
$6.63_{-2.88}^{+3.86}$ &
$-0.01_{-0.11}^{+0.09}$ &
$0.07_{-0.02}^{+0.02}$ &
$6.68\pm 0.29$ & --- & 2.6 & 0.42 \\
30 & AT2021uqv &
$10.14_{-0.11}^{+0.08}$ &
$1.65_{-0.03}^{+0.04}$ &
$2.18_{-1.03}^{+1.16}$ &
$7.70_{-2.87}^{+3.07}$ &
$-1.54_{-0.33}^{+0.42}$ &
$0.21_{-0.03}^{+0.02}$ &
$6.27\pm 0.39$ & $62.30 \pm 7.08$ & 1.4 & 0.49 \\
31 & AT2021utq &
$9.66_{-0.12}^{+0.09}$ &
$1.49_{-0.08}^{+0.11}$ &
$2.44_{-1.15}^{+1.32}$ &
$8.81_{-3.79}^{+2.64}$ &
$-0.94_{-0.55}^{+0.48}$ &
$0.09_{-0.06}^{+0.06}$ &
$5.84\pm 0.43$ & --- & 1.1 & 0.45 \\
32 & AT2021yzv &
$10.83_{-0.15}^{+0.12}$ &
$2.15_{-0.08}^{+0.08}$ &
$0.29_{-0.15}^{+0.38}$ &
$8.35_{-3.23}^{+2.87}$ &
$-1.13_{-0.55}^{+0.61}$ &
$0.13_{-0.08}^{+0.07}$ &
$7.90\pm 0.40$ & $146.38 \pm 20.78$ & 1.5 & 0.61 \\
33 & AT2021yte &
$9.17_{-0.21}^{+0.17}$ &
$1.38_{-0.17}^{+0.24}$ &
$3.40_{-2.60}^{+3.48}$ &
$6.38_{-3.57}^{+3.82}$ &
$-1.24_{-0.58}^{+0.77}$ &
$0.15_{-0.06}^{+0.06}$ &
$5.13\pm 0.45$ & $34.22 \pm 4.81$ & 1.6 & 0.29 \\
\enddata 
\tablecomments{
Columns 3--8 are host galaxy properties inferred with SED fitting (see \S\ref{subsubsec:host_sed_fit}). 
The black hole mass $M_{\rm BH}$ is inferred using the $M_{\rm BH}$--$\sigma_\ast$ scaling relation for the 19 objects with available $\sigma_\ast $ measurements, and using the $M_{\rm BH}$--$M_{\rm gal}$ scaling relation for the remaining 14 objects. 
$r_{1/2}$ is the mean of (seeing-corrected) half-light radii in the $g$-, $r$-, and $i$-band images as measured by \texttt{LAMBDAR}.
$z_{\rm max, h}$ is the maximum redshift out to which the host galaxy can be detected in the ZTF reference catalog (see details in \S\ref{subsec:Vmax}).
}
\end{deluxetable*}

For the TDE host galaxies, we retrieved science-ready coadded images from the \textit{Galaxy Evolution Explorer} (\galex) general release 6/7 \citep{Martin2005a}, the Sloan Digital Sky Survey data release 9 (SDSS DR9; \citealt{Ahn2012a}), the PS1, the Two Micron All Sky Survey (2MASS; \citealt{Skrutskie2006}), and the unWISE archive \citep{Lang2014}. 
We measured the brightness of the host galaxies using the \texttt{Lambda Adaptive Multi-Band Deblending Algorithm in R} (\texttt{LAMBDAR}; \citealt{Wright2016a}) and the methods described in \citet{Schulze2021a}. 

We note that some fields were observed more than once with \galex, while the \citet{Schulze2021a} pipeline only utilizes the deepest \galex exposure. Therefore, in two objects (IDs
8,
28), 
to make the most of \galex observations, we supplemented the \texttt{LAMBDAR} measurements with \galex photometry extracted by \texttt{gPhoton} \citep{Million2016}. 
We adopted an aperture of $10^{\prime\prime}$ and $5^{\prime\prime}$ for the host galaxies of AT2019dsg and AT2021qth, respectively. 
Appendix~\ref{sec:obs_log} presents the photometry in different bands.

\subsubsection{ESI Spectroscopy}

To measure the velocity dispersion of TDE host galaxies, we obtained medium-resolution spectra using the Echellette Spectrograph and Imager (ESI; \citealt{Sheinis2002}) on the Keck II telescope. 
In all observations, we used the Echelle mode. Spectra were obtained for the host galaxies of 17 TDEs (see Table~\ref{tab:spec_medres} in Appendix~\ref{sec:obs_log} for details).
A slit width of 0.3$^{\prime\prime}$, 0.5$^{\prime\prime}$, and 0.75$^{\prime\prime}$ gives an instrumental broadening of $\sigma_{\rm inst} =  9.5$, 15.8, and 23.7\,$\rm km\,s^{-1}$. We reduced the ESI spectra with the \texttt{makee} pipeline\footnote{\url{https://www2.keck.hawaii.edu/inst/esi/makee.html}}. We extracted the spectrum using a radius of $r_{\rm extract}$, which was implemented by specifying the \texttt{hw} and \texttt{uop} parameters in \texttt{makee}. For most objects, $r_{\rm extract}$ was chosen to match the half-light radius (see $r_{1/2}$ in Table~\ref{tab:host_pars}). For a few faint host galaxies, $r_{\rm extract}$ was chosen to enclose a larger aperture to maximize the S/N.

\subsection{Analysis}

\subsubsection{ESI Spectral Fitting}

\begin{figure*}[htbp!]
    \centering
    \includegraphics[width=0.9\textwidth]{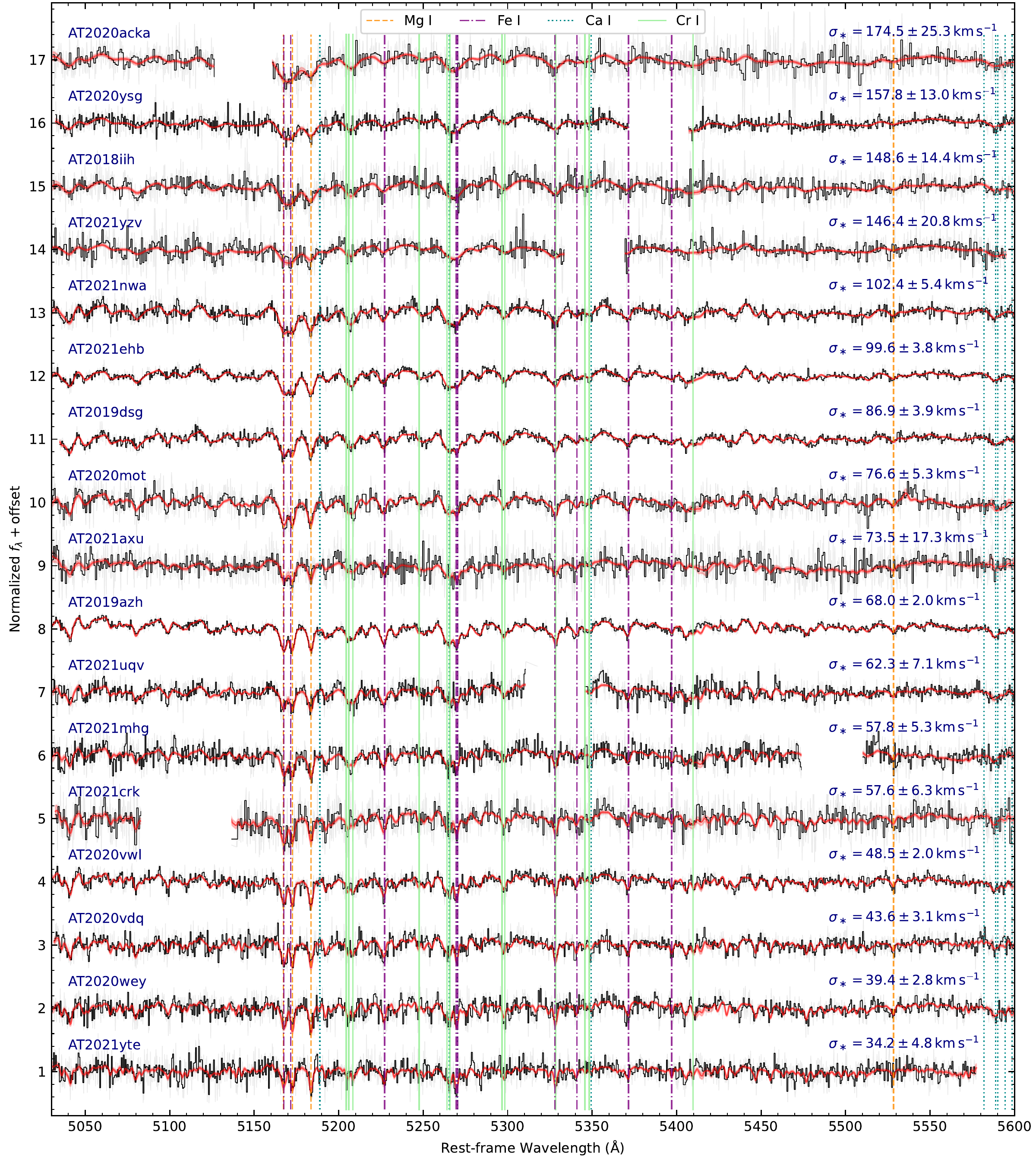}
    \caption{ESI spectra of 17 TDE host galaxies arranged in order of decreasing $\sigma_\ast$. The black lines are the data, and the red lines are the models. Prominent host galaxy absorption lines are indicated by the vertical lines. Masked regions are not plotted. The median S/N of the fitted wavelength range of each spectrum is given in Appendix~\ref{sec:obs_log} (Table~\ref{tab:spec_medres}). 
    \label{fig:ESI_spec}}
\end{figure*}

The galaxy central velocity dispersion $\sigma_\ast$ (i.e., the intensity weighted mean of the root-mean-square of the line-of-sight stellar velocity) is known to be correlated with the central massive BH mass \citep{Merritt2001, Pinkney2003, Gultekin2009, Kormendy2013}. Following previous works \citep{Wevers2017, Somalwar2022},  we measured $\sigma_\ast$ with the penalized pixel-fitting (\texttt{pPXF}) software \citep{Cappellari2004, Cappellari2017}, which fits the ESI absorption line spectrum by convolving a template stellar spectral library with Gauss-Hermite functions.

We used the ELODIE v3.1 high-resolution ($R=42000$) library \citep{Prugniel2001, Prugniel2007}. 
For all ESI spectra, we fit the rest-frame wavelength range from 5030 to 5600\,\AA. 
Prominent galaxy absorption lines\footnote{We take the strong lines table in the National Institute of Standards and Technology (NIST) atomic database.} of \ion{Mg}{I}, \ion{Fe}{I}, \ion{Ca}{I}, and \ion{Cr}{I} in this wavelength range are shown in Figure~\ref{fig:ESI_spec}.
We masked wavelength ranges of common galaxy emission lines, hydrogen Balmer lines, telluric regions, an instrument artifact feature at observer-frame $\sim 4510$\,\AA, and the \ion{Na}{I} D doublet at $z=0$ if Galactic absorption is strong. 

Following previous works \citep{Wevers2017, Wevers2019_Mbh, French2020}, we performed 1000 Monte Carlo (MC) simulations to robustly determine $\sigma_\ast$. In each MC simulation, the observed spectrum was resampled within its error spectrum and refitted with \texttt{pPXF}.
By visually examining results of the simulations, we confirmed that the distributions of the velocity dispersion are well-behaved (i.e., not double-peaked or skewed). 
We took the median of the distribution as the velocity dispersion, and the difference between the 84$^{\rm th}$/16$^{\rm th}$ percentiles as the uncertainty. 
The best-fit spectra and the measured $\sigma_\ast$ are shown in Figure~\ref{fig:ESI_spec}.

\subsubsection{SED Fitting} \label{subsubsec:host_sed_fit}

We modeled the photometric spectral energy distribution (SED) of host galaxies with the software package \texttt{prospector} version 1.1 \citep{Johnson2021a}. \texttt{prospector} uses the \texttt{Flexible Stellar Population Synthesis} (\texttt{FSPS}) code \citep{Conroy2009a} to generate the underlying physical model and \texttt{python-fsps} \citep{ForemanMackey2014a} to interface with \texttt{FSPS} in python. 
We assumed a Chabrier initial mass function \citep{Chabrier2003a} and approximated the star formation history (SFH) by a delayed exponentially declining function. The model was attenuated with the \citet{Calzetti2000a} model. 
The fitted parameters are presented in columns 3--8 of Table~\ref{tab:host_pars}, where $M_{\rm gal}$ is the host galaxy total stellar mass; ${}^{0,0}u-r$ is the Galactic extinction-corrected, synthetic rest-frame $u-r$ color; $\tau_{\rm SFH}$ is the characteristic e-folding timescale of the SFH; $t_{\rm age}$ is the stellar age; $Z$ is the metallicity; and $E(B-V)_{\rm h}$ is the host galaxy extinction. 

A fraction of our TDE host galaxies have been analyzed with similar approaches in the literature \citep{Ramsden2022, Hammerstein2023}. In Appendix~\ref{sec:host_compare}, we show that our estimates of $M_{\rm gal}$ and $^{0,0}u-r$ are mostly consistent with previous results, and point out possible reasons for the differences. The best-fit galaxy SEDs are also shown in Appendix~\ref{sec:host_compare}.

\subsubsection{Black Hole Mass Estimates} \label{subsubsec:ESI_Mbh}

Here, we estimate the BH mass $M_{\rm BH}$ of our TDE sample using host galaxy scaling relations.

For objects with $\sigma_\ast$ measurements, we use the \citet[][Eq.~3]{Kormendy2013} $M_{\rm BH}$--$\sigma_\ast$ relation:
\begin{align}
   {\rm log}&M_{\rm BH,9} = -(0.509\pm0.049)+(4.384\pm 0.287)\times \notag\\
   & {\rm log}\left( \frac{\sigma_\ast}{200\,{\rm km\,s^{-1}}}\right);\; {\rm intrinsic\; scatter=0.29},\label{eq:Mbhsigma}
\end{align}
where $M_{\rm BH,9} \equiv M_{\rm BH}/10^9\,M_\odot$. In addition to the 17 objects with ESI spectra (Table~\ref{tab:spec_medres}), we adopt $\sigma_\ast = 69\pm2\,{\rm km\,s^{-1}}$ for AT2019qiz \citep{Nicholl2020}, and $\sigma_\ast = 40\pm6\,{\rm km\,s^{-1}}$ for AT2020neh \citep{Angus2022}.

\begin{figure}[htbp!]
    \centering
    \includegraphics[width=\columnwidth]{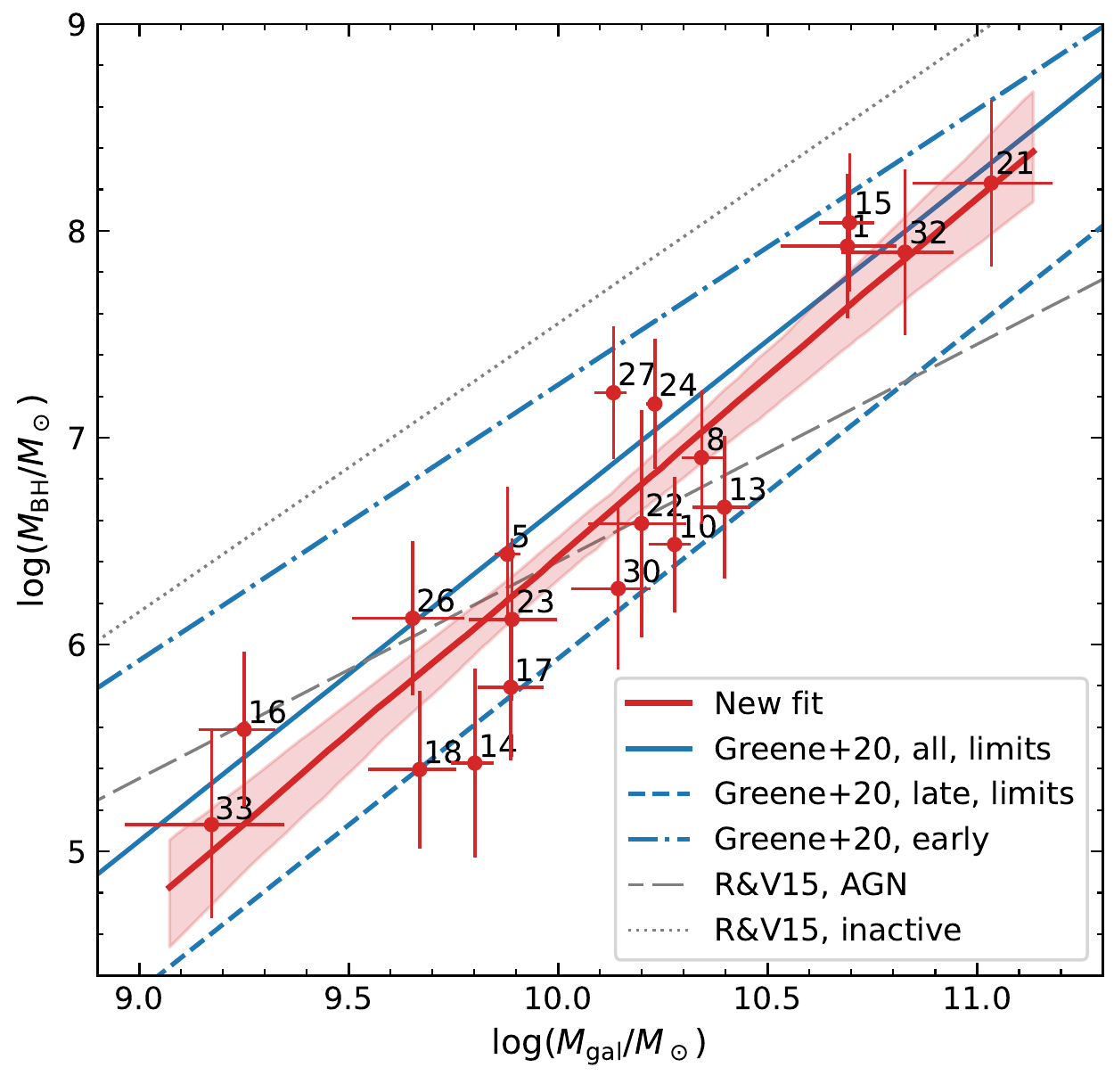}
    \caption{$M_{\rm BH}$ versus $M_{\rm gal}$ for 19 TDEs with $M_{\rm BH}$ measurements inferred from $\sigma_\ast$, labeled by IDs in Table~\ref{tab:tab_info}. The solid red line is a linear fit to these objects (Eq.~\ref{eq:MbhMgal_new}). The solid, dashed, and dash--dotted blue lines are relations presented in \citet[][supplemental Table~5]{Greene2020}, derived using all galaxies (with upper limits on $M_{\rm BH}$), late-type galaxies (with upper limits), and early-type galaxies. The thin dotted and long dashed gray lines are from \citet{Reines2015} using AGN and inactive galaxies.
    \label{fig:Mgal_Mbh}}
\end{figure}

Figure~\ref{fig:Mgal_Mbh} shows the inferred $M_{\rm BH}$ versus $M_{\rm gal}$ (derived from galaxy SED fitting; \S\ref{subsubsec:host_sed_fit}) of these 19 objects. 
We fit a linear relation to these objects:
\begin{align}
   {\rm log}&M_{\rm BH,9} = -(1.75\pm0.13) + (1.73\pm 0.23)\times \notag\\
   & {\rm log}\left( \frac{M_{\rm gal}}{3\times 10^{10}\,M_\odot}\right);\; {\rm intrinsic\; scatter=0.17}, \label{eq:MbhMgal_new}
\end{align}
which is shown as the solid red line. 
For reference, we also show empirical relations from the literature.
\citet{Reines2015} adopt dynamical BH masses for inactive galaxies \citep{Kormendy2013}, and use $M_{\rm BH}$ derived from the width and luminosity of the H$\alpha$ broad line for AGN.
\citet{Greene2020} adopt dynamical BH masses provided by \citet{Kormendy2013} and recent literatures (see details in Section 8.2 of \citealt{Greene2020}).
We use Eq.~(\ref{eq:MbhMgal_new}) to infer the $M_{\rm BH}$ for the remaining 14 objects without $\sigma_\ast$ measurements.

The inferred values of $M_{\rm BH}$ are shown in Table~\ref{tab:host_pars}.
The majority of events (25/33) in our sample are hosted by BHs with $M_{\rm BH} \in (10^5, 10^7)\,M_\odot$.
We computed the Eddington ratio of the UV and optical emitting component $\lambda_{\rm Edd} \equiv L_{\rm bb} / L_{\rm Edd}$, where $L_{\rm Edd} \equiv (M_{\rm BH}/M_\odot) \times 1.25\times  10^{38}\,{\rm erg\,s^{-1}}$. 

Among our sample, AT2020acka (ID 21) has the greatest value of $M_{\rm BH}$ at $10^{8.23\pm0.40}\,M_\odot$.
For a Schwarzschild BH, the maximum mass at which a star of mass $m_\ast$ (in $M_\odot$) and radius $r_\ast$ (in $R_\odot$) can be tidally disrupted outside the horizon is given by
\begin{align}
    M_{\rm Hills}(m_\ast) = 1.1\times 10^8\,M_\odot m_\ast^{-1/2} r_\ast^{3/2}.
\end{align}
Assuming $r_\ast \sim m_\ast^{0.6}$ for $m_\ast>1$ \citep{Demircan1991}, $M_{\rm Hills} = 10^{8.4} \, M_\odot (m_\ast/10)^{0.4}$. 
Therefore, the $M_{\rm BH}$ of AT2020acka is still below $M_{\rm Hills}$ of a massive star ($m_\ast\gtrsim 4$). The disruption of a low-mass main-sequence star requires a rapid BH spin \citep{Kesden2012}. Given that the $t_{\rm age}$ of its host galaxy is not young, the relatively large $M_{\rm BH}$ can also be explained by the disruption of evolved stars \citep{MacLeod2012, MacLeod2013}. 

\section{Survey Efficiency} \label{sec:method}

For an ideal survey that scans the entire sky to a given flux limit, the volumetric rate of a given type of transient can be estimated using the following \citep{Schmidt1968}:
\begin{align}
    \mathcal{R} = \sum_{i=1}^{N} \mathcal{R}_i = \sum_{i=1}^{N} \frac{1}{T_{{\rm span}, i}/(1+z_i)} \frac{1}{V_{{\rm max}, i}} \label{eq:rate},
\end{align}
where $T_{{\rm span}, i}/(1+z_i)$ is the rest-frame duration of the experiment within which the $i^{\rm th}$ transient is selected, 
$N$ is the number of transients that have passed the flux limit, the maximum volume $V_{{\rm max}, i} \equiv \frac{4\pi}{3}D^3_{{\rm max}, i}$ and $D_{\rm max}$ is the maximum luminosity distance (see \S\ref{subsec:Vmax}). 
In this work, $N=33$. For the 16 ZTF-I TDEs, 
$T_{{\rm span}, i}=2$\,yr (from 2018 October 1 to 2020 September 30); 
while for the 17 ZTF-II TDEs, 
$T_{{\rm span}, i}=1$\,yr (from 2020 October 1 to 2021 September 30).

\subsection{Loss Function} 

For a realistic sky survey, $V_{\rm max}$ in Eq.~(\ref{eq:rate}) needs to be replaced by the effective volume $\mathcal{V}_{\rm max} = V_{\rm max} f_{\rm loss}$ \citep{Perley2020}. Here, the loss factor $f_{\rm loss}$ takes into account the facts that the survey coverage is not all-sky, that the Galactic extinction reduces the survey volume, that the limiting magnitude of observations is not constant (it depends strongly on the moon phase, weather, and airmass), and that fast-evolving TDEs with fainter peak magnitudes are easier to be missed. 

To estimate $f_{\rm loss}$, we took the observation history of ZTF. 
We obtained the limiting magnitude for each observation (with a certain field ID and MJD) from the exposure table of ZTF DR14\footnote{Accessible at \url{https://irsa.ipac.caltech.edu/data/ZTF/docs/ztf_metadata_latest.db}}.
For each TDE, using the light curve model obtained in \S\ref{subsec:lc_result}, we simulated fake ZTF observations by inserting $10^5$ light curves uniformly across all sky and $T_{{\rm span}, i}$.
We then applied the cuts outlined in \S\ref{subsec:filtering} to compute the fraction of observations that would have passed our selection criteria.
The values of $f_{\rm loss}$ are given in the last column of Table~\ref{tab:lc_pars}. 

\subsection{Maximum Volume}  \label{subsec:Vmax}

If the TDE candidate selection only depends on transient photometric properties, then $D_{\rm max}=D_{\rm max, t}$, where $D_{\rm max, t}$ is the distance out to which a transient can be detected above the flux limit of our experiments (i.e., $m_{\rm peak}<18.75$ for ZTF-I TDEs, and $m_{g,{\rm peak}}<19.1$ for ZTF-II TDEs). $D_{\rm max, t}$ can be computed using the redshifts and the best-fit values of $T_{\rm bb}$, $L_{\rm bb}$ (\S\ref{sec:lc_fitting}). The results of $D_{\rm max, t}$ and the corresponding maximum redshift $z_{\rm max, t}$ are shown in Table~\ref{tab:lc_pars}.

However, in steps (1) and (2) of our TDE selection criteria (\S\ref{subsec:filtering}), we required the detection of each host galaxy in the ZTF reference image, the depth of which (for point sources) is $m\lesssim 23$ \citep{Masci2019}.
It is easy to imagine that TDEs hosted by lower-mass galaxies and galaxies with redder colors can only be selected out to a smaller volume (because at higher redshifts, these galaxies will not be cataloged in the ZTF reference, and the transient will appear as hostless). 

Therefore, for each of the TDE host galaxies, we estimated $z_{\rm max, h}$, which is the maximum redshift out to which the observer-frame PSF AB magnitude (in either $g$ or $r$ band) will be $<23$. 
We computed $z_{\rm max, h}$ using the best-fit \texttt{prospector} models derived in \S\ref{subsubsec:host_sed_fit}. 
To include the effects of PSF photometry on extended sources, we multiplied the model SED fluxes by a factor of $10^{-0.4( m_{\rm PSF} - m_\texttt{LAMBDAR})}$, where $m_{\rm PSF}$ is the \texttt{rPSFMag} column in the PS1 \texttt{StackObjectView} catalog \citep{Flewelling2020}, and $m_\texttt{LAMBDAR}$ is the PS1 $r$-band magnitude in the \texttt{LAMBDAR} photometry (see Table~\ref{tab:hostphot2} in Appendix~\ref{sec:obs_log}). 
The derived values of $z_{\rm max, h}$ are given in Table~\ref{tab:host_pars}. 

Taken together, 
\begin{align}
z_{{\rm max}, i} = {\rm min}(z_{{\rm max, t}, i}, z_{{\rm max, h}, i}). \label{eq:zmax}
\end{align}
We find that all 33 TDEs in our sample satisfy $z_i < z_{{\rm max, t},i} < z_{{\rm max, h}, i}$. 
Therefore, for this TDE sample, $z_{\rm max} = z_{\rm max, t}$.

\begin{figure*}[htbp!]
    \centering
    \includegraphics[width=\textwidth]{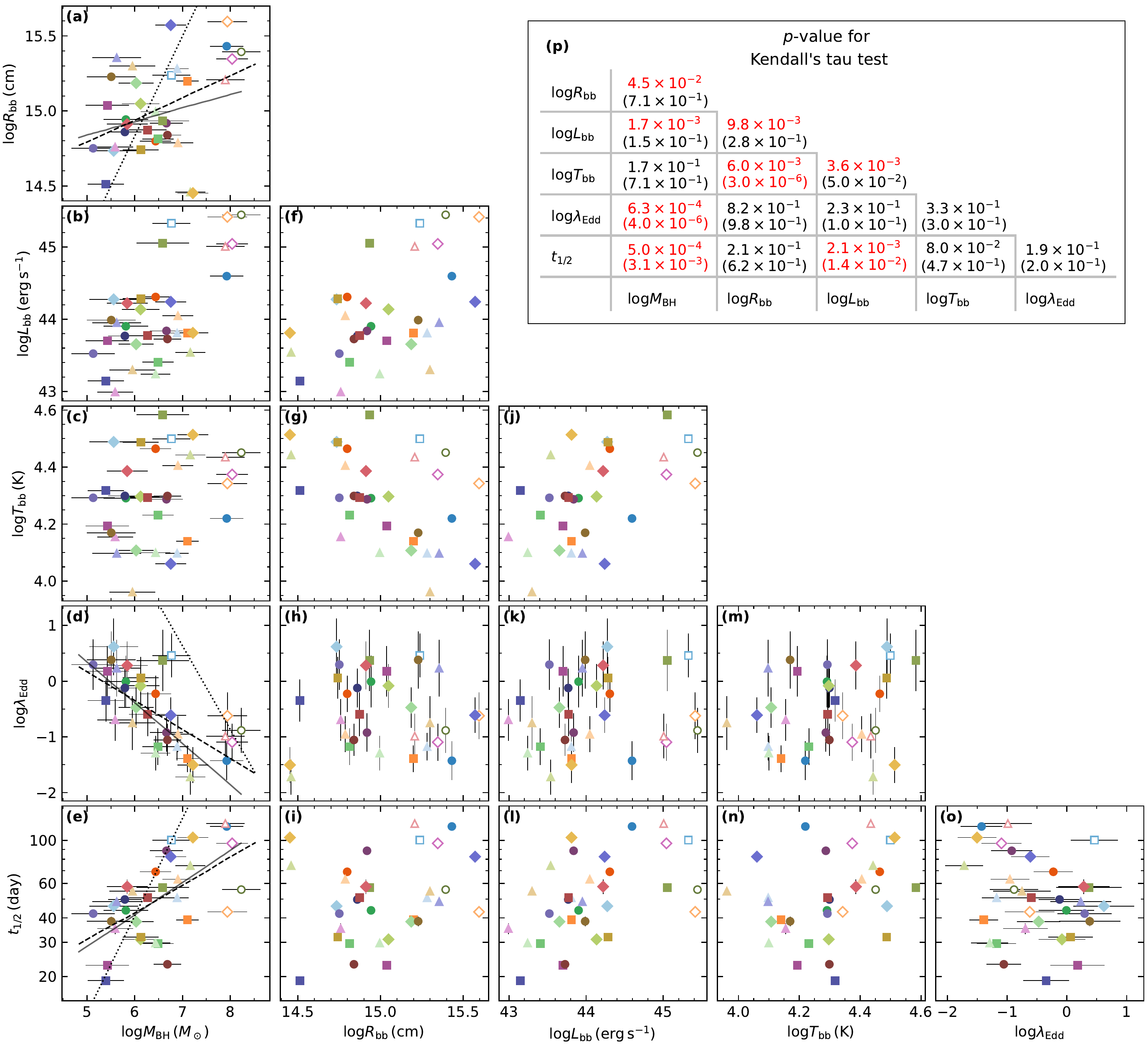}
    \caption{\textbf{Panels (a)-(o)}: correlations between TDE photometric properties, $\lambda_{\rm Edd}$, and $M_{\rm BH}$. 
    Symbol colors follow the same convention as in Figure~\ref{fig:color_evol} and Figure~\ref{fig:sample_lcs}. 
    Hollow markers show objects at $z>0.24$, where there is an observational bias toward selecting TDEs in higher-mass galaxies.
    \textbf{Panel (a)}: the dotted line shows the expected $R_{\rm bb} \propto M_{\rm BH}^{2/3}$ scaling relation in a fiducial cooling envelop model \citep{metzger22_colling_envelope}; the dashed ($R_{\rm bb}\propto M_{\rm BH}^{0.15}$) and solid ($R_{\rm bb}\propto M_{\rm BH}^{0.09}$) lines show the best-fit power laws using all markers and filled markers, respectively (see \S\ref{subsec:rate_Rbb}).
    \textbf{Panel (d)}: the dotted line shows the expected Eddington ratio of peak fall-back accretion rate $\lambda_{\rm Edd}\propto M_{\rm BH}^{-3/2}$; the dashed ($\lambda_{\rm Edd}\propto M_{\rm BH}^{-0.52}$) and solid ($\lambda_{\rm Edd}\propto M_{\rm BH}^{-0.74}$) lines show the best-fit power laws using all markers and filled markers, respectively (see \S\ref{subsubsec:Eddington_limited}). 
    \textbf{Panel (e)}: the dotted line shows the expected fall-back timescale of $t_{1/2}\propto M_{\rm BH}^{1/2}$; the dashed ($t_{1/2}\propto M_{\rm BH}^{0.14}$) and solid ($t_{1/2}\propto M_{\rm BH}^{0.16}$) lines show the best-fit power laws using all markers and filled markers, respectively (see \S\ref{subsubsec:t1/2_Mbh}). 
    \textbf{Panel (p)}: $p$-value of Kendall's tau test for 15 pairs of parameters. 
    The results using 33 TDEs are shown outside the parenthesis, 
    and the results using 28 TDEs at $z<0.24$ are shown in the parenthesis. Significant correlations with $p<0.05$ are highlighted in red colors. 
    \label{fig:see_pars}
    }
\end{figure*}

\section{Results and Discussion} \label{sec:rate}

\subsection{Correlations between TDE Photometric and Galaxy Properties} \label{subsec:correlation}

Here, we investigate the correlations between the TDE photometric and host galaxy properties.
We focus on the three blackbody parameters ($L_{\rm bb}$, $T_{\rm bb}$, $R_{\rm bb}$), $t_{1/2}$ (defined in \S\ref{subsec:lc_result}), $\lambda_{\rm Edd}$, and $M_{\rm BH}$.
We did not include $M_{\rm gal}$ since it is strongly correlated with $M_{\rm BH}$ (Figure~\ref{fig:Mgal_Mbh}). 
We also did not include $t_{1/2,{\rm rise}}$ and $t_{1/2, {\rm decline}}$, because both parameters are strongly correlated with $t_{1/2}$ (this can be seen in Figure~\ref{fig:sample_lcs}, where TDEs that rise fast generally also decline fast). The $p$-value of a Kendall’s tau test between $t_{1/2,{\rm rise}}$ and $t_{1/2, {\rm decline}}$ is $1.29\times 10^{-5}$. This result is in agreement with \citet{Hammerstein2023}. We note that the first ZTF TDE sample study found no correlation between the TDE rise and decline rates \citep{vanVelzen2021}, which possibly results from the smaller sample size. 

Figure~\ref{fig:see_pars} shows the distribution of our sample on various diagrams. 
Panel (p) shows the $p$-values of a Kendall’s tau test between any two of the six quantities of interest, using the total sample of 33 TDEs and the subset of 28 TDEs at $z<0.24$ (see reasons for this cut in \S\ref{subsubsec:selection}). 

\subsubsection{The Selection Effects}  \label{subsubsec:selection}
Considering the whole sample of 33 TDEs, the correlations between eight pairs of parameters appear to be significant. While a few similar correlations have also been reported by \citet{Hammerstein2023}, we note that such correlations might be promoted by selection effects. To be in our sample, the host galaxies need to be bright enough to be detected in the ZTF reference catalog (\S\ref{subsec:filtering}). Since $M_{\rm BH}\propto M_{\rm gal}^{1.6}$ (see Eq.~\ref{eq:MbhMgal_new}) and $M_{\rm gal} \propto L_{\rm gal}$, we can find luminous TDEs hosted by higher-mass BHs even at high redshifts. 

Based on the the values of $z_{\rm max, h}$ computed in \S\ref{subsec:Vmax} (see Table~\ref{tab:host_pars}), within $z<0.24$, even the faintest host galaxy of our sample (i.e., the host of AT2020wey) can be detected in the ZTF reference catalog. Therefore, within this volume, there should be no observational bias toward bright galaxies\footnote{Note that, here, we do not consider galaxies with an absolute $r$-band PSF magnitude fainter than that of AT2020wey.}. 

Restricting ourselves to the 28 TDEs at $z<0.24$, the correlation between a few pairs of parameters becomes statistically less insignificant. The correlation between $R_{\rm bb}$ and $T_{\rm bb}$ becomes even more significant, as expected in a flux-limited sample if many TDEs have a similar peak blackbody luminosity.
In \S\ref{subsubsec:t1/2_Mbh} and \S\ref{subsubsec:Eddington_limited}, we discuss the other two strong correlations. 

\subsubsection{Duration above Half-max Versus Black Hole Mass} \label{subsubsec:t1/2_Mbh}

The correlation between the light curve evolutionary speed and BH mass has been reported in the literature \citep{vanVelzen2020, Gezari2021, Hammerstein2023}, which we confirm in panel (e) of Figure~\ref{fig:see_pars}.
We note that the $p$-values between $t_{1/2,{\rm rise}}$ and log$M_{\rm BH}$ ($2.3\times 10^{-3}$) and between $t_{1/2,{\rm decline}}$ and log$M_{\rm BH}$ ($1.0\times 10^{-3}$) are comparable to (but slightly greater than) the $p$-value between $t_{1/2}$ and log$M_{\rm BH}$ ($5.0\times 10^{-4}$). 

We define $M_6 \equiv M_{\rm BH} / (10^6\,M_\odot)$.
A log-linear fit between $t_{1/2}$ and $M_{\rm BH}$ for 33 TDEs yields the following (see the dashed line):
\begin{align}
    \frac{t_{1/2}}{ 42.5_{-3.5}^{+3.9}\,{\rm days}} = M_6^{0.14\pm0.04}, \label{eq:t1/2_1}
\end{align}
which has an intrinsic scatter of 0.17\,dex. 
Restricting to the 28 TDEs at $z<0.24$, we obtain a similar power-law relation of the following (see the solid line):
\begin{align}
    \frac{t_{1/2}}{41.6_{-3.5}^{+3.8}\,{\rm days}} = M_6^{0.16\pm0.05},\label{eq:t1/2_2}
\end{align}
which has an intrinsic scatter of 0.15\,dex. 

Equations~(\ref{eq:t1/2_1}, \ref{eq:t1/2_2}) can be compared with the fall-back timescale of the most bound debris (see the dotted line): 
\begin{align}
    \frac{t_{\rm fb}}{ 41\,{\rm days} } =M_6^{1/2} m_\ast^{-1} r_\ast^{3/2}.
\end{align}
The observed shallow power-law index may be caused by other processes. For example, the circularization of the stellar debris has been shown to be more rapid around higher-mass BHs \citep{Bonnerot2016, Bonnerot2020}.

\subsubsection{Eddington Ratio Versus Black Hole Mass} \label{subsubsec:Eddington_limited}
The distribution of our sample on the Eddington ratio and BH mass diagram is shown in panel (d) of Figure~\ref{fig:see_pars}.
A log-linear fit between $\lambda_{\rm Edd}$ and $M_{\rm BH}$ for 33 TDEs yields the following (see the dashed line):
\begin{align}
    \frac{\lambda_{\rm Edd}}{0.45_{-0.10}^{+0.12}}=M_6^{-0.52\pm0.11}, \label{eq:lambda_Edd_1}
\end{align}
which has an intrinsic scatter of 0.28\,dex. 
To correct for the selection bias, we also fit for the 28 TDEs at $z<0.24$, obtaining a steeper power-law as follows (see the solid line):
\begin{align}
   \frac{ \lambda_{\rm Edd}}{0.41_{-0.09}^{+0.11}}=M_6^{-0.74\pm0.12}, \label{eq:lambda_Edd_2}
\end{align}
which has an intrinsic scatter of 0.11\,dex.
This relatively tight correlation is not surprising since by definition ${\rm log}\lambda_{\rm Edd}\equiv {\rm log}L_{\rm bb} - {\rm log} M_{\rm BH}-38.10$. And Eq.~(\ref{eq:lambda_Edd_2}) comes from the fact that $L_{\rm bb}$ is only weakly positively correlated with $M_{\rm BH}$ (see the filled markers in panel b). 

Eq.~(\ref{eq:lambda_Edd_2}) can also be compared with the expected peak fall-back rate of $\dot M_{\rm fb} \approx M_\ast / (3 t_{\rm fb})$ relative to the Eddington accretion rate (see the dotted line):
\begin{align}
    \frac{\dot M_{\rm fb}}{\dot M_{\rm Edd}} 
    =136 \,\eta_{-1} m_\ast^2 r_\ast^{-3/2} M_6^{-3/2} \label{eq:lambda_fb}
\end{align}
where $\eta$ is the accretion radiative efficiency, and $\eta_{-1}\equiv \eta / 0.1$. The observed power law is much shallower than Eq.~(\ref{eq:lambda_fb}). In fact, the majority of TDEs in panel (d) lie well below the dotted line. One likely reason might be Eddington-limited accretion. Indeed, none of the TDEs in our sample appear to have a peak blackbody luminosity that is significantly super-Eddington. Another natural explanation is that the UV and optical peak blackbody luminosity only captures a fraction of the total bolometric luminosity, with the EUV and X-ray luminosity unaccounted for. 

\subsection{Luminosity Functions} \label{subsec:lfs}

While theoretical calculations show that the TDE rate may decline by a factor of 5 from $z=0$ to $z=1$ \citep{Kochanek2016}, a detailed discussion of the redshift evolution of TDE rates is beyond the scope of this work. 
Hereafter, we assume that the TDE rate remains the same out to the highest redshift object in our sample (i.e., $z<0.519$). 

\subsubsection{Rest-frame $g$-band LF} \label{subsubsec:rate_Lg}

\begin{figure}[htbp!]
    \centering
    \includegraphics[width=\columnwidth]{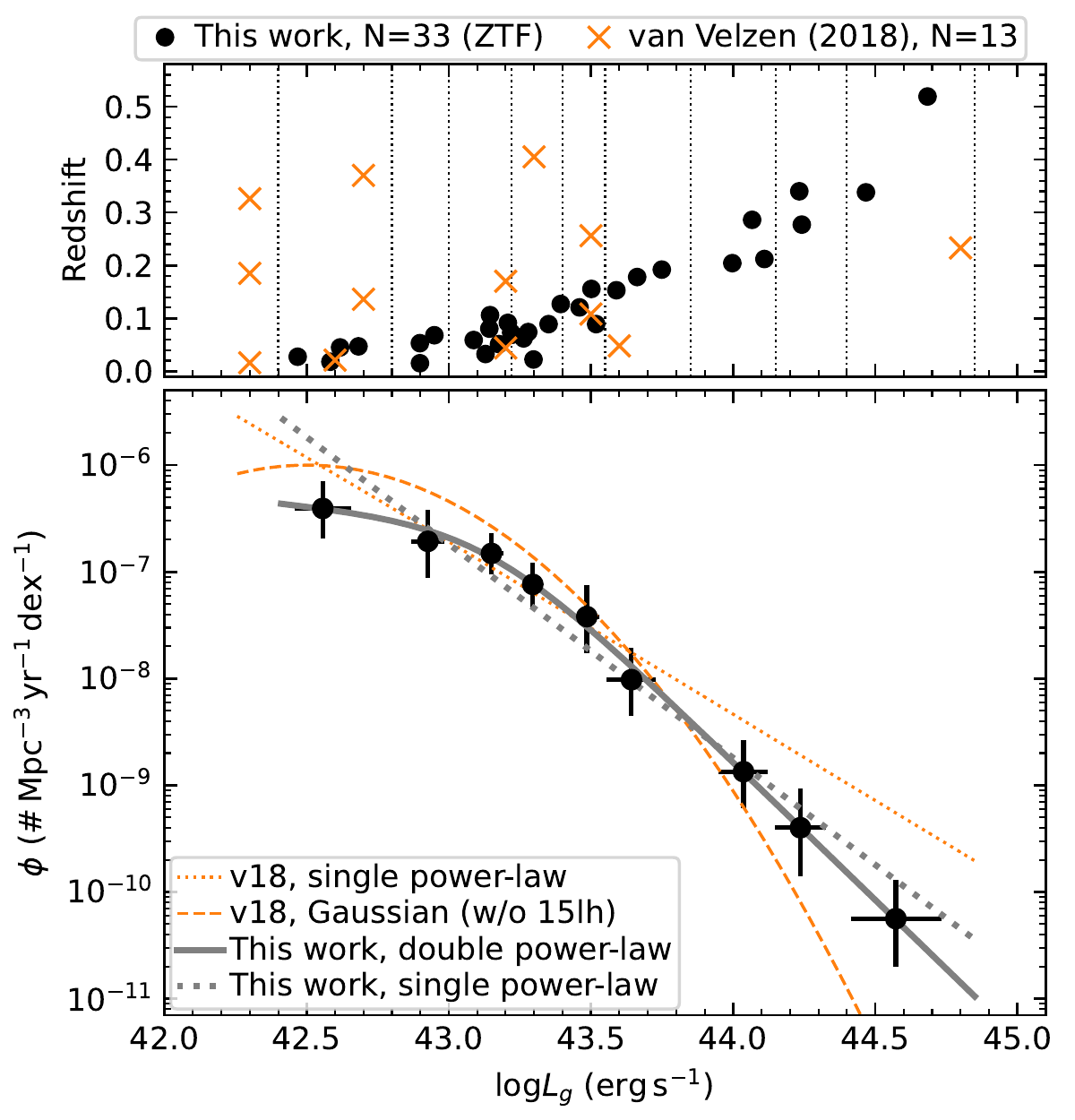}
    \caption{
    \textit{Upper}: redshift versus log$L_{\rm g}$ for 33 TDEs in this work (circles) and 13 TDEs used by \citet[][crosses]{vanVelzen2018}.
    The boundaries of the 9 luminosity bins used in this work are indicated by the vertical dotted lines. 
    \textit{Lower}: TDE LF in rest-frame $g$ band. We show the single and double power-law fits as well as the two LFs presented in \citet{vanVelzen2018}.
    \label{fig:Lg_func}}
\end{figure}

In the upper panel of Figure~\ref{fig:Lg_func}, we show the distribution of the 33 TDEs in the observed redshift versus peak rest-frame $g$-band luminosity diagram, where the boundaries of the nine log$L_g$ bins are indicated with vertical lines. 
For a certain bin $j$ with $n_j$ TDEs and width $\Delta_j {\rm log}L_{g}$, the rate $\phi_j = \left[ \sum_{i=1}^{n_j} 1/(T_{{\rm span}, i}\mathcal{V}_{{\rm max}, i}) \right ] / \Delta_j {\rm log}L_{g}$, and we compute the corresponding uncertainty of $\phi_j$ based on the Poisson error \citep{Gehrels1986}.
For example, when $n_j=4$, the upper and lower limits of $\phi_j$ are
$\phi_j^{u} = \phi_j\times 7.163/4$ and 
$\phi_j^{l} = \phi_j\times 2.086/4$.

First, we fit the seven solid data points in the lower panel of Figure~\ref{fig:Lg_func} with a single power law of
\begin{align}
    \phi(L_{g}) = \frac{d \mathcal{R}(L_g)}{d {\rm log}L_g} = \dot N_0 \left( \frac{L_g}{L_0}\right)^{-\gamma}.
\end{align}
For $L_0 = 10^{43}\,{\rm erg\,s^{-1}}$, we have 
$\dot N_0 = 1.82^{+0.48}_{-0.39}\times 10^{-7}\,{\rm Mpc^{-3}\,yr^{-1}}$ and 
$\gamma = 2.00_{-0.14}^{+0.15}$. 
The best-fit model, shown as the dotted gray line in Figure~\ref{fig:Lg_func}, is steeper than the power-law model with $\gamma=1.6\pm0.2$ presented by \citet{vanVelzen2018}. 

Next, we describe the LF with a double power law of the following:
\begin{align}
    \phi(L_{g}) = 
    \dot N_0 \left[ \left( \frac{L_g}{L_{\rm bk}}\right)^{\gamma_1 } + \left( \frac{L_g}{L_{\rm bk}}\right)^{\gamma_2 } \right]^{-1}
\end{align}
where $-\gamma_1$ is the faint-end slope, $-\gamma_2$ is the bright-end slope, and $L_{\rm bk}$ is the characteristic break luminosity. 
We perform the fit with MCMC, obtaining
$L_{\rm bk}=1.36_{-0.48}^{+0.89}\times 10^{43}\,{\rm erg\,s^{-1}}$, 
$\dot N_0 = 2.87^{+2.98}_{-1.68}\times 10^{-7}\,{\rm Mpc^{-3}\,yr^{-1}}$, 
$\gamma_1 = 0.26^{+0.61}_{-0.80}$, and 
$\gamma_2 = 2.58^{+0.27}_{-0.25}$. 
This model is shown as the solid gray line in Figure~\ref{fig:Lg_func}. 

The BIC value of the double power-law fit is smaller than the single power-law fit by 6.07. 
According to \citet{Raftery1995}, a BIC difference of 0--2 is weak, a difference of 2--6 is positive, and a difference of 6--10 is strong.
Therefore, we conclude that a double power-law LF provides a better description of the data.

Our result of $\phi(L_g)$ is consistent with that provided by \citet{vanVelzen2018} at $L_g \sim 10^{43.5}\,{\rm erg\,s^{-1}}$. For overluminous events, ASASSN-15lh is the only object with $L_g>10^{43.6}\,{\rm erg\,s^{-1}}$ in the \citet{vanVelzen2018} sample. 
The fact that nine objects in our sample have $L_g > 10^{43.6}\,{\rm erg\,s^{-1}}$ allows us to constrain the upper end of the LF more precisely. 

For subluminous events, the LF measured with the ZTF sample is shallower, and the rate is about a factor of two smaller than that measured by \citet{vanVelzen2018}. 
No objects in our sample have $L_g<10^{42.4}\,{\rm erg\,s^{-1}}$, while three objects in the \citet{vanVelzen2018} sample (\galex-D1-9, \galex-D23H-1, and iPTF16fnl) have $L_g \approx 10^{42.3}\,{\rm erg\,s^{-1}}$. 
However, the two \galex events have relatively sparse light curves (note the lack of data points on the rise in Fig.~15 of \citealt{Gezari2008} and Fig.~2 of \citealt{Gezari2009}), which can possibly lead to an underestimation of their peak $g$-band luminosity. 

\subsubsection{UV and Optical Blackbody LF} \label{subsubsec:rate_Lbb}

\begin{figure}[htbp!]
    \centering
    \includegraphics[width=\columnwidth]{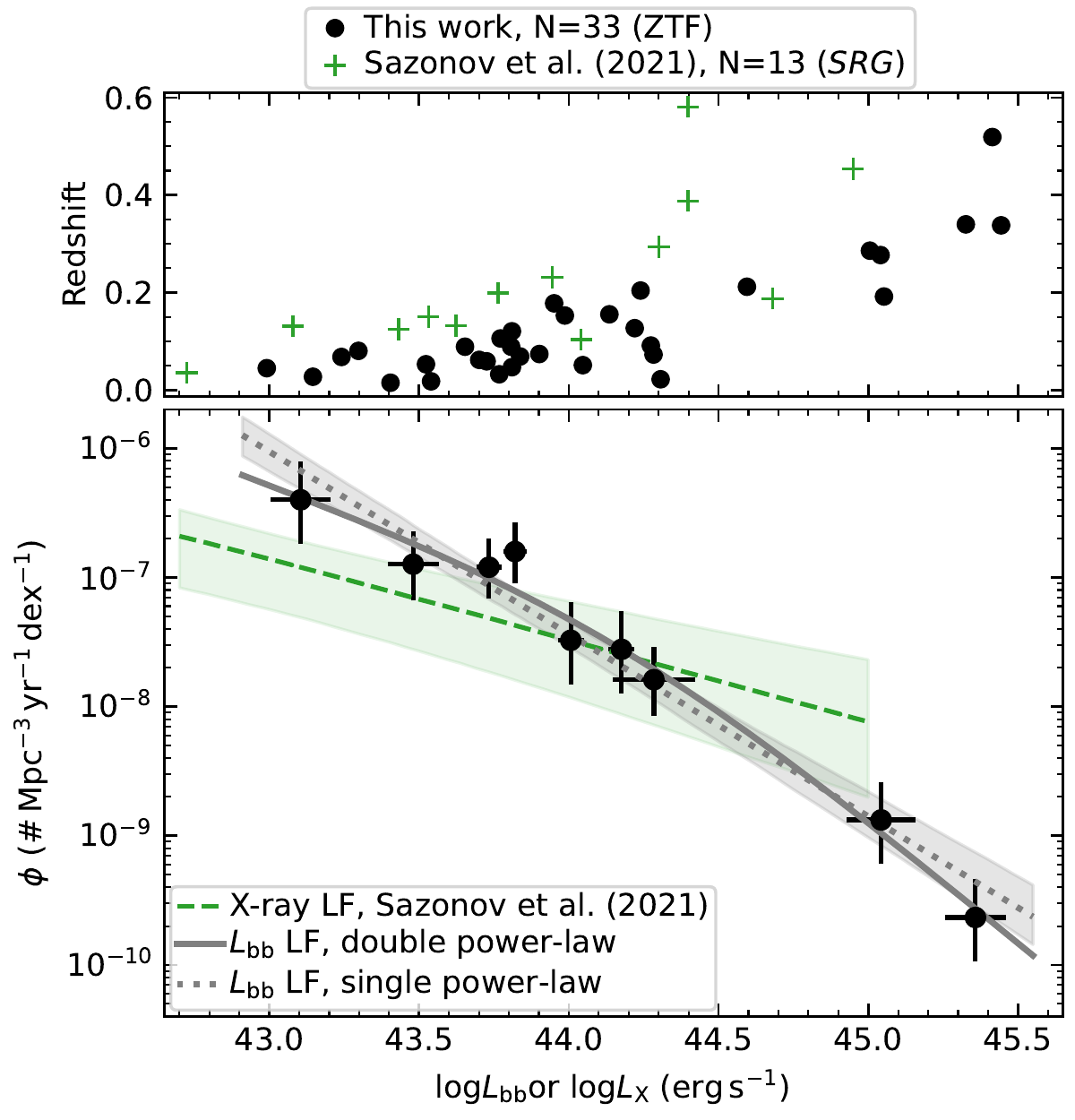}
    \caption{\textit{Upper}: redshift versus log$L_{\rm bb}$ for 33 TDEs in this work (circles), and versus the peak X-ray luminosity for 13 \srg-selected TDEs presented by \citet{Sazonov2021}. 
    \textit{Lower}: TDE LF in terms of peak UV and optical blackbody luminosity or peak 0.2--6\,keV X-ray luminosity.
    The dotted and solid gray lines show the single power-law (Eq.~\ref{eq:Lbb_LF_single}) and double power-law (Eq.~\ref{eq:Lbb_LF_double}) fits. The dashed green line shows the X-ray LF given by \citet{Sazonov2021}. 
    For the dashed and dotted lines, 1$\sigma$ uncertainties are indicated with the semitransparent regions.
    \label{fig:Lbb_func}}
\end{figure}

Following the procedures outlined in \S\ref{subsubsec:rate_Lg}, we compute the TDE rate as a function of the peak UV and optical blackbody luminosity (see Figure~\ref{fig:Lbb_func}). 

With $L_0 = 10^{43}\,{\rm erg\,s^{-1}}$, a single power-law fit yields 
\begin{align}
    \phi (L_{\rm bb}) = (9.43^{+4.53}_{-3.04} \times 10^{-7}\,{\rm Mpc^{-3}\,yr^{-1}}) \left( \frac{L_{\rm bb}}{L_0} \right)^{-1.41\pm 0.14}. \label{eq:Lbb_LF_single}
\end{align}
A double power-law fit yields 
\begin{align}
    \phi (L_{\rm bb}) = & (5.72^{+7.08}_{-3.29} \times 10^{-8}\,{\rm Mpc^{-3}\,yr^{-1}}) \times \notag \\
    &  \left[ \left( \frac{L_{\rm bb}}{L_{\rm bk}} \right)^{0.84^{+0.30}_{-0.36}} + \left( \frac{L_{\rm bb}}{L_{\rm bk}} \right)^{1.93_{-0.27}^{+0.32}} \right]^{-1},\label{eq:Lbb_LF_double}
\end{align}
where $L_{\rm bk} = 1.46_{-0.64}^{+1.20}\times 10^{44}\,{\rm erg\,s^{-1}}$.
The BIC value of the double power-law fit is greater than that of the single power-law fit by 2.2.
Therefore, the single power-law fit is slightly favored.

With Eq.~(\ref{eq:Lbb_LF_single}), the integrated volumetric rate of optical TDEs with $L_{\rm bb} > 10^{43}\,{\rm erg\,s^{-1}}$ is $3.1^{+0.6}_{-1.0}\times 10^{-7} \,{\rm Mpc^{-3}\,yr^{-1}}$. This can be compared with the volumetric rate of X-ray selected TDEs.
Using a sample of 13 TDEs selected from \srg/eROSITA, \citet{Sazonov2021} found that the majority of X-ray selected events are intrinsically faint in the optical. 
Previous studies also implied that the majority of ZTF-selected TDEs are intrinsically faint in the X-ray band (see Fig.~8 of \citealt{Hammerstein2023}).
Using the LF provided by \citet{Sazonov2021}, the rate of X-ray TDEs with $L_{\rm X}>10^{43}\,{\rm erg\,s^{-1}}$ is $\sim 2.3\times 10^{-7}\,{\rm Mpc^{-3}\,yr^{-1}}$. 
Therefore, we conclude that the rates of optically loud and X-ray loud TDEs are comparable to each other. 

\subsection{Rate Dependence on $R_{\rm bb}$} \label{subsec:rate_Rbb}

\begin{figure}[htbp!]
    \centering
    \includegraphics[width=\columnwidth]{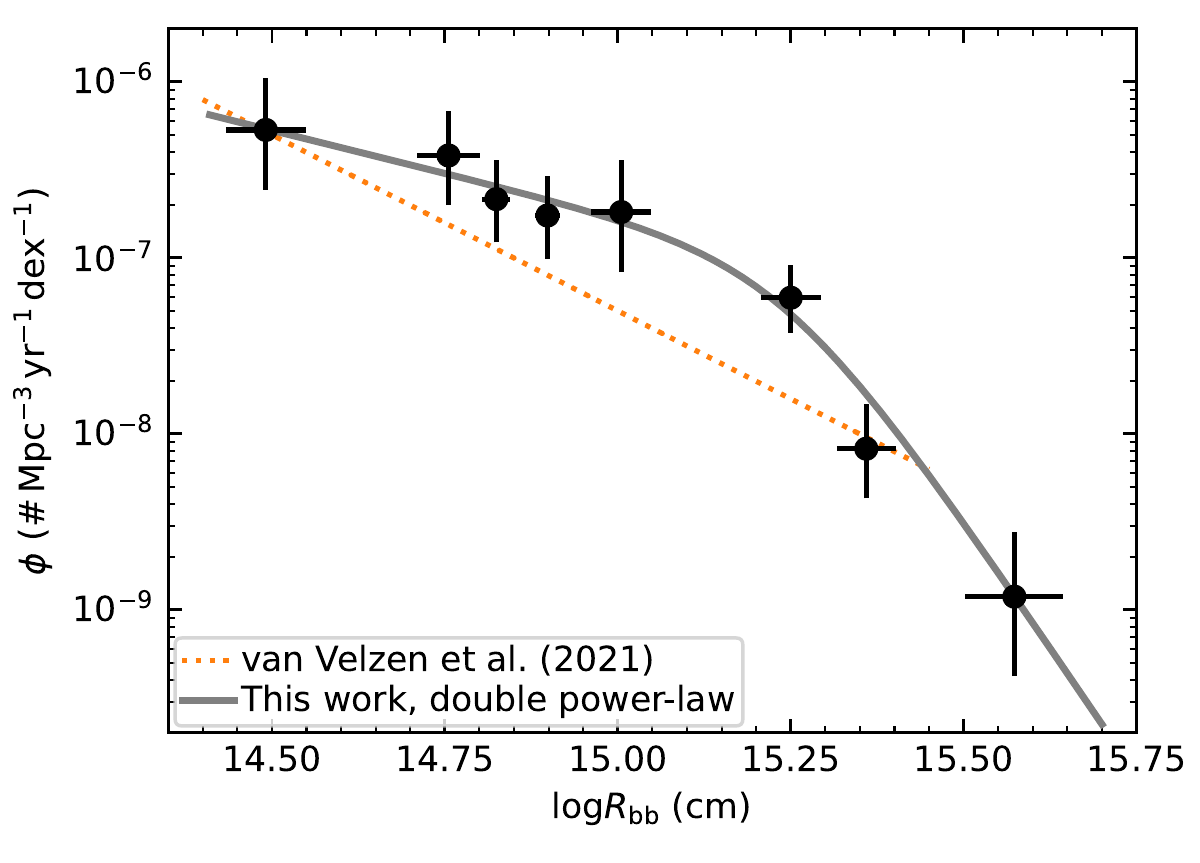}
    \caption{TDE rate as a function of $R_{\rm bb}$. \label{fig:Rbb_func}}
\end{figure}

Following the procedures outlined in \S\ref{subsubsec:rate_Lg}, we compute the TDE rate as a function of the peak blackbody radius $R_{\rm bb}$ (see Figure~\ref{fig:Rbb_func}). 

A double power-law fit gives
\begin{align}
    \phi (R_{\rm bb}) = & (1.00^{+1.33}_{-0.62} \times 10^{-7}\,{\rm Mpc^{-3}\,yr^{-1}}) \times \notag \\
    &  \left[ \left( \frac{R_{\rm bb}}{R_{\rm bk}} \right)^{0.97^{+0.59}_{-0.67}} + \left( \frac{R_{\rm bb}}{R_{\rm bk}} \right)^{5.81_{-1.57}^{+2.16}} \right]^{-1} \label{eq:Rbb},
\end{align}
where $R_{\rm bk} = 1.75_{-0.41}^{+0.53}\times 10^{15}\,{\rm cm}$.
Compared with the $\phi(R_{\rm bb}) \propto R_{\rm bb}^{-2}$ relation found by \citet{vanVelzen2021}, our results indicate a slope that is much shallower at small radii and much steeper at large radii. 

\citet{vanVelzen2021} suggested that the observed $R_{\rm bb}$ in the majority of TDEs can be explained by the self-intersection radius ($R_{\rm I}$) of the debris stream for disruptions of stars with $0.2\lesssim m_\ast \lesssim 3$ and impact parameter $R_{\rm p}/R_{\rm T} \approx 1$ \citep{Dai2015}. 
For TDEs hosted by the most massive BHs, we find $R_{\rm bb}\gg R_{\rm I}$ because the self-intersection radius decreases with $M_{\rm BH}$ for $M_{\rm BH}\gtrsim 10^{6.5}\,M_\odot$ (see Fig. 8 of \citealt{Gezari2021}). 
In fact, we find that TDEs at a given $M_{\rm BH}$ show a broad range of $R_{\rm bb}$.
As suggested by \citet{Nicholl2022}, $R_{\rm bb}$ can vary a lot even for the same $M_{\rm BH}$ depending on the impact parameter --- it could be set by the collision-induced outflow in shallow encounters, but by the disk wind in deep encounters.

In the TDE cooling envelope model \citep{loeb97_spherical_envelope, metzger22_colling_envelope}, the stellar debris promptly form a quasi-spherical envelope. The ``virial radius'' of the envelope, which is bound to the massive BH by the energy spread imparted by the disruption process, is $R_{\rm v}\approx6.8\times 10^{13}\,{\rm cm}\,m_\ast^{2/15}M_6^{2/3}(M_{\rm e}/0.2\,M_\odot)$, where $M_{\rm e}$ is the mass of the envelope (see Eq.~7 of \citealt{metzger22_colling_envelope}).
The photosphere radius is greater than this $R_{\rm v}$ by a factor of $\sim\! 10$, which is shown as the dotted line in panel (a) of Figure~\ref{fig:see_pars}. The above scaling relation is derived assuming a lower main sequence star mass--radius relationship. 
The observed $R_{\rm bb}$ dependence on $M_{\rm BH}$ is much shallower with huge scatter, which might be accounted for with a broader range of stellar properties.

The steep upper power-law index ($\gamma_2\sim 5.8$) in Eq.~(\ref{eq:Rbb}) suggests that there is a physical maximum blackbody radius for TDEs: $R_{\rm bb,max}\sim \rm few\times 10^{15}$\,cm. 
One possibility is that this maximum radius corresponds to the semimajor axis of the most bound tidal debris $a\simeq 0.5 R_\ast(M_{\rm BH}/M_\ast)^{2/3}\simeq 3\times10^{15}\mathrm{\,cm}\,(M_{\rm BH}/10^{7.5}M_\odot)^{2/3}$, where we have taken the mass--radius relation $R_\ast \propto M_\ast^{\approx 2/3}$ for main-sequence stars. 
Under this hypothesis, the fact that the TDE rate is strongly suppressed at $M_{\rm BH}\gtrsim10^{7.5}M_\odot$ (see \S\ref{subsec:Mbh_func}) would lead to a maximum blackbody radius that is in reasonable agreement with observations. However, we leave detailed theoretical considerations to future works.

\subsection{Optical TDE Black Hole Mass Function} \label{subsec:Mbh_func}

Since the uncertainty of ${\rm log}M_{\rm BH}$ is relatively large (0.1--0.4\,dex), instead of the binning method utilized in \S\ref{subsec:lfs} and \S\ref{subsec:rate_Rbb}, we compute the optical TDE black hole mass function using kernel density estimation. We adopt a Gaussian kernel with the same variance as the uncertainties of the log$M_{\rm BH}$ measurements.

\begin{figure}[htbp!]
    \centering
    \includegraphics[width=\columnwidth]{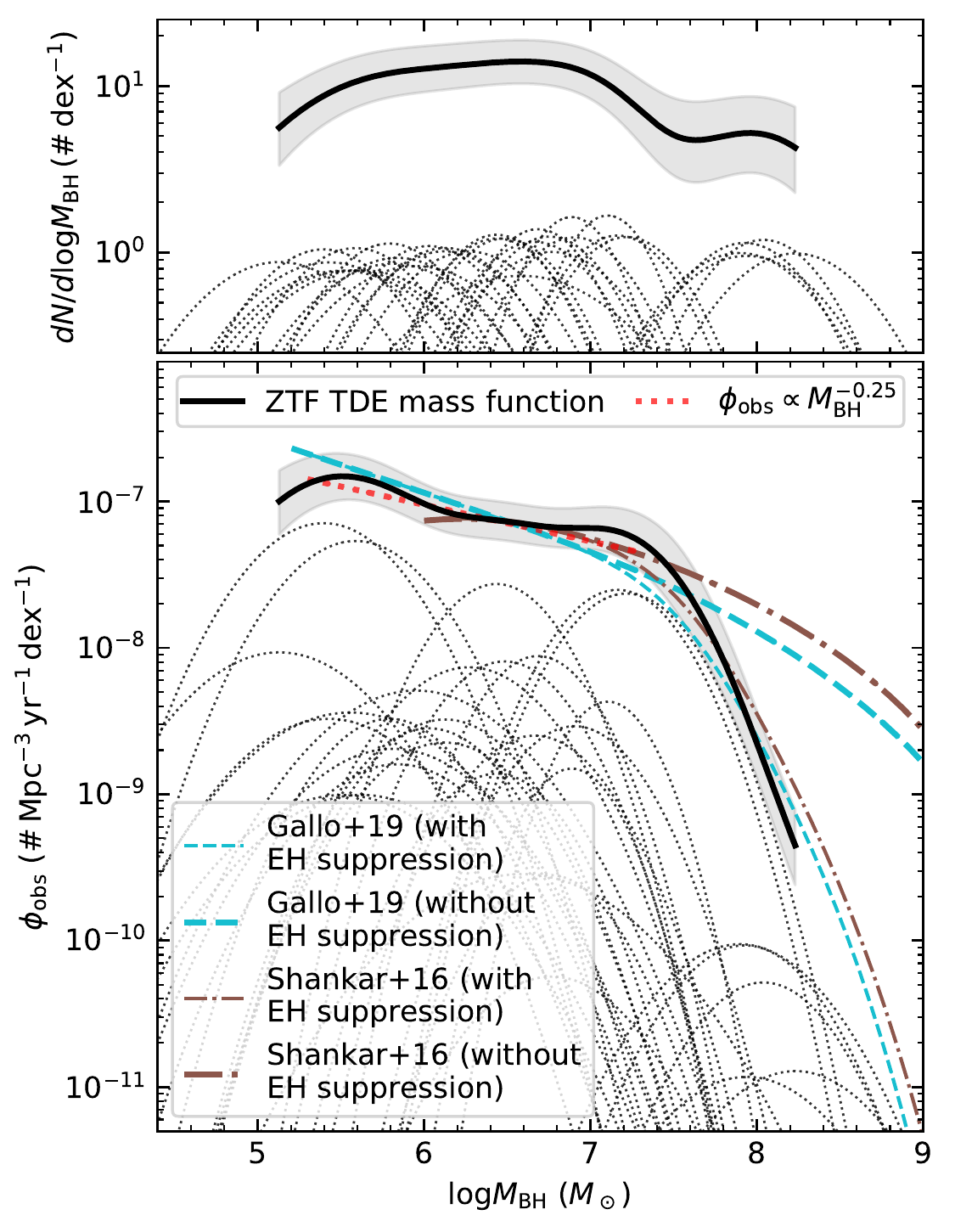}
    \caption{
    \textit{Upper}: the thin lines are the log$M_{\rm BH}$ PDFs of the 33 TDE host galaxies.
    The think black line shows the total number of detected TDEs per dex, computed by summing over the individual Gaussians and plotted between the peak of the PDF of the lowest BH mass ($10^{5.13}\,M_\odot$) and highest BH mass ($10^{8.23}\,M_\odot$). 
    The semitransparent region represents the 1$\sigma$ uncertainties. 
    \textit{Lower}:
    the thin dotted lines are the PDFs (in the upper panel) multiplied by $\mathcal{R}_i$. 
    The solid black curve shows the total optical TDE rate as a function of $M_{\rm BH}$.
    From $10^{5.3}\,M_\odot$ to $10^{7.3}\,M_\odot$, the slope follows a power law of $\phi \propto {\rm log}M_{\rm BH}^{-0.25}$ (red dotted line). 
    We show predictions of two BHMFs \citep{Shankar2016, Gallo2019} with and without the event horizon (EH) suppression factor $g(M_{\rm BH})$ (see Eq.~\ref{eq:dRdMbh}), normalized to match the black curve at $M_{\rm BH}=10^{6.5}\,M_\odot$.  
    \label{fig:Mbh_func}}
\end{figure}

The upper panel of Figure~\ref{fig:Mbh_func} shows the raw observed number of TDEs per dex $d N/d{\rm log}M_{\rm BH}$, which peaks at $M_{\rm BH}\approx 10^{6.6}\,M_\odot$. We estimated the 1$\sigma$ Poisson single-sided upper and lower limits by interpolating Tab.~1 and Tab.~2 of \citet{Gehrels1986}. 

The lower panel of Figure~\ref{fig:Mbh_func} shows the optical TDE rate with respect to $M_{\rm BH}$. 
We observed a significant drop of $\phi(M_{\rm BH})$ from $10^{7.4}\,M_\odot$ to $10^{8.2}\,M_\odot$. 
This roughly corresponds to $M_{\rm Hills}$ for main-sequence stars. 
A similar result was first reported by \citet[Fig.~3]{vanVelzen2018} and later updated by \citet[Fig.~13]{vanVelzen2020}.
While more massive galaxies exhibit shallower (``cored'') stellar density profiles that can also lead to a suppression of TDE rates by a factor of $\lesssim 10$ (see Fig.~5 of \citealt{Magorrian1999} and Fig.~4 of \citealt{Stone2016}), this effect alone does not account for the observed (much steeper) rate suppression. 

To compare our observations to theoretical predictions, we write the mass function for the BHs that are causing TDEs as
\begin{align}
    \phi(M_{\rm BH}) = \dot N_0 \times M_6^{\beta} \times  \frac{d n_{\rm BH}}{d{\rm log}M_{\rm BH}} g(M_{\rm BH}), \label{eq:dRdMbh}
\end{align}
where $\dot N_0 \times M_6^{\beta}$ is the rate at which stars are scattered into the loss cone ($\dot{N}_0$ being a normalization constant, and $\beta$ will be explained shortly), 
$dn_{\rm BH}/d {\rm log}M_{\rm BH}$ is the local BHMF,  and $g(M_{\rm BH})$ is the event-horizon suppression factor that describes the fraction of stars that produce observable optical flares. 
The observed optical TDE mass function, $\phi_{\rm obs}$, is computed by convolving Eq.~(\ref{eq:dRdMbh}) with a Gaussian kernel of the typical log$M_{\rm BH}$ measurement uncertainty of 0.3\,dex.
The convolution is needed since the measurement error blurs and broadens the distribution of quantities \citep{Kelly2012}. 

Most TDEs originate from the BH's sphere of influence $R_{\rm infl}$ \citep{Wang2004}, where the number of stars within $R_{\rm infl}$ is $N \sim  M_{\rm BH}/M_\ast$. 
Since $R_{\rm infl} \approx G M_{\rm BH}/\sigma_\ast^2 \propto \sigma_\ast^2 \sim M_{\rm BH}^{1/2}$, 
the orbital period at $R_{\rm infl}$ is $P_{\rm orb}\propto  R_{\rm infl}^{3/2} / M_{\rm BH}^{1/2} \propto M_{\rm BH}^{1/4}$.
The two-body relaxation timescale at $R_{\rm infl}$ is $t_{\rm rel}\propto (P_{\rm orb}/N) ( \frac{M_{\rm BH}}{M_\ast})^2 \propto M_{\rm BH}^{5/4}$ \citep{Alexander2017}.
The TDE rate is expected to be the total number of stars within the sphere of influence divided by $t_{\rm rel}$, 
which is $\sim N/t_{\rm rel} \propto M_{\rm BH}/t_{\rm rel} \propto M_{\rm BH}^{-1/4}$. 
Therefore, in Eq.~(\ref{eq:dRdMbh}), we adopt $\beta=-0.25$. 

The rate suppression factor $g(M_{\rm BH})\sim1$ at $M_{\rm BH} \lesssim 10^7\,M_\odot$, and drops at higher BH masses because stars are swallowed by the event horizon. The shape of $g(M_{\rm BH})$ depends on the stellar age, the stellar metallicity, the BH spin distribution, the stellar density structure (how centrally concentrated the star is), the exact boundary between full and partial TDEs, and the rate at which stars of different masses are scattered into the loss cone (see more detailed theoretical calculations in \citealt{Huang2023}).
We compute $g(M_{\rm BH})$ as the fraction of stars in a given stellar population that satisfies $M_{\rm Hills}(m_*, M_{\rm BH}) < M_{\rm BH}$. The stellar population we consider has metallicity ${\rm [Fe/H]} = 0.3$ (twice solar, appropriate for stars near galactic centers) and a single age of 100 Myr. Our small sample is insufficient to differentiate models of different stellar ages, BH spins, and loss-cone filling mechanisms.

Using two BHMFs \citep{Shankar2016, Gallo2019}, the predictions of $\phi_{\rm obs}$ are shown as the dashed cyan and dashed-dotted brown lines in the lower panel of Figure~\ref{fig:Mbh_func}. To demonstrate the effect of event horizon suppression, we show the results with and without the $g(M_{\rm BH})$ factor in thin and thick lines, respectively.
All curves are scaled at $M_{\rm BH}=10^{6.5}\,M_\odot$ to match the observation (the thick black line). 
We confirm that the observed high-mass rate drop is consistent with the theoretical expectation of the event horizon effect. 

A novel result in Figure~\ref{fig:Mbh_func} is that the optical TDE mass function roughly follows a power law of $\phi_{\rm obs} \propto M_{\rm BH}^{-0.25}$ over two orders of magnitude in BH mass ($10^{5.3}\,M_\odot\lesssim M_{\rm BH} \lesssim 10^{7.3}\,M_\odot$). 
In \S\ref{subsec:local_bhmf}, we discuss the implications of this result for the local BHMF.

\subsection{Rate Enhancement in Green Galaxies and Suppression in Blue Galaxies} \label{subsec:rate_Mgal}

\begin{figure}[htbp!]
    \centering
    \includegraphics[width=\columnwidth]{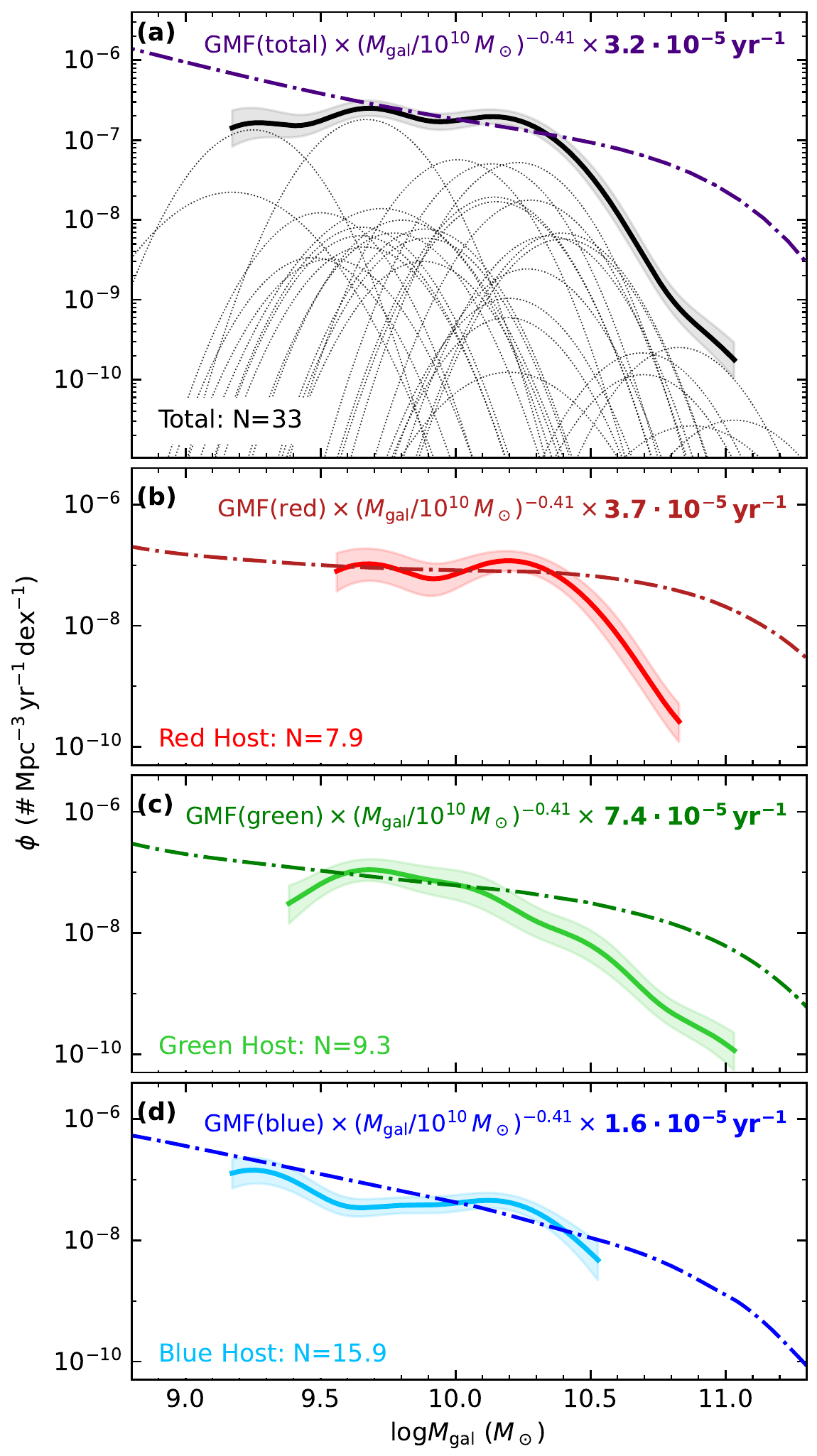}
    \caption{
    Panel (a): the dotted thin lines represent the values of $\mathcal{R}_i$ (Eq.~\ref{eq:rate}) multiplied by the individual PDFs of log$M_{\rm gal}$.
    The solid thick curve shows the total optical TDE rate as a function of $M_{\rm gal}$, plotted between the peak of the PDF of the lowest galaxy mass ($10^{9.17}\,M_\odot$) and highest galaxy mass ($10^{11.03}\,M_\odot$). 
    The semitransparent region represents the 1-$\sigma$ uncertainties. 
    Panels (b)--(d): the observed optical TDE galaxy mass functions in three bins of $\mathcal{C}$ (Eq.~\ref{eq:color_gal}).
    The dashed-dotted lines show the local GMFs multiplied by $M_{\rm gal}^{-0.41}$ and scaled to match the observation at $M_{\rm gal}=10^{10}\,M_\odot$.
    \label{fig:Mgal_func}}
\end{figure}

Following the procedures outlined in \S\ref{subsec:Mbh_func}, we compute the TDE rate as a function of $M_{\rm gal}$. 
We limit the minimum kernel bandwidth to be 0.15.
In panel (a) of Figure~\ref{fig:Mgal_func}, the thin lines show the probability density function (PDF) of each host's log$M_{\rm gal}$ multiplied by $\mathcal{R}_i$, and the thick line shows the observed optical TDE galaxy mass function $\phi(M_{\rm gal})$.

Using Eq.~(\ref{eq:MbhMgal_new}) and Eq.~(\ref{eq:dRdMbh}) and assuming that the occupation fraction of BHs is close to unity, the observed TDE galaxy mass function should follow
\begin{align}
\phi(M_{\rm gal}) & \approx \dot N_0^\prime M_{\rm gal}^{-0.41} \frac{d n_{\rm gal}}{d{\rm log}M_{\rm gal}} g(M_{\rm gal}), \label{eq:dRdMgal}
\end{align}
where $d n_{\rm gal}/d{\rm log}M_{\rm gal}$ is the local galaxy mass function (GMF). 
We took the GMF given by \citet{Baldry2012}, which is similar to the most recent GMF \citep{Wright2017} at $M_{\rm gal}\gtrsim 10^9\,M_\odot$. 
At a typical galaxy mass of $M_{\rm gal}=10^{10}\,M_\odot$, the optical TDE rate is $3.2_{-0.6}^{+0.8}\times 10^{-5}$\,galaxy$^{-1}$\,yr$^{-1}$, as shown by the dashed purple line in panel (a) of Figure~\ref{fig:Mgal_func}.

\begin{figure}[htbp!]
    \centering
    \includegraphics[width=\columnwidth]{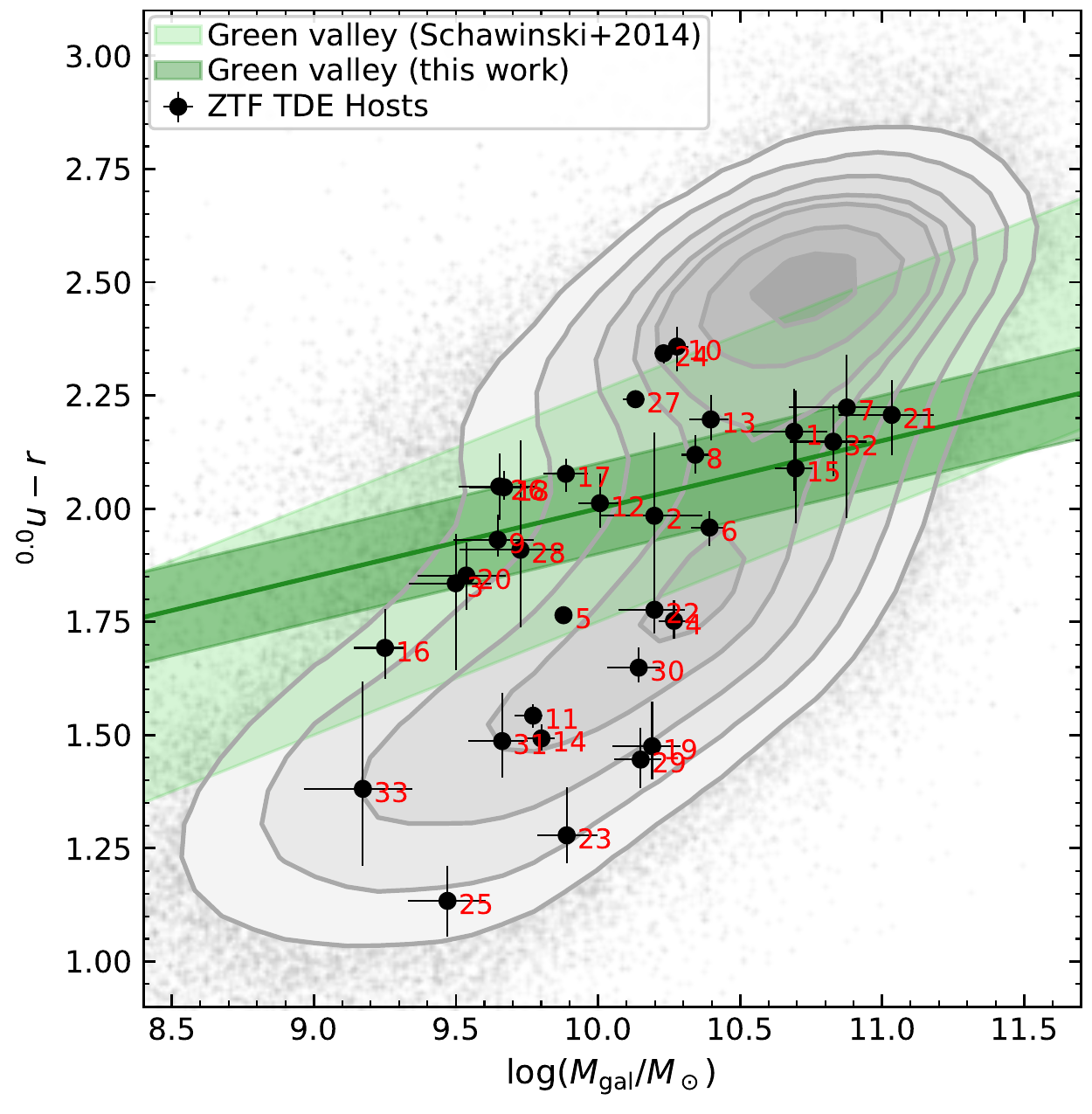}
    \caption{Host galaxies of the TDE sample on the ${}^{0,0}u-r$ versus $M_{\rm gal}$ diagram, labeled by the IDs in Table~\ref{tab:tab_info}. 
    The background contours represent a comparison sample of galaxies from SDSS (see text).
    The region of green valley defined by \citet{Schawinski2014} is denoted by the light green band. 
    In this work, we define a narrower region of green valley (dark green band) by following the contour of the SDSS comparison sample.
    The solid green line marks the middle of the new green valley (Eq.~\ref{eq:green}).
    \label{fig:color_Mgal}}
\end{figure}

Next, we aim to quantify the relative optical TDE rate in galaxies with different colors.
In Figure~\ref{fig:color_Mgal}, we show the host galaxy distribution on the ${}^{0,0}u-r$ versus $M_{\rm gal}$ diagram. 
To compare the properties of TDE hosts to the population of local galaxies, we started with the flux-limited ($14 \leq m_r \leq 17.77$) sample of $\sim 6.6\times10^5$ spectroscopically classified SDSS galaxies \citep{Strauss2002} with $M_{\rm gal}$ estimated by \citet[][Tab.~4]{Mendel2014}. We computed $^{0,0}u-r$ using the rest-frame absolute magnitude in $u$ and $r$ bands provided by the \texttt{Photoz} table in SDSS DR7 \citep{Abazajian2009}.
To build a comparison sample representative of galaxies that our ZTF TDE selection is sensitive to, for each TDE in our sample, we randomly select $10^{3}$ galaxies with $z<z_{\rm max}$, where $z$ is the redshift of the SDSS galaxy, and $z_{\rm max}$ is computed in \S\ref{subsec:Vmax}.
The gray contours in Figure~\ref{fig:color_Mgal} are regions encircling 6.7\%, 16\%, 31\%, 50\%, 69\%, 84\%, and 93.3\% (i.e., in steps of 0.5$\sigma$) of the final sample of $3.3\times10^{4}$ galaxies.

The region of green valley galaxies defined by \citet{Schawinski2014} is marked by the light green band in Figure~\ref{fig:color_Mgal}, which already enclosed galaxies in the ``red sequence'' and ``blue cloud'' loci of the SDSS comparison sample. Therefore, we define a new green valley locus (shown as the solid green line):
\begin{align}
   {}^{0,0}u-r=0.5+0.15\times {\rm log}(M_{\rm gal}/M_\odot) . \label{eq:green}
\end{align}
Based on Eq.~(\ref{eq:green}), we define a new quantity of $M_{\rm gal}$-corrected color:
\begin{align}
    \mathcal{C} \equiv {}^{0,0}u-r-0.5-0.15\times {\rm log}(M_{\rm gal}/M_\odot), \label{eq:color_gal}
\end{align} 
which represents the vertical distance to the green valley loci on the color--mass diagram.
We define red, green, and blue galaxies to be those with $\mathcal{C}> 0.1$, $|\mathcal{C}|\leq 0.1$, and $\mathcal{C}<- 0.1$, respectively. 

We compute $\phi(M_{\rm gal})$ for red, green, and blue galaxies. 
Note that the uncertainty of $\mathcal{C}$ is not negligible and is dominated by the uncertainty of ${}^{0,0}u-r$.
Therefore, for each TDE host, we computed the PDF of its $\mathcal{C}$ (assuming Gaussian distributions), and calculated the probabilities of it being a red or green or blue galaxy. 
For example, the host position of AT2018iih/ZTF18acaqdaa (ID 1) is in the green valley, but the probability of it being a red, green, and blue galaxy is 0.40, 0.52, and 0.08, respectively. 
The resulting $\phi(M_{\rm gal})$ for three $\mathcal{C}$ bins are shown as the solid thick curves in panels (b)--(d) of Figure~\ref{fig:Mgal_func}.

The GMFs for the three $\mathcal{C}$ bins are computed using the \citet{Mendel2014} sample. By definition, ${\rm GMF(red) + GMF(green) + GMF(blue) = GMF(total)}$. 
We compute $M_{\rm gal}^{-0.41} \times {\rm GMF}$, and scale it to match the observed optical TDE galaxy mass function at the typical galaxy mass of $10^{10}\,M_\odot$. 
Considering red, green, and blue galaxies, the per-galaxy TDE rate is 
$3.7_{-1.5}^{+2.3}\times 10^{-5}\,{\rm galaxy^{-1}\, yr^{-1}}$, 
$7.4_{-3.2}^{+5.0}\times 10^{-5}\,{\rm galaxy^{-1}\, yr^{-1}}$, and 
$1.6_{-0.4}^{+0.6}\times 10^{-5}\,{\rm galaxy^{-1}\, yr^{-1}}$, respectively. At a typical galaxy mass of $M_{\rm gal}=10^{10}\,M_\odot$, the relative ratio of optical TDE rate in red, green, and blue galaxies is $1:\frac{7.4_{-3.2}^{+5.0}}{3.7_{-1.5}^{+2.3}}:\frac{1.6_{-0.4}^{+0.6}}{3.7_{-1.5}^{+2.3}} =1:2.0_{-0.7}^{+1.1}:0.4_{-0.1}^{+0.2}$.

The rate suppression in blue galaxies may come from the fact that star-forming galaxies exhibit larger amounts of dust in the galaxy nuclei. It is expected that optical searches, which generally select blue transients, will be biased against TDEs, which are intrinsically redder due to dust extinction \citep{Roth2021}. 
The rate enhancement in green-valley galaxies can be attributed to the higher number density of stars scattered into the loss cone following recent star formation or galaxy mergers (e.g., \citealt{French2020, Hammerstein2021}).
We note that the rate enhancement we found appears to be smaller than previous observational constraints \citep{Law-Smith2017, French2020, Hammerstein2021}, although, instead of using the ``green-valley'' definition, some other studies focus on the overrepresentation factor in E+A galaxies.

\subsection{TDE Rates: The Tension between Observations and Loss Cone Models}
Our new results have brought back to life a tension between observationally inferred TDE rates and those computed using quasi-empirical loss cone models \citep{Wang2004, Stone2016}. For example, in \citet{Wang2004}, the volumetric rate is estimated to be $\sim10^{-5}$\,Mpc$^{-3}$\,yr$^{-1}$, or ${\rm few}\times 10^{-4}$\,galaxy$^{-1}$\,yr$^{-1}$ in galaxies similar to our Milky Way (MW). 
\citet{Stone2016} investigated ways to bring theory and observation into alignment, adopting conservative assumptions that would push the loss cone rates down; yet, the rate was calculated to be $\sim 3\times 10^{-6}$\,Mpc$^{-3}$\,yr$^{-1}$, or (1--2)\,$\times 10^{-4}$\,galaxy$^{-1}$\,yr$^{-1}$ in MW-like galaxies. Both studies suggest an expected rate that is significantly higher than the observed value of ${\rm few}\times 10^{-5}$\,galaxy$^{-1}$\,yr$^{-1}$ (see \S\ref{subsec:rate_Mgal}). 

One possible resolution of this issue could be substantial dust obscuration in most galactic nuclei (as suggested for blue galaxies in \S\ref{subsec:rate_Mgal}). A more theoretical resolution would be a tangentially anisotropic velocity distribution in galactic nuclei, namely a preferential destruction of stars on radial orbits.  
If this kind of tangential bias is put in by hand and then the nucleus is allowed to evolve, the velocity anisotropy will be washed away too quickly to solve a TDE rate discrepancy \citep{Lezhnin2015}. 
However, \citet{Teboul2022} recently showed that it can be sustained for longer periods of time if most galactic nuclei have steep (``strongly segregated'') cusps of stellar mass BHs; in this case, the ejection in strong scatterings will eliminate stars on the most radial orbits and effectively ``shield'' the SMBH loss cone.

\subsection{Implications of the Local BHMF} \label{subsec:local_bhmf}
Here, we aim to independently measure the shape of the local BHMF in the mass range of $10^{5.3}\,M_\odot \leq M_{\rm BH}\leq 10^{7.3}\,M_\odot$. 
We assume $g(M_{\rm BH})=1$, and use the observed optical TDE black hole mass function (lower panel of Figure~\ref{fig:Mbh_func}).
To correct for the relative rate differences in red/green/blue galaxies (\S\ref{subsec:rate_Mgal}), we compute the corrected $\phi_{\rm corr}(M_{\rm BH}) = \phi_{\rm red}(M_{\rm BH}) \times\frac{3.2}{3.7} + \phi_{\rm green}(M_{\rm BH})\times\frac{3.2}{7.4} + \phi_{\rm blue}(M_{\rm BH})\times\frac{3.2}{1.6}$. 

Parameterizing the BHMF as $dn_{\rm BH}/d{\rm log}M_{\rm BH} \propto M_{\rm BH}^{p}$, we obtain $p = 0.014\pm0.059$.
Note that this value is subject to the uncertainty of $\beta$ in Eq.~(\ref{eq:dRdMbh}). For example, \citet{Stone2016} performed the most recent detailed theoretical calculations by applying loss cone dynamics to observations of nearby galactic nuclei, finding $\beta=-0.247$ for core nuclei, and $\beta=-0.223$ for cusp nuclei. A greater value of $\beta=-0.22$ would render a lower value of $p=-0.016\pm0.059$.
Generally speaking, our result favors a flat BHMF in the mass range of $10^{5.3}\,M_\odot\leq M_{\rm BH}\leq 10^{7.3}\,M_\odot$. Below, we compare it with literature estimates and model predictions in \S\ref{subsubsec:BHMF_review}, and comment on some caveats in our assessment in \S\ref{subsubsec:caveats}. 

\subsubsection{Comparison with Literature Estimates and Model Predictions} \label{subsubsec:BHMF_review} 

The traditional approach to calculate the local BHMF is to convert the observed galaxy distribution $\Phi(y)$ into the BHMF using a $M_{\rm BH}$--$y$ scaling relation (see reviews by \citealt{Kelly2012, Shankar2013}). 
A key assumption here is that BHs exist ubiquitously in galaxy nuclei, which has been justified in high-mass galaxies ($M_{\rm gal}\gtrsim 10^{10}\,M_\odot$; \citealt{Miller2015}). This approach has been widely applied to compute the BHMF at $M_{\rm BH}\gtrsim 10^6\,M_\odot$ \citep{Marconi2004, Merloni2008, Yu2008, Shankar2009, Vika2009, Shankar2016}.

In a few nearby dwarf galaxies, however, stellar dynamical measurements have placed stringent upper limits on $M_{\rm BH}$ (e.g., \citealt{Gebhardt2001, Valluri2005}), suggesting that the occupation fraction in low-mass galaxies is $<$100\%. 
An empirical method to constrain the occupation fraction is to use high spatial resolution \chandra X-ray observations \citep{Gallo2008, Gallo2010, Miller2012}. 
By assuming that the nuclear X-ray luminosity $L_{\rm X}$ is a power-law function of $M_{\rm gal}$ with Gaussian scatter \citep{Gallo2019_wp}, and that the occupation fraction $f_{\rm occ}(M_{\rm gal})$ follows
\begin{align}
    0.5 + 0.5\times {\rm tanh} \left[ 2.5^{|8.9-{\rm log}M_{\rm gal, 0}|}{{\rm log}(M_{\rm gal}/M_{\rm gal, 0})}\right],\label{eq:focc}
\end{align}
one can simultaneously constrain the $L_{\rm X}$--$M_{\rm gal}$ relation and the critical galaxy mass $M_{\rm gal, 0}$ at which $f_{\rm occ}=0.5$. 
This approach was first adopted by \citet{Miller2015} using 194 early-type galaxies, and later updated by \citet{Gallo2019} using 326 early-type galaxies. The latter study found a BHMF slope of $p=-0.16\pm0.04$ (see Fig.~2 of \citealt{Gallo2019}).

The actual $f_{\rm occ}(M_{\rm gal})$ does not necessarily follow the functional form of Eq.~(\ref{eq:focc}). \citet{Greene2020} assumed two different shapes of $f_{\rm occ}$, with the pessimistic case drawn as a linear curve and the optimistic case provided by the fraction of nuclear star cluster (NSC) from \citet{Sanchez-Janssen2019}. The authors then converted the GMF of \citet{Wright2017} into the local BHMF using the $M_{\rm BH}$--$M_{\rm gal}$ relation (gray lines in Figure~\ref{fig:Mgal_Mbh}). 
The BHMFs thus derived exhibit $p=0.00\pm0.03$ and $p=-0.05\pm0.03$ in the pessimistic case and optimistic case, respectively.

\begin{figure}[htbp!]
    \includegraphics[width=\columnwidth]{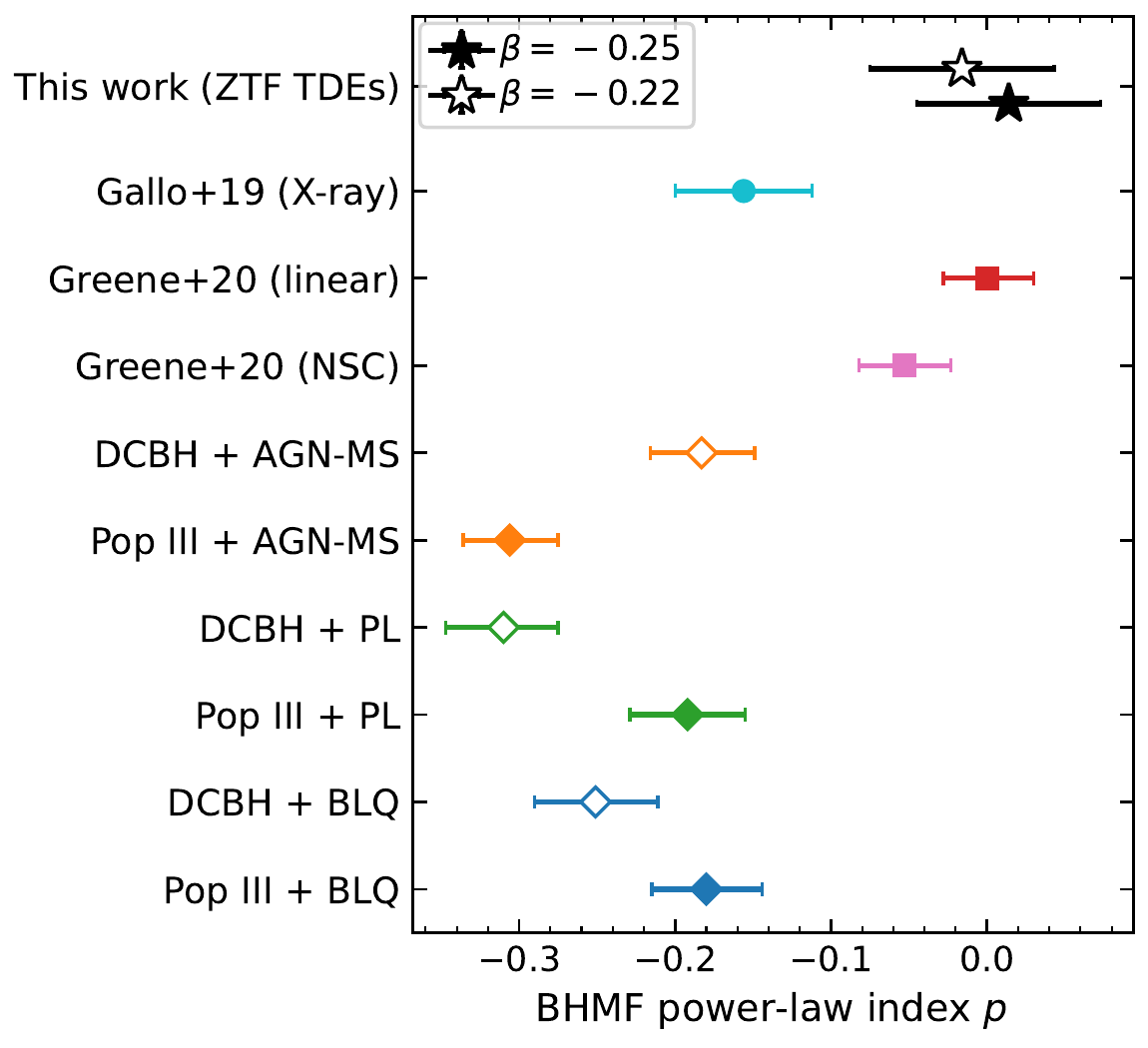}
    \caption{Power-law index of the local BHMF ($dn_{\rm BH}/d{\rm log}M_{\rm BH} \propto M_{\rm BH}^{p}$) from the optical TDE mass function (asterisks) with two assumptions on $\beta$ (see Eq.~\ref{eq:dRdMbh}), X-ray nuclei observations (circle, \citealt{Gallo2019}), GMF+scaling relations (squares, \citealt{Greene2020}), and SAMs (diamonds, \citealt{Chadayammuri2023}). \label{fig:p_bhmf}} 
\end{figure}

The slope of the BHMF inferred with optical TDEs is consistent with that of the \citealt{Greene2020} method, whereas the \citealt{Gallo2019} value is $\approx \! 2\sigma$ lower than our result (see Figure~\ref{fig:p_bhmf}).
Among the two BHMFs presented in Fig.~6 of \citealt{Greene2020}, we are not able to differentiate the nuances under various $f_{\rm occ}$ assumptions with the current sample size.

Next, we compare our result with physically motivated BHMFs from the semi-analytic models (SAMs) presented in \citet{Ricarte2018b}, \citet{Ricarte2018a}, \citet{Ricarte2019}, and \cite{Chadayammuri2023}, which include halo masses down to $10^7\, M_\odot$ from redshifts $0<z<20$. We explore two different BH seeding models, and three different BH growth prescriptions. Population III (Pop III) models place a \textit{light} seed initialized at approximately $10^2\, M_\odot$ in almost all dwarf galaxies by $z=0$, while the direct collapse black hole (DCBH) models place a \textit{heavy} seed of approximately $10^5\, M_\odot$ in a subset of these halos. 

These SAMs do not model the astrophysics of galaxy formation, and instead use empirical relations to determine the BH growth rate across cosmic time. Each of them includes a \textit{burst} mode triggered during a major merger until the BH reaches the $M_{\rm BH}$--$\sigma_\ast$ relation, and a \textit{steady} mode that operates otherwise.
Under the power-law (PL) growth, BHs grow at the Eddington rate during the burst mode, and otherwise draw from a universal power-law Eddington ratio distribution. 
Under the AGN main-sequence (AGN-MS) growth, BHs grow at the Eddington rate during the burst mode, and otherwise accrete at a fixed fraction of the star formation rate. 
The broad-line quasar (BLQ) growth only contains a burst mode, where BHs grow at an Eddington ratio drawn from a log-normal distribution that was fit to BLQs \citep{Kelly&Shen2013, Tucci&Volonteri2017}. 

These SAMs all match the $M_{\rm BH}$--$\sigma_\ast$ relation at high masses but deviate at lower masses depending on the seeding and accretion prescriptions. Figure~\ref{fig:p_bhmf} shows the measured power-law slope of the resulting BHMFs in the mass range of $10^{5.3}<M_{\rm BH}<10^{7.3}\,M_\odot$.
Interestingly, the SAMs generally show a higher fraction of lower-mass BHs that are not seen in the BHMF shape determined by optical TDEs.

\subsubsection{Caveats} \label{subsubsec:caveats}

The above analysis only includes the optical TDE sample. Therefore, the implications for the local BHMF are only robust if the $M_{\rm BH}$ distribution of optical TDEs is representative of the underlying $M_{\rm BH}$ distribution of all TDEs. 
While previous studies do not find a significant difference in the $M_{\rm BH}$ distributions between optically and X-ray selected TDE samples \citep{Wevers2019_Mbh, French2020}, we note that the literature samples consist of events from various surveys with different sensitivity and selection criteria. A robust assessment requires detailed understanding of how TDE emission properties (across the electromagnetic spectrum from X-ray to radio) depend on $M_{\rm BH}$ in a way that biases the sample $M_{\rm BH}$ distributions under different selection criteria.

We also note that, in order to obtain the BHMF, we assumed that the $M_{\rm BH}$--$\sigma_\ast$ relation remains valid down to $M_{\rm BH}\sim 10^5\,M_\odot$. There are two caveats associated with this assumption: (i) the number of dynamical $M_{\rm BH}$ measurements at $M_{\rm BH}\lesssim 10^6\,M_\odot$ is still insufficient to robustly test the $M_{\rm BH}$--$\sigma_\ast$ relation in the IMBH regime \citep{Greene2020}, (ii) the \citet[][Eq.~3]{Kormendy2013} relation is mainly based on massive elliptical galaxies. If using the $M_{\rm BH}$--$\sigma_\ast$ relations derived by \citet{Gultekin2009, Greene2020}, the inferred $M_{\rm BH}$ will be lower by by 0.2--0.4\,dex across the range of $\sigma_\ast$ measurements, whereas the shape of the inferred BHMF remains flat.

\section{Summary} \label{sec:conclusion}

We present a complete flux-limited sample of 55 blue nuclear transients systematically selected with ZTF. Among the 55 objects, 33 are classified as TDEs. Their BH masses are inferred with host galaxy scaling relations (using central velocity dispersion $\sigma_\ast$ for 19 objects, and using galaxy total stellar mass $M_{\rm gal}$ for the other 14 objects).
We recovered a number of correlations between $M_{\rm BH}$ and photometric properties (\S\ref{subsec:correlation}).

For rate inferences, we develop a survey efficiency corrected maximum volume method (\S\ref{sec:method}).
We present the rest-frame $g$-band LF (\S\ref{subsubsec:rate_Lg}), precisely constrain the upper end ($10^{43.5}\lesssim L_g \lesssim 10^{44.7}\,{\rm erg\,s^{-1}}$) for the first time, and observe a shallower slope (compared to \citealt{vanVelzen2018}) at the low end  ($10^{42.5}\lesssim L_g \lesssim 10^{43.1}$\,erg\,s$^{-1}$) that drives a $\approx 2\times $ reduction in the inferred volumetric rate. 
Using a newly determined LF in terms of the peak UV and optical blackbody luminosity (\S\ref{subsubsec:rate_Lbb}), we find the rates of optically loud and X-ray loud TDEs are comparable. 

We construct the optical TDE mass function (\S\ref{subsec:Mbh_func}), confirming the previous result of rate suppression due to event horizon, and revealing a $\phi(M_{\rm BH})\propto M_{\rm BH}^{-0.25}$ dependence at $10^{5.3}\,M_\odot\lesssim M_{\rm BH} \lesssim 10^{7.3}\,M_\odot$.
This indicates that the local BHMF is relatively flat  (\S\ref{subsec:local_bhmf}). 
At a typical galaxy mass of $10^{10}\,M_\odot$, we constrain the per-galaxy TDE rate to be [3.7, 7.4, and $1.6]\times 10^{-5}\,{\rm galaxy^{-1}\,yr^{-1}}$ in galaxies with red, green, and blue colors, respectively (\S\ref{subsec:rate_Mgal}).

While we have mainly focused on TDE demographics in this paper, the TDE sample presented here can also be used to address the origin of TDE's UV and optical emission, and to train machine-learning algorithms (e.g., \citealt{Gomez2023}) for real-time photometric selection of TDE candidates. The luminosity and mass functions of optical TDEs should ultimately be compared to that of X-ray-, infrared- and radio-selected TDEs.

Over the next few years, we expect substantial progresses to be made in studies of TDE demographics. 
The excellent angular resolution and depth of the Vera Rubin Observatory Legacy Survey of Space and Time \citep{Ivezic2019} will enable the creation of a reference galaxy catalog that is complete to low-mass galaxies out to higher redshifts. 
Since TDE BH mass scales positively with transient duration [see Eq.~(\ref{eq:t1/2_2}) and panel (e) of Figure~\ref{fig:see_pars}], the selection of fast-evolving TDEs will rely on high-cadence wide-field experiments such as those conducted by ZTF, the La Silla Schmidt Southern Survey (LS4), and the wide-field (200\,deg$^2$) \textit{Ultraviolet Transient Astronomy Satellite} \citep{Ben-Ami2022}.

\vspace{1cm}

\textit{Acknowledgements} -- 
We thank Jenny Greene for insightful discussions, and Ryan Chornock for providing the spectrum of PS1-11af.
We thank Morgan MacLeod, Nick Stone, and Brian Metzger for constructive comments.

Y.Y. and S.R.K. acknowledge support from Heising-Simons Foundation.
E.K.H. acknowledges support from NASA under award No.~80GSFC21M0002.
M.N. is supported by the European Research Council (ERC) under the European Union’s Horizon 2020 research and innovation program (grant agreement No.~948381) and by funding from the UK Space Agency.
This research benefited from interactions at workshops funded by the Gordon and Betty Moore Foundation through grant GBMF5076. 
This research was made possible in part through the support of grants from the Gordon and Betty Moore Foundation and the John Templeton Foundation. The opinions expressed in this publication are those of the author(s) and do not necessarily reflect the views of the Moore or Templeton Foundations.

This work is based on observations obtained with the Samuel Oschin Telescope 48-inch and the 60-inch Telescope at the Palomar Observatory as part of the Zwicky Transient Facility project. ZTF is supported by the National Science Foundation under grants No. AST-1440341 and AST-2034437 and a collaboration including current partners Caltech, IPAC, the Weizmann Institute of Science, the Oskar Klein Center at Stockholm University, the University of Maryland, Deutsches Elektronen-Synchrotron and Humboldt University, the TANGO Consortium of Taiwan, the University of Wisconsin at Milwaukee, Trinity College Dublin, Lawrence Livermore National Laboratories, IN2P3, University of Warwick, Ruhr University Bochum, Northwestern University and former partners the University of Washington, Los Alamos National Laboratories, and Lawrence Berkeley National Laboratories. Operations are conducted by Caltech Optical Observatories, IPAC, and University of Washington.

The ZTF forced-photometry service was funded under the Heising-Simons Foundation grant No.~12540303 (PI: Graham).
SED Machine is based upon work supported by the National Science Foundation under grant No.~1106171.

This work has made use of data from the Asteroid Terrestrial-impact Last Alert System (ATLAS) project. The ATLAS project is primarily funded to search for near-earth asteroids through NASA grants NN12AR55G, 80NSSC18K0284, and 80NSSC18K1575; byproducts of the NEO search include images and catalogs from the survey area. This work was partially funded by Kepler/K2 grant J1944/80NSSC19K0112 and Hubble Space Telescope GO-15889, and STFC grants ST/T000198/1 and ST/S006109/1. The ATLAS science products have been made possible through the contributions of the University of Hawaii Institute for Astronomy, the Queen’s University Belfast, the Space Telescope Science Institute, the South African Astronomical Observatory, and The Millennium Institute of Astrophysics (MAS), Chile.


\appendix
\section{Supplementary Tables} \label{sec:obs_log}

\begin{deluxetable}{cccccc}[htbp!]
	\tablecaption{UV and Optical Photometry of 33 TDEs.
            \label{tab:tdephot}}
	\tablehead{
        \colhead{IAU Name} &
		\colhead{MJD} &
		\colhead{Instrument} &
		\colhead{Filter} &
		\colhead{$f_\nu$ ($\mu$Jy)} &
		\colhead{$\sigma_{f_\nu}$ ($\mu$Jy)} 
	}
\startdata
AT2021mhg & 59421.5384 & ATLAS & $o$ & 23.2366 & 8.7284 \\
AT2021mhg & 59422.3478 & ZTF & $i$ & 6.4130 & 11.8362 \\
AT2021mhg & 59422.4213 & ZTF & $r$ & 29.2828 & 3.0782 \\
AT2021mhg & 59422.4560 & ZTF & $g$ & 50.9955 & 4.8532 \\
AT2021mhg & 59424.3924 & ZTF & $r$ & 23.6341 & 3.0218 \\
AT2021uqv & 59454.7033 & UVOT & $uvw1$ & 92.3872 & 8.6091 \\
AT2021uqv & 59454.7044 & UVOT & $U$ & 80.0122 & 13.5501 \\
AT2021uqv & 59454.7062 & UVOT & $uvw2$ & 89.1477 & 6.2627 \\
AT2021uqv & 59454.7097 & UVOT & $uvm2$ & 91.1047 & 6.2769 \\
AT2021uqv & 59455.3383 & ZTF & $g$ & 70.7813 & 4.0499 \\
AT2021yzv & 59524.3409 & ZTF & $g$ & 88.7688 & 2.8362 \\
AT2021yzv & 59524.3631 & ZTF & $r$ & 72.1670 & 2.9713 \\
AT2021yzv & 59524.5512 & ATLAS & $c$ & 79.4877 & 3.7740 \\
AT2021yzv & 59526.3054 & ZTF & $i$ & 58.4191 & 5.1675 \\
AT2021yzv & 59526.3680 & ZTF & $g$ & 84.3688 & 2.6204 \\
\enddata
\tablecomments{$f_\nu$ is observed flux density before extinction correction. 
Only 15 observations of three objects are shown to present the format of this table, which is available in its entirety in the machine-readable version online.}
\end{deluxetable}

The UV and optical photometry of 33 TDEs is presented in Table~\ref{tab:tdephot}.
The observing logs of low-resolution spectroscopy and ESI spectroscopy are provided in Table~\ref{tab:spec_lowres} and Table~\ref{tab:spec_medres}, respectively.
The pre-flare host galaxy photometry is provided in Table~\ref{tab:hostphot1} and Table~\ref{tab:hostphot2}.

\begin{deluxetable*}{rlcrccccr}[htbp!]
\tabletypesize{\scriptsize}
\tablecaption{Log of Low-Resolution Optical Spectroscopy. \label{tab:spec_lowres}}
\tablehead{
	\colhead{ID}
    & \colhead{IAU Name}
    & \colhead{Start Date}  
	& \colhead{$t$ (days)}
	& \colhead{Telescope}
	& \colhead{Instrument}
	& \colhead{Wavelength Range (\AA)}
	& \colhead{Slit Width ($^{\prime\prime}$)}
	& \colhead{Exposure Time (s)} 
	}
\startdata
17 & AT2020vwl & 2021-01-11.5 & $+54$ & LDT & DeVeny & 3586--8034 & 1.5 & 2700 \\
\hline
\multirow{2}{*}{19} & \multirow{2}{*}{AT2020yue} & 2022-11-17.6$^{\dagger}$ &  $+599$ & Keck I & LRIS & 3200--10250 & 1.0 & 2700 \\
 &  & 2022-11-25.6$^{\dagger}$ &  $+605$ & Keck I & LRIS & 3200--10250 & 1.0 & 2400 \\
\hline
20 & AT2020abri & 2022-04-07.5 & $+395$ & Keck I & LRIS & 3200--10250 & 1.0 & 1500 \\
\hline
\multirow{9}{*}{21} & \multirow{9}{*}{AT2020acka} & 2021-01-14.5 & $+7$ & P60 & SEDM & 3770--9223 & --- & 2700 \\
 &  & 2021-01-16.5 & $+9$ & P60 & SEDM & 3770--9223 & --- & 2700 \\
 &  & 2021-02-08.5 & $+26$ & P60 & SEDM & 3770--9223 & --- & 2700 \\
 &  & 2021-02-20.5 & $+31$ & P200 & DBSP & 3410--5550, 5750--9995 & 1.5 & 1200 \\
 &  & 2021-04-14.5 & $+70$ & Keck I & LRIS & 3200--10250 & 1.0 & 400 \\
 &  & 2021-06-07.5 & $+111$ & Keck I & LRIS & 3200--10250 & 1.0 & 430 \\
 &  & 2021-08-13.3 & $+161$ & Keck I & LRIS & 3200--10250 & 1.0 & 430 \\
 &  & 2021-09-07.3 & $+179$ & Keck I & LRIS & 3200--10250 & 1.0 & 900 \\
 &  & 2022-02-06.6 & $+293$ & Keck I & LRIS & 3200--10250 & 1.0 & 900 \\
\hline
22 & AT2021axu & 2021-06-07.3 & $+100$ & Keck I & LRIS & 3200--10250 & 1.0 & 485 \\
\hline
23 & AT2021crk & 2021-04-09.4 & $+34$ & P200 & DBSP & 3410--5550, 5750--9995 & 1.5 & 1200 \\
\hline
25 & AT2021jjm & 2021-05-13.5 & $+17$ & Keck I & LRIS & 3200--10250 & 1.0 & 300 \\
\hline
26 & AT2021mhg & 2021-08-01.4 & $+51$ & P200 & DBSP & 3410--5550, 5750--9995 & 1.5 & 1800 \\
\hline
27 & AT2021nwa & 2021-07-06.3 & $-1$ & Keck I & LRIS & 3200--10250 & 1.0 & 300 \\
\hline
\multirow{2}{*}{28} & \multirow{2}{*}{AT2021qth} & 2021-08-04.2 & $+27$ & P200 & DBSP & 3410--5550, 5750--9995 & 1.5 & 900 \\
 &  & 2022-05-26.3 & $+300$ & Keck I & LRIS & 3200--10250 & 1.0 & 900 \\
\hline
29 & AT2021sdu & 2021-08-13.4 & $+18$ & Keck I & LRIS & 3200--10250 & 1.0 & 750 \\
\hline
30 & AT2021uqv & 2021-09-07.5 & $+25$ & Keck I & LRIS & 3200--10250 & 1.0 & 600 \\
\hline
31 & AT2021utq & 2022-10-03.2 & $+353$ & P200 & DBSP & 3410--5550, 5750--9995 & 1.0 & 1500 \\
\hline
\multirow{3}{*}{32} & \multirow{3}{*}{AT2021yzv} & 2021-10-04.6 & $-15$ & Keck I & LRIS & 3200--10250 & 1.0 & 600 \\
 &  & 2022-02-05.3 & $+80$ & Keck I & LRIS & 3200--10250 & 1.0 & 900 \\
& & 2023-01-16.4 & $+349$ & Keck I & LRIS & 3200--10250 & 1.0 & 1200 \\
\hline
33 & AT2021yte & 2021-10-14.5 & $+14$ & P200 & DBSP & 3410--5550, 5750--9995 & 1.5 & 900 \\
\enddata 
\tablecomments{
$^{\dagger}$: On 2022 November 17, one exposure (900\,s) on the red CCD is badly affected by cosmic rays and is therefore not included in spectral extraction. We stack the observations on 2022 November 17 and 2022 November 25 together to create a deep spectrum for analysis.  
}
\end{deluxetable*}

\begin{deluxetable*}{rlccccccc}[htbp!]
\tabletypesize{\scriptsize}
\tablecaption{Details of ESI Spectroscopy. \label{tab:spec_medres}}
\tablehead{
	\colhead{ID}
    & \colhead{IAU Name}
    & \colhead{Start Date}  
	& \colhead{Slit Width}
	& \colhead{Exposure Time} 
    & \colhead{$r_{\rm extract}$}
    & \colhead{Fitted $\lambda_{\rm rest}$}
	& \colhead{$\sigma_\ast$}
    & \colhead{S/N}\\
	\colhead{}
	& \colhead{}
	& \colhead{}
	& \colhead{($^{\prime\prime}$)}
	& \colhead{(s)}
    & \colhead{(pixel)}
    & \colhead{(\AA)}
	& \colhead{($\rm km\, s^{-1}$)}
    & \colhead{}
	}
\startdata
1 & AT2018iih & 2022-07-04.5 & 0.5 & 1200 & 4.2 & 5030--5600 & $148.6\pm 14.4$ & 6.9\\
5 & AT2019azh & 2022-10-21.6 & 0.5 & 1200 & 5.7 & 5030--5600 & $68.0\pm 2.0$ & 33.3 \\
8 & AT2019dsg & 2022-08-24.4 & 0.5 & 900 & 4.3 &5030--5600 &  $86.9\pm 3.9$ & 16.9\\
13 & AT2020mot & 2022-10-21.4 & 0.5 & 1200 & 9.3 & 5030--5600 & $76.6\pm 5.3$ &  8.8\\
15 & AT2020ysg & 2023-03-26.4 & 0.75 & 2400 & 7.8 & 5030--5392, 5407--5600 & $157.8 \pm 13.0$ &  13.6\\
16 & AT2020vdq & 2022-11-25.5 & 0.3 & 2700 & 5.8 & 5030--5600 & $43.6\pm 3.1$ & 12.0  \\
17 & AT2020vwl & 2022-03-07.6 & 0.5 & 600 & 4.2 & 5030--5600 & $48.5\pm 2.0$ & 11.6 \\
18 & AT2020wey & 2022-10-22.6 & 0.5 & 600 & 8.2 & 5030--5600 & $40.1\pm 3.1$ & 7.4 \\
21 & AT2020acka & 2022-03-07.6 & 0.5 & 2400 & 6.0 & 5030--5127, 5159--5600 & $174.5\pm 25.3$ & 9.1 \\
\hline
\multirow{2}{*}{22} & \multirow{2}{*}{AT2021axu} & 2022-03-07.3 & 0.5 & 1500 & \multirow{2}{*}{4.3} & \multirow{2}{*}{5030--5600} & \multirow{2}{*}{$73.5\pm 17.3$} & \multirow{2}{*}{7.2} \\
& & 2022-11-25.6 & 0.5 & 2400 \\
\hline
\multirow{2}{*}{23} & \multirow{2}{*}{AT2021crk} & 2022-03-07.3 & 0.5 & 1600 & \multirow{2}{*}{5.6} & \multirow{2}{*}{5030--5083, 5137--5600}  &\multirow{2}{*}{$57.6\pm 6.3$} & \multirow{2}{*}{6.8}\\
 &  & 2022-11-25.6 & 0.5 & 2400 &  \\
\hline
24 & AT2021ehb & 2021-12-28.4 & 0.75 & 300 & 5.0 & 5030--5600 & $99.6\pm 3.8$ & 18.4\\
26 & AT2021mhg & 2022-10-22.3 & 0.5 & 1800 & 4.2 & 5030--5196, 5200--5600 & $57.8\pm 5.3$ & 8.1 \\
27 & AT2021nwa & 2022-03-07.7 & 0.5 & 600 & 4.6 & 5030--5600 & $102.4\pm 5.4$ & 11.3\\
30 & AT2021uqv & 2022-08-24.5 & 0.5 & 1200 & 5.0 & 5030--5310, 5346--5600& $62.3\pm 7.1$ & 10.6\\
32 & AT2021yzv & 2023-03-26.3 & 0.75 & 2400 & 8.2 & 4900--5335, 5369--5600 & $146.4\pm20.8$ & 8.6 \\
33 & AT2021yte & 2022-03-07.2 & 0.5 & 1120 & 3.8 & 5030--5578 & $34.2\pm 4.8$ & 7.3 \\
\enddata 
\tablecomments{
All ESI spectra were obtained after the optical TDE flux has faded to $<10$\% of the host galaxy flux. 
$r_{\rm extract}$ can be converted to angular scale using a conversion factor of 0.154$^{\prime\prime}$ per pixel.}
\end{deluxetable*}

\begin{deluxetable*}{cccccccccc}[htbp!]
\scriptsize
	\tablecaption{\galex, SDSS, and WISE Photometry of TDE Host Galaxies.\label{tab:hostphot1}}
	\tablehead{
		\colhead{ID}   
		& \colhead{FUV}
		& \colhead{NUV}
        & \colhead{SDSS/$u$}
        & \colhead{SDSS/$g$}
        & \colhead{SDSS/$r$}
        & \colhead{SDSS/$i$}
        & \colhead{SDSS/$z$}
        & \colhead{WISE/$W1$}
        & \colhead{WISE/$W2$}
	}
	\startdata
1  &  &  &  & $20.55\pm0.19$ & $19.24\pm0.16$ & $18.80\pm0.17$ & $18.28\pm0.17$ &  &  \\
2  &  & $23.73\pm0.89$ &  & $22.73\pm0.28$ & $21.32\pm0.13$ & $20.95\pm0.14$ & $20.23\pm0.17$ & $20.59\pm0.57$ & $20.66\pm0.45$ \\
3  &  &  &  &  &  &  &  & $19.57\pm0.19$ & $20.91\pm0.62$ \\
4  & $20.91\pm0.27$ & $20.19\pm0.12$ & $19.63\pm0.13$ & $18.37\pm0.03$ & $17.69\pm0.01$ & $17.28\pm0.02$ & $17.11\pm0.19$ & $17.02\pm0.04$ & $17.47\pm0.04$ \\
5  & $19.24\pm0.18$ & $17.83\pm0.03$ & $16.51\pm0.08$ & $15.01\pm0.02$ & $14.49\pm0.01$ & $14.20\pm0.01$ & $14.04\pm0.04$ & $14.60\pm0.01$ & $15.23\pm0.02$ \\
6  & $22.51\pm1.04$ & $21.12\pm0.23$ & $20.13\pm0.28$ & $19.00\pm0.04$ & $18.24\pm0.02$ & $17.81\pm0.03$ & $17.71\pm0.10$ & $17.78\pm0.04$ & $18.36\pm0.05$ \\
7  &  &  &  &  &  &  &  & $19.22\pm0.13$ & $19.89\pm0.17$ \\
8  & $21.19\pm0.32$ & $21.22\pm0.26$ &  &  &  &  &  & $15.65\pm0.02$ & $16.16\pm0.02$ \\
9  &  & $22.54\pm0.19$ & $20.29\pm0.30$ & $19.28\pm0.06$ & $18.52\pm0.07$ & $18.24\pm0.07$ & $17.96\pm0.18$ & $18.50\pm0.07$ & $19.08\pm0.09$ \\
10  &  &  &  &  &  &  &  & $13.95\pm0.02$ & $14.60\pm0.04$ \\
11  & $21.12\pm0.09$ & $20.87\pm0.04$ & $20.04\pm0.10$ & $19.07\pm0.02$ & $18.55\pm0.01$ & $18.23\pm0.02$ & $17.97\pm0.06$ & $18.40\pm0.05$ & $18.90\pm0.07$ \\
12  &  &  & $20.05\pm0.09$ & $18.63\pm0.02$ & $17.90\pm0.01$ & $17.50\pm0.02$ & $17.33\pm0.04$ & $17.56\pm0.04$ & $18.13\pm0.04$ \\
13  & $22.65\pm0.64$ & $21.56\pm0.25$ &  &  &  &  &  & $16.66\pm0.03$ & $17.21\pm0.03$ \\
14  & $20.91\pm0.34$ & $19.61\pm0.11$ & $19.18\pm0.09$ & $18.20\pm0.01$ & $17.68\pm0.02$ & $17.35\pm0.02$ & $17.23\pm0.05$ & $17.53\pm0.05$ & $18.01\pm0.05$ \\
15  &  & $21.97\pm0.25$ & $21.79\pm0.70$ & $21.14\pm0.26$ & $19.81\pm0.07$ & $19.22\pm0.07$ & $19.46\pm0.35$ & $18.59\pm0.07$ & $19.10\pm0.08$ \\
16  &  &  & $19.80\pm0.14$ & $18.87\pm0.02$ & $18.26\pm0.02$ & $18.05\pm0.02$ & $17.94\pm0.12$ & $18.48\pm0.11$ & $18.98\pm0.12$ \\
17  &  &  & $18.81\pm0.12$ & $17.24\pm0.02$ & $16.53\pm0.02$ & $16.18\pm0.01$ & $15.90\pm0.05$ & $16.48\pm0.03$ & $17.16\pm0.04$ \\
18  & $21.82\pm0.34$ & $21.61\pm0.09$ & $18.88\pm0.09$ & $17.40\pm0.01$ & $16.70\pm0.01$ & $16.34\pm0.01$ & $16.11\pm0.02$ & $16.63\pm0.03$ & $17.22\pm0.03$ \\
19  & $21.98\pm0.34$ & $21.09\pm0.15$ &  &  &  &  &  & $18.49\pm0.07$ & $19.10\pm0.09$ \\
20  &  &  & $23.14\pm0.64$ & $21.83\pm0.14$ & $21.00\pm0.07$ & $20.63\pm0.07$ & $20.83\pm0.35$ & $20.75\pm0.20$ &  \\
21  &  &  & $22.84\pm0.89$ & $21.07\pm0.11$ & $19.71\pm0.07$ & $19.09\pm0.08$ & $18.66\pm0.17$ & $18.18\pm0.14$ & $18.65\pm0.16$ \\
22  &  &  &  & $20.34\pm0.04$ & $19.57\pm0.03$ & $19.29\pm0.06$ & $18.77\pm0.15$ & $19.33\pm0.10$ & $20.15\pm0.20$ \\
23  &  &  & $20.52\pm0.21$ & $19.51\pm0.04$ & $19.04\pm0.05$ & $18.68\pm0.07$ & $18.50\pm0.28$ & $18.98\pm0.10$ & $19.80\pm0.18$ \\
24  &  &  & $17.66\pm0.06$ & $15.86\pm0.01$ & $14.98\pm0.01$ & $14.50\pm0.01$ & $14.18\pm0.02$ & $14.57\pm0.02$ & $15.25\pm0.02$ \\
25  &  &  &  &  &  &  &  & $19.98\pm0.19$ & $20.11\pm0.17$ \\
26  &  &  &  &  &  &  &  & $18.70\pm0.10$ & $19.30\pm0.10$ \\
27  & $23.29\pm0.17$ & $22.05\pm0.09$ & $19.22\pm0.10$ & $17.67\pm0.01$ & $16.90\pm0.01$ & $16.51\pm0.01$ & $16.24\pm0.03$ & $16.69\pm0.03$ & $17.29\pm0.03$ \\
28  &  & $22.51\pm0.38$ &  &  &  &  &  &  &  \\
29  &  & $20.01\pm0.14$ &  &  &  &  &  & $16.11\pm0.04$ & $16.62\pm0.03$ \\
30  & $21.76\pm0.39$ & $20.76\pm0.13$ & $20.33\pm0.39$ & $18.79\pm0.05$ & $18.14\pm0.05$ & $17.78\pm0.04$ & $17.62\pm0.07$ & $17.72\pm0.06$ & $18.16\pm0.07$ \\
31  & $21.88\pm0.51$ & $21.61\pm0.31$ &  &  &  &  &  & $19.55\pm0.44$ & $20.49\pm1.05$ \\
32  &  &  &  &  &  &  &  & $18.68\pm0.11$ & $19.19\pm0.16$ \\
33  &  &  & $20.62\pm0.36$ &  &  &  &  & $18.03\pm0.25$ & $18.73\pm0.30$ \\
\enddata
\end{deluxetable*}

\begin{deluxetable*}{ccccccccc}[htbp!]
\scriptsize
	\tablecaption{PS1 and 2MASS Photometry of TDE Host Galaxies.\label{tab:hostphot2}}
	\tablehead{
        \colhead{ID}   
        & \colhead{PS1/$g$}
        & \colhead{PS1/$r$}
        & \colhead{PS1/$i$}
        & \colhead{PS1/$z$}
        & \colhead{PS1/$y$}
        & \colhead{2MASS/$J$}
        & \colhead{2MASS/$H$}
        & \colhead{2MASS/$K_{\rm s}$}
}
\startdata
 1  & $20.42\pm0.22$ & $19.18\pm0.15$ & $18.74\pm0.15$ & $18.54\pm0.16$ & $18.55\pm0.27$ & $18.66\pm0.41$ & $17.99\pm0.32$ & $17.66\pm0.28$ \\
2  & $23.02\pm0.67$ & $21.55\pm0.17$ & $20.96\pm0.15$ & $20.72\pm0.18$ & $20.78\pm0.34$ &  &  &  \\
3  & $20.41\pm0.26$ & $19.57\pm0.14$ & $19.20\pm0.13$ & $19.16\pm0.22$ & $18.93\pm0.15$ &  &  &  \\
4  & $18.34\pm0.05$ & $17.70\pm0.03$ & $17.29\pm0.01$ & $17.10\pm0.03$ & $16.82\pm0.06$ &  &  &  \\
5  & $14.99\pm0.03$ & $14.48\pm0.01$ & $14.26\pm0.01$ & $14.09\pm0.02$ & $13.88\pm0.03$ & $13.71\pm0.01$ & $13.61\pm0.02$ & $13.77\pm0.03$ \\
6  & $18.94\pm0.03$ & $18.24\pm0.02$ & $17.85\pm0.03$ & $17.57\pm0.05$ & $17.34\pm0.05$ &  & $17.17\pm0.12$ &  \\
7  &  & $21.53\pm0.15$ & $20.63\pm0.10$ & $20.63\pm0.14$ & $20.11\pm0.22$ &  &  &  \\
8  & $17.03\pm0.04$ & $16.20\pm0.02$ & $15.83\pm0.02$ & $15.57\pm0.03$ & $15.39\pm0.06$ & $15.02\pm0.02$ & $15.02\pm0.03$ & $15.13\pm0.04$ \\
9  & $19.30\pm0.07$ & $18.65\pm0.10$ & $18.28\pm0.06$ & $18.21\pm0.08$ & $18.03\pm0.11$ &  &  &  \\
10  & $15.01\pm0.05$ & $14.33\pm0.03$ & $13.91\pm0.06$ & $13.69\pm0.04$ & $13.44\pm0.05$ & $13.26\pm0.02$ & $12.95\pm0.02$ & $13.26\pm0.03$ \\
11  & $19.03\pm0.02$ & $18.52\pm0.01$ & $18.27\pm0.02$ & $18.05\pm0.03$ & $17.97\pm0.06$ & $17.62\pm0.12$ & $17.69\pm0.16$ &  \\
12  & $18.54\pm0.03$ & $17.92\pm0.02$ & $17.51\pm0.02$ & $17.34\pm0.03$ & $17.06\pm0.05$ & $16.72\pm0.05$ & $16.83\pm0.10$ & $17.04\pm0.12$ \\
13  & $17.99\pm0.03$ & $17.20\pm0.01$ & $16.76\pm0.01$ & $16.53\pm0.02$ & $16.39\pm0.05$ & $16.05\pm0.04$ &  & $15.98\pm0.06$ \\
14  & $18.10\pm0.04$ & $17.68\pm0.04$ & $17.33\pm0.02$ & $17.20\pm0.02$ & $17.07\pm0.06$ &  &  &  \\
15  & $21.44\pm0.26$ & $19.88\pm0.09$ & $19.37\pm0.04$ & $19.10\pm0.08$ & $19.26\pm0.28$ &  &  &  \\
16  & $18.79\pm0.06$ & $18.30\pm0.03$ & $18.03\pm0.02$ & $17.88\pm0.03$ & $17.82\pm0.09$ &  &  &  \\
17  & $17.17\pm0.05$ & $16.51\pm0.03$ & $16.16\pm0.03$ & $16.03\pm0.03$ & $15.87\pm0.06$ & $15.77\pm0.05$ & $15.38\pm0.05$ & $15.67\pm0.08$ \\
18  & $17.32\pm0.01$ & $16.69\pm0.01$ & $16.36\pm0.01$ & $16.15\pm0.01$ & $16.00\pm0.03$ &  &  &  \\
19  & $19.74\pm0.11$ & $19.33\pm0.05$ & $18.89\pm0.09$ & $18.71\pm0.21$ & $18.40\pm0.16$ &  &  &  \\
20  & $22.00\pm0.15$ & $20.87\pm0.05$ & $20.64\pm0.06$ & $20.63\pm0.09$ & $20.26\pm0.19$ &  &  &  \\
21  &  & $19.84\pm0.13$ & $19.17\pm0.07$ & $18.89\pm0.07$ & $18.69\pm0.24$ & $18.32\pm0.17$ & $18.01\pm0.21$ & $17.67\pm0.15$ \\
22  & $20.32\pm0.05$ & $19.53\pm0.05$ & $19.17\pm0.04$ & $19.01\pm0.07$ &  &  &  &  \\
23  & $19.66\pm0.07$ & $19.09\pm0.06$ & $18.86\pm0.05$ & $18.63\pm0.07$ & $18.62\pm0.15$ &  &  &  \\
24  & $15.73\pm0.02$ & $14.93\pm0.01$ & $14.49\pm0.01$ & $14.21\pm0.01$ & $13.98\pm0.02$ & $13.83\pm0.01$ & $13.62\pm0.01$ & $13.80\pm0.01$ \\
25  & $20.44\pm0.06$ & $20.08\pm0.04$ & $19.82\pm0.05$ & $19.58\pm0.04$ & $19.67\pm0.12$ &  &  &  \\
26  & $19.63\pm0.07$ & $18.93\pm0.03$ & $18.55\pm0.08$ & $18.33\pm0.07$ & $18.25\pm0.09$ & $18.26\pm0.18$ & $18.30\pm0.26$ & $18.07\pm0.22$ \\
27  & $17.56\pm0.02$ & $16.88\pm0.02$ & $16.52\pm0.01$ & $16.27\pm0.02$ & $16.17\pm0.04$ & $15.87\pm0.04$ & $15.59\pm0.04$ & $15.91\pm0.07$ \\
28  & $19.84\pm0.09$ & $19.01\pm0.09$ & $18.58\pm0.13$ & $18.33\pm0.09$ & $18.09\pm0.12$ &  &  &  \\
29  & $17.49\pm0.02$ & $16.84\pm0.02$ & $16.39\pm0.05$ & $16.18\pm0.05$ & $15.93\pm0.06$ & $15.73\pm0.03$ & $15.79\pm0.05$ & $15.46\pm0.04$ \\
30  & $18.79\pm0.03$ & $18.19\pm0.03$ & $17.88\pm0.01$ & $17.78\pm0.04$ & $17.53\pm0.11$ & $17.35\pm0.13$ & $17.29\pm0.17$ &  \\
31  & $20.03\pm0.06$ & $19.53\pm0.08$ & $19.19\pm0.07$ & $18.98\pm0.11$ & $18.91\pm0.11$ & $19.08\pm0.52$ &  &  \\
32  & $20.96\pm0.15$ & $19.90\pm0.14$ & $19.25\pm0.06$ & $19.04\pm0.12$ & $18.80\pm0.13$ &  & $18.09\pm0.29$ &  \\
33  & $19.34\pm0.22$ & $18.65\pm0.26$ & $18.18\pm0.25$ & $18.09\pm0.33$ & $17.77\pm0.24$ & $17.90\pm0.14$ & $17.92\pm0.23$ &  \\
 \enddata
\end{deluxetable*}

\section{Details of Sample Selection} \label{sec:details_selection}
Here, we justify a few selection cuts adopted in \S\ref{subsec:filtering}. 

\subsection{sgscore1}
\begin{figure}[htbp!]
    \centering
    \includegraphics[width=\columnwidth]{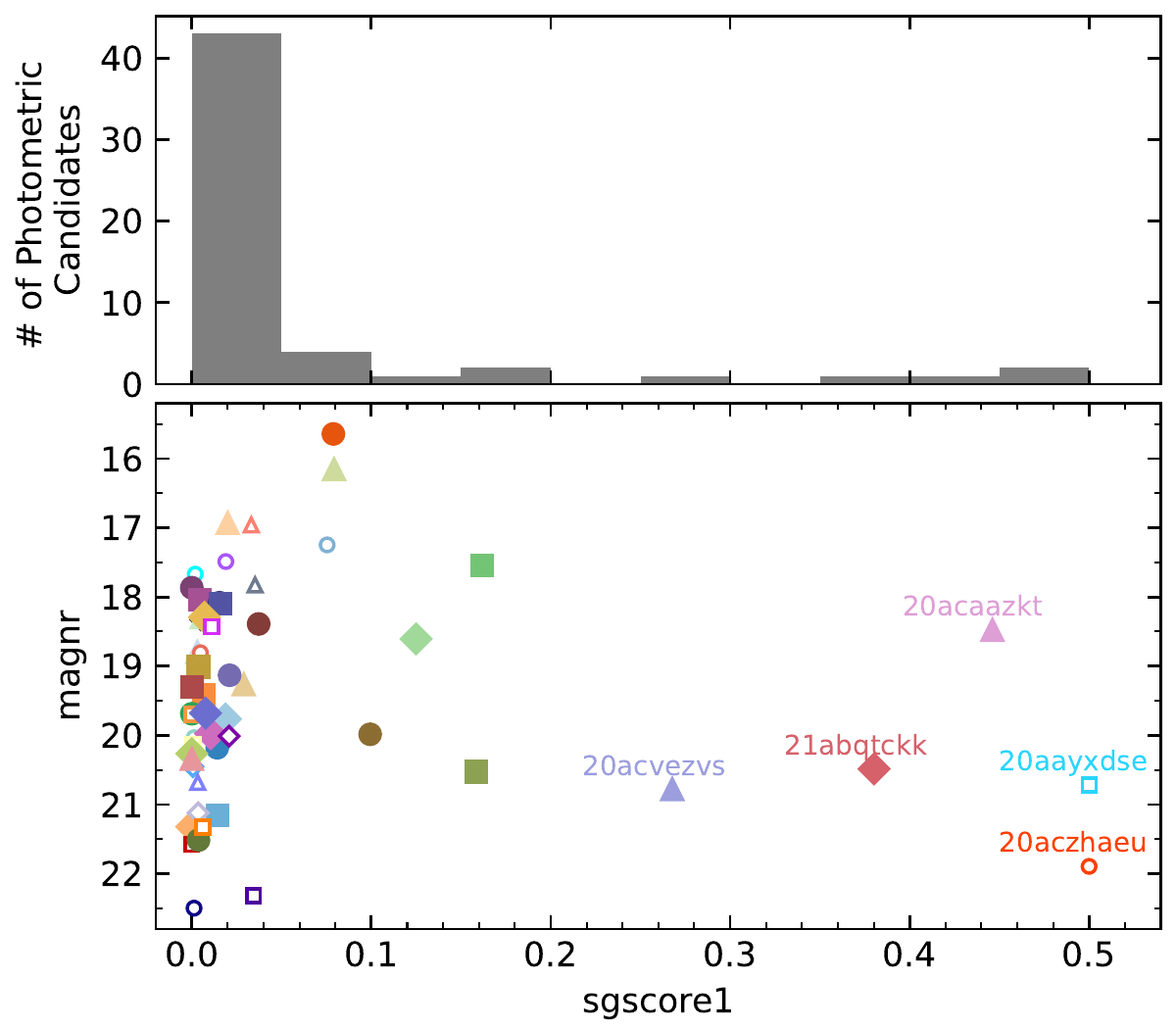}
    \caption{The \texttt{sgscore1} (star--galaxy classification score) and \texttt{magnr} (magnitude of the nearest object in the ZTF reference image) parameters of 55 photometric TDE candidates. 
    Symbol colors follow the same convention as in Figure~\ref{fig:color_evol}. The 33 TDEs are shown in solid markers, and the 22 false positives are shown in hollow markers.
    We show the ZTF names for objects with \texttt{sgscore1>0.2}. \label{fig:sgscore_hist}}
\end{figure}

The \texttt{sgscore} parameter is close to one (zero) for a star-like (galaxy-like) morphology.
Its value is set to 0.5 if the PS1 counterpart is not ``detected'' in the PS1 \texttt{StackObjectAttributes} table (see details in \citealt{Tachibana2018, Miller2021}).
In Figure~\ref{fig:sgscore_hist}, we show the distribution of the 55 photometric TDE candidates (after step 7 in \S\ref{subsec:filtering}) on the \texttt{magnr} versus \texttt{sgscore1} diagram.
The highest value of \texttt{sgscore} is 0.5, implying that our selection cut of \texttt{sgscore1} is sufficiently liberal. 

\subsection{Rise and Decline Timescales} 
\begin{figure}[htbp!]
    \centering
    \includegraphics[width=\columnwidth]{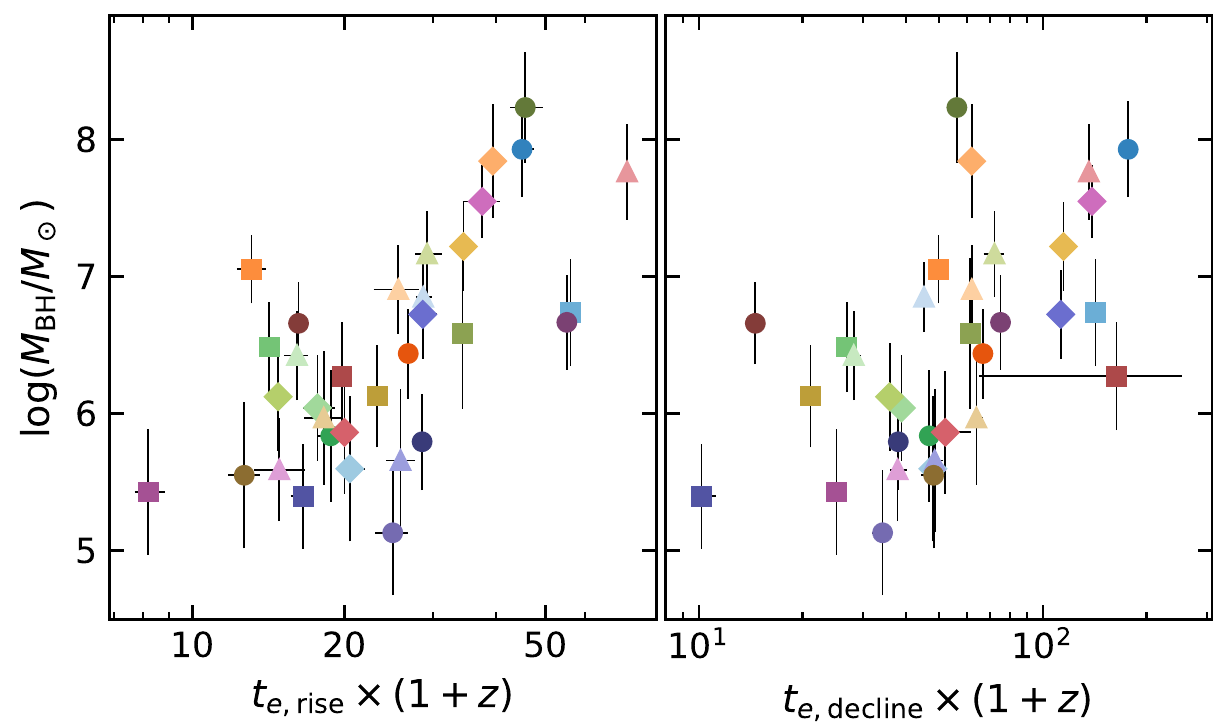}
    \caption{The black hole mass and observer-frame e-folding rise and decline timescales of 33 TDEs.
    \label{fig:te_selection}}
\end{figure}

In Figure~\ref{fig:te_selection}, the observer-frame e-folding rise and decline timescales (computed using the best-fit models derived in \S\ref{sec:lc_fitting}) are shown versus $M_{\rm BH}$. The values are well within the boundaries of 2 and 300 days, implying that our criteria adopted in steps (5) and (6) of \S\ref{subsec:filtering} are not at the boundaries.

\section{Host Galaxy SEDs and Comparison with Previous Studies} \label{sec:host_compare}

\begin{figure*}[htbp!]
    \centering
    \includegraphics[width=\textwidth]{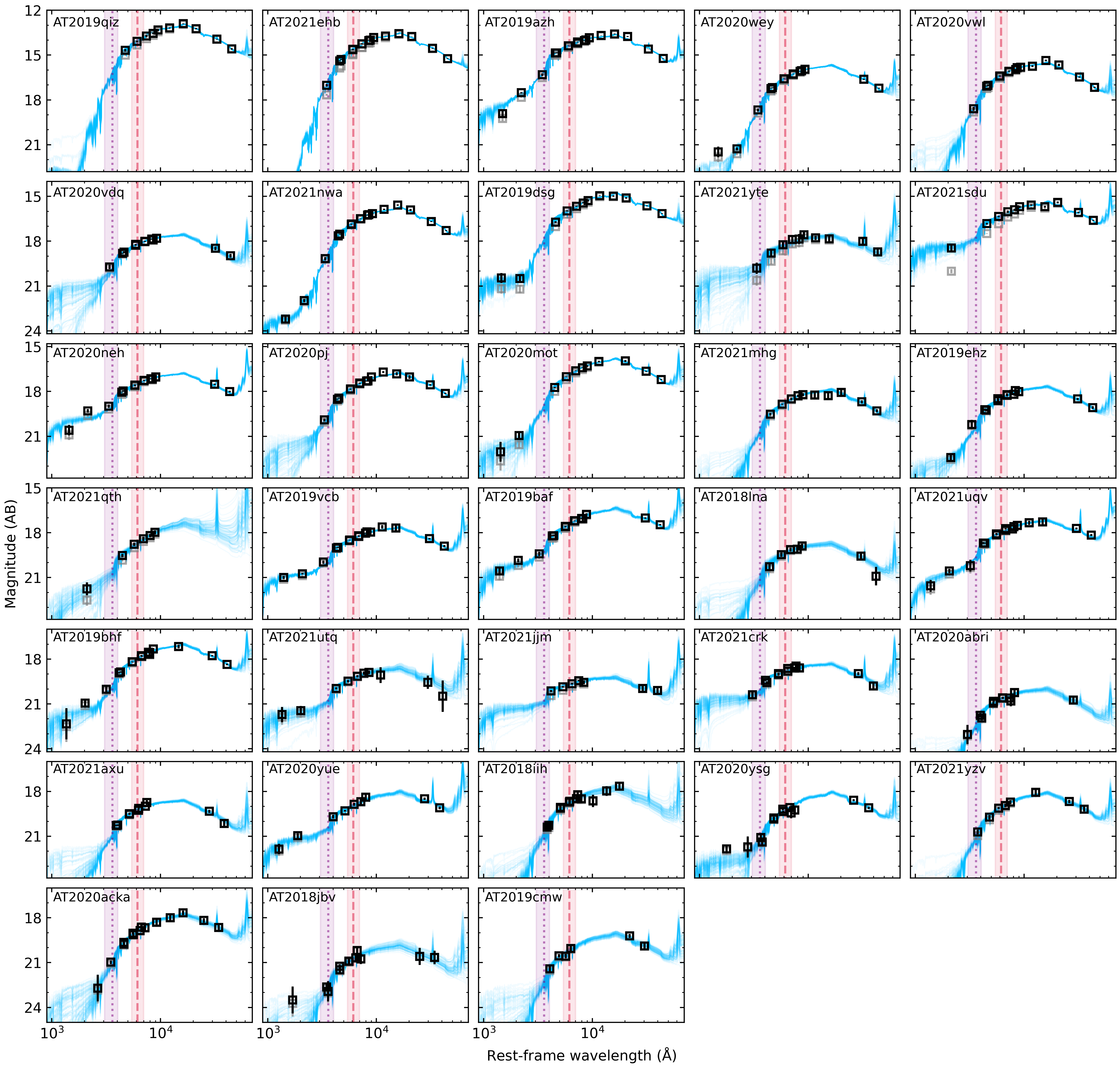}
    \caption{Host galaxy SEDs (sorted by redshift). The gray and black squares show the observed and Galactic extinction-corrected photometry, respectively. The blue lines show models of the 100 walkers in the MCMC sampler. The dotted and dashed vertical lines mark rest-frame wavelength of the SDSS $u$ and $r$ filters. 
    \label{fig:SED_library}}
\end{figure*}

Figure~\ref{fig:SED_library} shows the SEDs of 33 TDE host galaxies. 

\begin{figure}[htbp!]
    \centering
    \includegraphics[width=0.54\columnwidth]{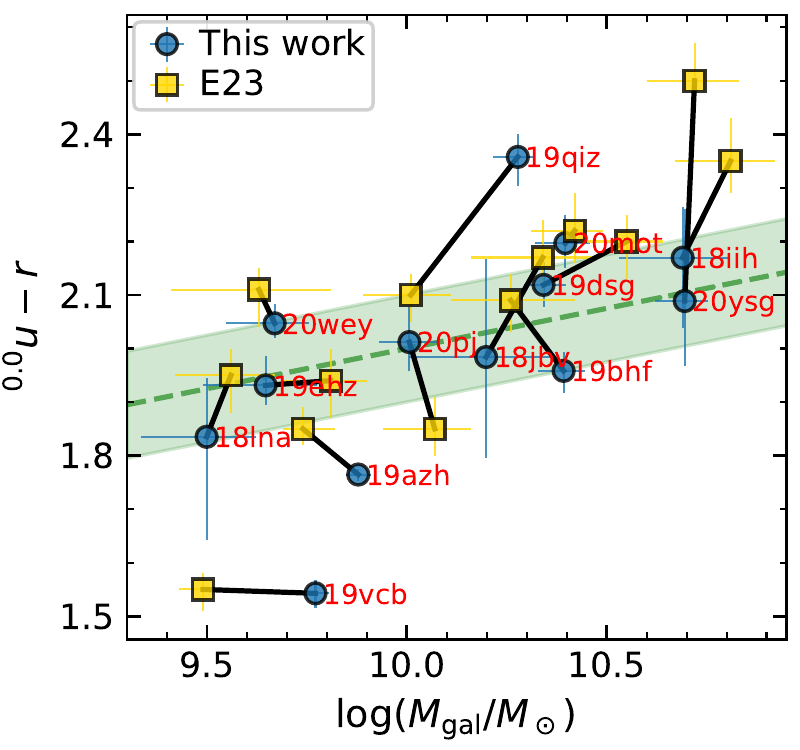}
    \includegraphics[width=0.44\columnwidth]{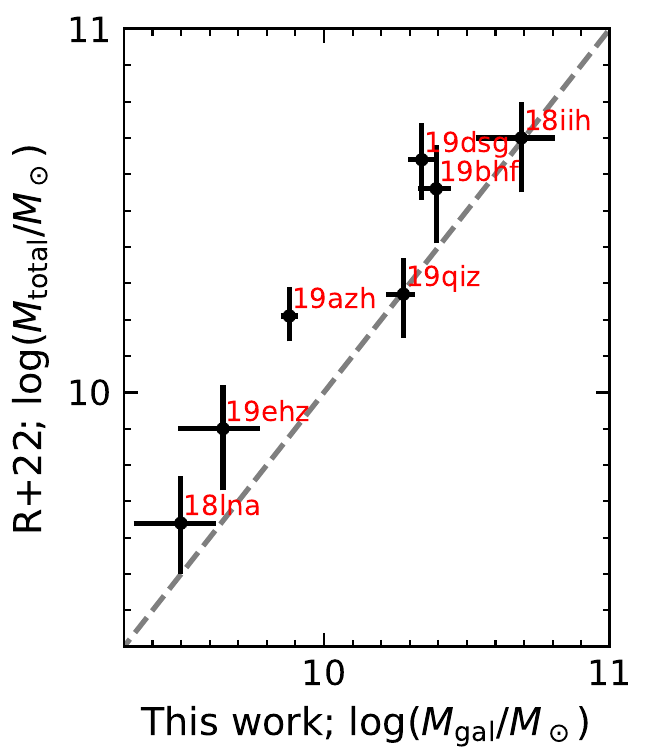}
    \caption{
    \textit{Left}: distributions of 13 galaxies on the color--mass diagram in both this work and \citet{Hammerstein2023}.
    Data points for the same object are connected with the solid black lines. The dashed green line shows the green valley defined by Eq.~(\ref{eq:green}).
    \textit{Right}: comparison of host galaxy total stellar mass derived in this work and \citet{Ramsden2022}. 
     \label{fig:compare_galaxies}}
\end{figure}

There are 13 galaxies in common between our sample and \citet{Hammerstein2023}.
The left panel of Figure~\ref{fig:compare_galaxies} shows the distributions of these objects on the galaxy color--mass diagram, using values derived in this work and \citet{Hammerstein2023}. 
For nine of the 13 objects, the log($M_{\rm gal}/M_\odot$) and $^{0,0}u-r$ parameters are consistent with each other (to within 2$\sigma$). For the other four objects (AT2019qiz, AT2019vcb, AT2019azh, and AT2020ysg), the difference probably comes from the different source of photometry: \citet{Hammerstein2023} obtained photometry from various catalogs whereas we measured the host brightness using \texttt{LAMBDAR} (see \S\ref{sec:host}).

There are 7 galaxies in common between our sample and \citet{Ramsden2022}. 
The mean offset in log($M_{\rm gal}/M_\odot$) between this work and \citet{Ramsden2022} is $-0.17$\,dex (see the right panel of Figure~\ref{fig:compare_galaxies}). 
The difference could be because \citet{Ramsden2022} used a nonparametric SFH, whereas we assumed a delayed exponentially declining function.

\bibliography{main}{}

\begin{thebibliography}{}
\expandafter\ifx\csname natexlab\endcsname\relax\def\natexlab#1{#1}\fi
\providecommand{\url}[1]{\href{#1}{#1}}
\providecommand{\dodoi}[1]{doi:~\href{http://doi.org/#1}{\nolinkurl{#1}}}
\providecommand{\doeprint}[1]{\href{http://ascl.net/#1}{\nolinkurl{http://ascl.net/#1}}}
\providecommand{\doarXiv}[1]{\href{https://arxiv.org/abs/#1}{\nolinkurl{https://arxiv.org/abs/#1}}}

\bibitem[{{Abazajian} {et~al.}(2009){Abazajian}, {Adelman-McCarthy},
  {Ag{\"u}eros}, {Allam}, {Allende Prieto}, {An}, {Anderson}, {Anderson},
  {Annis}, {Bahcall}, {Bailer-Jones}, {Barentine}, {Bassett}, {Becker},
  {Beers}, {Bell}, {Belokurov}, {Berlind}, {Berman}, {Bernardi}, {Bickerton},
  {Bizyaev}, {Blakeslee}, {Blanton}, {Bochanski}, {Boroski}, {Brewington},
  {Brinchmann}, {Brinkmann}, {Brunner}, {Budav{\'a}ri}, {Carey}, {Carliles},
  {Carr}, {Castander}, {Cinabro}, {Connolly}, {Csabai}, {Cunha}, {Czarapata},
  {Davenport}, {de Haas}, {Dilday}, {Doi}, {Eisenstein}, {Evans}, {Evans},
  {Fan}, {Friedman}, {Frieman}, {Fukugita}, {G{\"a}nsicke}, {Gates},
  {Gillespie}, {Gilmore}, {Gonzalez}, {Gonzalez}, {Grebel}, {Gunn},
  {Gy{\"o}ry}, {Hall}, {Harding}, {Harris}, {Harvanek}, {Hawley}, {Hayes},
  {Heckman}, {Hendry}, {Hennessy}, {Hindsley}, {Hoblitt}, {Hogan}, {Hogg},
  {Holtzman}, {Hyde}, {Ichikawa}, {Ichikawa}, {Im}, {Ivezi{\'c}}, {Jester},
  {Jiang}, {Johnson}, {Jorgensen}, {Juri{\'c}}, {Kent}, {Kessler}, {Kleinman},
  {Knapp}, {Konishi}, {Kron}, {Krzesinski}, {Kuropatkin}, {Lampeitl},
  {Lebedeva}, {Lee}, {Lee}, {French Leger}, {L{\'e}pine}, {Li}, {Lima}, {Lin},
  {Long}, {Loomis}, {Loveday}, {Lupton}, {Magnier}, {Malanushenko},
  {Malanushenko}, {Mandelbaum}, {Margon}, {Marriner}, {Mart{\'\i}nez-Delgado},
  {Matsubara}, {McGehee}, {McKay}, {Meiksin}, {Morrison}, {Mullally}, {Munn},
  {Murphy}, {Nash}, {Nebot}, {Neilsen}, {Newberg}, {Newman}, {Nichol},
  {Nicinski}, {Nieto-Santisteban}, {Nitta}, {Okamura}, {Oravetz}, {Ostriker},
  {Owen}, {Padmanabhan}, {Pan}, {Park}, {Pauls}, {Peoples}, {Percival}, {Pier},
  {Pope}, {Pourbaix}, {Price}, {Purger}, {Quinn}, {Raddick}, {Re Fiorentin},
  {Richards}, {Richmond}, {Riess}, {Rix}, {Rockosi}, {Sako}, {Schlegel},
  {Schneider}, {Scholz}, {Schreiber}, {Schwope}, {Seljak}, {Sesar}, {Sheldon},
  {Shimasaku}, {Sibley}, {Simmons}, {Sivarani}, {Allyn Smith}, {Smith},
  {Smol{\v{c}}i{\'c}}, {Snedden}, {Stebbins}, {Steinmetz}, {Stoughton},
  {Strauss}, {SubbaRao}, {Suto}, {Szalay}, {Szapudi}, {Szkody}, {Tanaka},
  {Tegmark}, {Teodoro}, {Thakar}, {Tremonti}, {Tucker}, {Uomoto}, {Vanden
  Berk}, {Vandenberg}, {Vidrih}, {Vogeley}, {Voges}, {Vogt}, {Wadadekar},
  {Watters}, {Weinberg}, {West}, {White}, {Wilhite}, {Wonders}, {Yanny},
  {Yocum}, {York}, {Zehavi}, {Zibetti}, \& {Zucker}}]{Abazajian2009}
{Abazajian}, K.~N., {Adelman-McCarthy}, J.~K., {Ag{\"u}eros}, M.~A., {et~al.}
  2009, \apjs, 182, 543, \dodoi{10.1088/0067-0049/182/2/543}

\bibitem[{{Ahn} {et~al.}(2012){Ahn}, {Alexandroff}, {Allende Prieto},
  {Anderson}, {Anderton}, {Andrews}, {Aubourg}, {Bailey}, {Balbinot}, {Barnes},
  \& et~al.}]{Ahn2012a}
{Ahn}, C.~P., {Alexandroff}, R., {Allende Prieto}, C., {et~al.} 2012, \apjs,
  203, 21, \dodoi{10.1088/0067-0049/203/2/21}

\bibitem[{{Aird} {et~al.}(2012){Aird}, {Coil}, {Moustakas}, {Blanton},
  {Burles}, {Cool}, {Eisenstein}, {Smith}, {Wong}, \& {Zhu}}]{Aird2012}
{Aird}, J., {Coil}, A.~L., {Moustakas}, J., {et~al.} 2012, \apj, 746, 90,
  \dodoi{10.1088/0004-637X/746/1/90}

\bibitem[{{Alexander}(2017)}]{Alexander2017}
{Alexander}, T. 2017, \araa, 55, 17,
  \dodoi{10.1146/annurev-astro-091916-055306}

\bibitem[{{Amaro Seoane}(2022)}]{Amaro-Seoane2022}
{Amaro Seoane}, P. 2022, in Handbook of Gravitational Wave Astronomy. Edited by
  C. Bambi, 17, \dodoi{10.1007/978-981-15-4702-7_17-1}

\bibitem[{{Amaro-Seoane} {et~al.}(2017){Amaro-Seoane}, {Audley}, {Babak},
  {Baker}, {Barausse}, {Bender}, {Berti}, {Binetruy}, {Born}, {Bortoluzzi},
  {Camp}, {Caprini}, {Cardoso}, {Colpi}, {Conklin}, {Cornish}, {Cutler},
  {Danzmann}, {Dolesi}, {Ferraioli}, {Ferroni}, {Fitzsimons}, {Gair}, {Gesa
  Bote}, {Giardini}, {Gibert}, {Grimani}, {Halloin}, {Heinzel}, {Hertog},
  {Hewitson}, {Holley-Bockelmann}, {Hollington}, {Hueller}, {Inchauspe},
  {Jetzer}, {Karnesis}, {Killow}, {Klein}, {Klipstein}, {Korsakova}, {Larson},
  {Livas}, {Lloro}, {Man}, {Mance}, {Martino}, {Mateos}, {McKenzie},
  {McWilliams}, {Miller}, {Mueller}, {Nardini}, {Nelemans}, {Nofrarias},
  {Petiteau}, {Pivato}, {Plagnol}, {Porter}, {Reiche}, {Robertson},
  {Robertson}, {Rossi}, {Russano}, {Schutz}, {Sesana}, {Shoemaker}, {Slutsky},
  {Sopuerta}, {Sumner}, {Tamanini}, {Thorpe}, {Troebs}, {Vallisneri},
  {Vecchio}, {Vetrugno}, {Vitale}, {Volonteri}, {Wanner}, {Ward}, {Wass},
  {Weber}, {Ziemer}, \& {Zweifel}}]{Amaro-Seoane2017}
{Amaro-Seoane}, P., {Audley}, H., {Babak}, S., {et~al.} 2017, arXiv e-prints,
  arXiv:1702.00786.
\newblock \doarXiv{1702.00786}

\bibitem[{{Angus}(2021)}]{Angus2021_ZTF18abavruc}
{Angus}, C.~R. 2021, Transient Name Server Classification Report, 2021-2291, 1

\bibitem[{{Angus} {et~al.}(2022){Angus}, {Baldassare}, {Mockler}, {Foley},
  {Ramirez-Ruiz}, {Raimundo}, {French}, {Auchettl}, {Pfister}, {Gall},
  {Hjorth}, {Drout}, {Alexander}, {Dimitriadis}, {Hung}, {Jones}, {Rest},
  {Siebert}, {Taggart}, {Terreran}, {Tinyanont}, {Carroll}, {DeMarchi}, {Earl},
  {Gagliano}, {Izzo}, {Villar}, {Zenati}, {Arendse}, {Cold}, {de Boer},
  {Chambers}, {Coulter}, {Khetan}, {Lin}, {Magnier}, {Rojas-Bravo},
  {Wainscoat}, \& {Wojtak}}]{Angus2022}
{Angus}, C.~R., {Baldassare}, V.~F., {Mockler}, B., {et~al.} 2022, Nature
  Astronomy, 6, 1452, \dodoi{10.1038/s41550-022-01811-y}

\bibitem[{{Arcavi} {et~al.}(2020){Arcavi}, {Burke}, {Nyiha}, {Howell},
  {Hiramatsu}, {McCully}, {Pellegrino}, \& {Gonzalez}}]{Arcavi2020}
{Arcavi}, I., {Burke}, J., {Nyiha}, I., {et~al.} 2020, Transient Name Server
  Classification Report, 2020-3228, 1

\bibitem[{{Arcavi} {et~al.}(2014){Arcavi}, {Gal-Yam}, {Sullivan}, {Pan},
  {Cenko}, {Horesh}, {Ofek}, {De Cia}, {Yan}, {Yang}, {Howell}, {Tal},
  {Kulkarni}, {Tendulkar}, {Tang}, {Xu}, {Sternberg}, {Cohen}, {Bloom},
  {Nugent}, {Kasliwal}, {Perley}, {Quimby}, {Miller}, {Theissen}, \&
  {Laher}}]{Arcavi2014}
{Arcavi}, I., {Gal-Yam}, A., {Sullivan}, M., {et~al.} 2014, \apj, 793, 38,
  \dodoi{10.1088/0004-637X/793/1/38}

\bibitem[{{Bade} {et~al.}(1996){Bade}, {Komossa}, \& {Dahlem}}]{Bade1996}
{Bade}, N., {Komossa}, S., \& {Dahlem}, M. 1996, \aap, 309, L35

\bibitem[{{Bailer-Jones} {et~al.}(2018){Bailer-Jones}, {Rybizki}, {Fouesneau},
  {Mantelet}, \& {Andrae}}]{Bailer-Jones2018}
{Bailer-Jones}, C.~A.~L., {Rybizki}, J., {Fouesneau}, M., {Mantelet}, G., \&
  {Andrae}, R. 2018, \aj, 156, 58, \dodoi{10.3847/1538-3881/aacb21}

\bibitem[{{Baldry} {et~al.}(2012){Baldry}, {Driver}, {Loveday}, {Taylor},
  {Kelvin}, {Liske}, {Norberg}, {Robotham}, {Brough}, {Hopkins}, {Bamford},
  {Peacock}, {Bland-Hawthorn}, {Conselice}, {Croom}, {Jones}, {Parkinson},
  {Popescu}, {Prescott}, {Sharp}, \& {Tuffs}}]{Baldry2012}
{Baldry}, I.~K., {Driver}, S.~P., {Loveday}, J., {et~al.} 2012, \mnras, 421,
  621, \dodoi{10.1111/j.1365-2966.2012.20340.x}

\bibitem[{{Beck} {et~al.}(2021){Beck}, {Szapudi}, {Flewelling}, {Holmberg},
  {Magnier}, \& {Chambers}}]{Beck2021}
{Beck}, R., {Szapudi}, I., {Flewelling}, H., {et~al.} 2021, \mnras, 500, 1633,
  \dodoi{10.1093/mnras/staa2587}

\bibitem[{{Bellm} {et~al.}(2019{\natexlab{a}}){Bellm}, {Kulkarni}, {Graham},
  {Dekany}, {Smith}, {Riddle}, {Masci}, {Helou}, {Prince}, \&
  {Adams}}]{Bellm2019b}
{Bellm}, E.~C., {Kulkarni}, S.~R., {Graham}, M.~J., {et~al.}
  2019{\natexlab{a}}, \pasp, 131, 018002, \dodoi{10.1088/1538-3873/aaecbe}

\bibitem[{{Bellm} {et~al.}(2019{\natexlab{b}}){Bellm}, {Kulkarni}, {Barlow},
  {Feindt}, {Graham}, {Goobar}, {Kupfer}, {Ngeow}, {Nugent}, {Ofek}, {Prince},
  {Riddle}, {Walters}, \& {Ye}}]{Bellm2019a}
{Bellm}, E.~C., {Kulkarni}, S.~R., {Barlow}, T., {et~al.} 2019{\natexlab{b}},
  \pasp, 131, 068003, \dodoi{10.1088/1538-3873/ab0c2a}

\bibitem[{{Ben-Ami} {et~al.}(2022){Ben-Ami}, {Shvartzvald}, {Waxman}, {Netzer},
  {Yaniv}, {Algranatti}, {Gal-Yam}, {Lapid}, {Ofek}, {Topaz}, {Arcavi}, {Asif},
  {Azaria}, {Bahalul}, {Barschke}, {Bastian-Querner}, {Berge}, {Berlea},
  {Buehler}, {Dittmar}, {Gelman}, {Giavitto}, {Guttman}, {Haces Crespo},
  {Heilbrunn}, {Kachergincky}, {Kaipachery}, {Kowalski}, {Kulkarni}, {Kumar},
  {K{\"u}sters}, {Liran}, {Miron-Salomon}, {Mor}, {Nir}, {Nitzan}, {Philipp},
  {Porelli}, {Sagiv}, {Schliwinski}, {Sprecher}, {De Simone}, {Stern}, {Stone},
  {Trakhtenbrot}, {Vasilev}, {Watson}, \& {Zappon}}]{Ben-Ami2022}
{Ben-Ami}, S., {Shvartzvald}, Y., {Waxman}, E., {et~al.} 2022, in Society of
  Photo-Optical Instrumentation Engineers (SPIE) Conference Series, Vol. 12181,
  Society of Photo-Optical Instrumentation Engineers (SPIE) Conference Series,
  ed. J.-W.~A. {den Herder}, S.~{Nikzad}, \& K.~{Nakazawa}, 1218105,
  \dodoi{10.1117/12.2629850}

\bibitem[{{Blagorodnova} {et~al.}(2018){Blagorodnova}, {Neill}, {Walters},
  {Kulkarni}, {Fremling}, {Ben-Ami}, {Dekany}, {Fucik}, {Konidaris}, {Nash},
  {Ngeow}, {Ofek}, {O' Sullivan}, {Quimby}, {Ritter}, \&
  {Vyhmeister}}]{Blagorodnova2018}
{Blagorodnova}, N., {Neill}, J.~D., {Walters}, R., {et~al.} 2018, \pasp, 130,
  035003, \dodoi{10.1088/1538-3873/aaa53f}

\bibitem[{{Blanchard} {et~al.}(2017){Blanchard}, {Nicholl}, {Berger},
  {Guillochon}, {Margutti}, {Chornock}, {Alexander}, {Leja}, \&
  {Drout}}]{Blanchard2017}
{Blanchard}, P.~K., {Nicholl}, M., {Berger}, E., {et~al.} 2017, \apj, 843, 106,
  \dodoi{10.3847/1538-4357/aa77f7}

\bibitem[{{Bonnerot} \& {Lu}(2020)}]{Bonnerot2020}
{Bonnerot}, C., \& {Lu}, W. 2020, \mnras, 495, 1374,
  \dodoi{10.1093/mnras/staa1246}

\bibitem[{{Bonnerot} {et~al.}(2016){Bonnerot}, {Rossi}, {Lodato}, \&
  {Price}}]{Bonnerot2016}
{Bonnerot}, C., {Rossi}, E.~M., {Lodato}, G., \& {Price}, D.~J. 2016, \mnras,
  455, 2253, \dodoi{10.1093/mnras/stv2411}

\bibitem[{{Brown} {et~al.}(2017){Brown}, {Holoien}, {Auchettl}, {Stanek},
  {Kochanek}, {Shappee}, {Prieto}, \& {Grupe}}]{Brown2017_14li}
{Brown}, J.~S., {Holoien}, T.~W.~S., {Auchettl}, K., {et~al.} 2017, \mnras,
  466, 4904, \dodoi{10.1093/mnras/stx033}

\bibitem[{{Bruch} {et~al.}(2021){Bruch}, {Strotjohann}, {Gal-Yam}, {Schulze},
  {Sollerman}, {Perley}, \& {Tzanidakis}}]{Bruch2021_ZTF21abdmevk}
{Bruch}, R., {Strotjohann}, N., {Gal-Yam}, A., {et~al.} 2021, Transient Name
  Server AstroNote, 200, 1

\bibitem[{{Burke} {et~al.}(2021){Burke}, {Arcavi}, {Lam}, {Hiramatsu},
  {Howell}, {McCully}, {Newsome}, {Gonzalez}, \& {Pellegrino}}]{Burke2021}
{Burke}, J., {Arcavi}, I., {Lam}, M., {et~al.} 2021, Transient Name Server
  Classification Report, 2021-3259, 1

\bibitem[{{Calzetti} {et~al.}(2000){Calzetti}, {Armus}, {Bohlin}, {Kinney},
  {Koornneef}, \& {Storchi-Bergmann}}]{Calzetti2000a}
{Calzetti}, D., {Armus}, L., {Bohlin}, R.~C., {et~al.} 2000, \apj, 533, 682,
  \dodoi{10.1086/308692}

\bibitem[{{Cappellari}(2017)}]{Cappellari2017}
{Cappellari}, M. 2017, \mnras, 466, 798, \dodoi{10.1093/mnras/stw3020}

\bibitem[{{Cappellari} \& {Emsellem}(2004)}]{Cappellari2004}
{Cappellari}, M., \& {Emsellem}, E. 2004, \pasp, 116, 138,
  \dodoi{10.1086/381875}

\bibitem[{{Cardelli} {et~al.}(1989){Cardelli}, {Clayton}, \&
  {Mathis}}]{Cardelli1989}
{Cardelli}, J.~A., {Clayton}, G.~C., \& {Mathis}, J.~S. 1989, \apj, 345, 245,
  \dodoi{10.1086/167900}

\bibitem[{{Cenko} {et~al.}(2006){Cenko}, {Fox}, {Moon}, {Harrison}, {Kulkarni},
  {Henning}, {Guzman}, {Bonati}, {Smith}, \& {Thicksten}}]{Cenko2006}
{Cenko}, S.~B., {Fox}, D.~B., {Moon}, D.-S., {et~al.} 2006, \pasp, 118, 1396,
  \dodoi{10.1086/508366}

\bibitem[{{Chabrier}(2003)}]{Chabrier2003a}
{Chabrier}, G. 2003, \pasp, 115, 763, \dodoi{10.1086/376392}

\bibitem[{{Chadayammuri} {et~al.}(2023){Chadayammuri}, {Bogd{\'a}n}, {Ricarte},
  \& {Natarajan}}]{Chadayammuri2023}
{Chadayammuri}, U., {Bogd{\'a}n}, {\'A}., {Ricarte}, A., \& {Natarajan}, P.
  2023, \apj, 946, 51, \dodoi{10.3847/1538-4357/acbea6}

\bibitem[{{Chambers} {et~al.}(2016){Chambers}, {Magnier}, {Metcalfe},
  {Flewelling}, {Huber}, {Waters}, {Denneau}, {Draper}, {Farrow}, {Finkbeiner},
  {Holmberg}, {Koppenhoefer}, {Price}, {Rest}, {Saglia}, {Schlafly}, {Smartt},
  {Sweeney}, {Wainscoat}, {Burgett}, {Chastel}, {Grav}, {Heasley}, {Hodapp},
  {Jedicke}, {Kaiser}, {Kudritzki}, {Luppino}, {Lupton}, {Monet}, {Morgan},
  {Onaka}, {Shiao}, {Stubbs}, {Tonry}, {White}, {Ba{\~n}ados}, {Bell},
  {Bender}, {Bernard}, {Boegner}, {Boffi}, {Botticella}, {Calamida},
  {Casertano}, {Chen}, {Chen}, {Cole}, {Deacon}, {Frenk}, {Fitzsimmons},
  {Gezari}, {Gibbs}, {Goessl}, {Goggia}, {Gourgue}, {Goldman}, {Grant},
  {Grebel}, {Hambly}, {Hasinger}, {Heavens}, {Heckman}, {Henderson}, {Henning},
  {Holman}, {Hopp}, {Ip}, {Isani}, {Jackson}, {Keyes}, {Koekemoer}, {Kotak},
  {Le}, {Liska}, {Long}, {Lucey}, {Liu}, {Martin}, {Masci}, {McLean}, {Mindel},
  {Misra}, {Morganson}, {Murphy}, {Obaika}, {Narayan}, {Nieto-Santisteban},
  {Norberg}, {Peacock}, {Pier}, {Postman}, {Primak}, {Rae}, {Rai}, {Riess},
  {Riffeser}, {Rix}, {R{\"o}ser}, {Russel}, {Rutz}, {Schilbach}, {Schultz},
  {Scolnic}, {Strolger}, {Szalay}, {Seitz}, {Small}, {Smith}, {Soderblom},
  {Taylor}, {Thomson}, {Taylor}, {Thakar}, {Thiel}, {Thilker}, {Unger},
  {Urata}, {Valenti}, {Wagner}, {Walder}, {Walter}, {Watters}, {Werner},
  {Wood-Vasey}, \& {Wyse}}]{Chambers2016a}
{Chambers}, K.~C., {Magnier}, E.~A., {Metcalfe}, N., {et~al.} 2016, arXiv
  e-prints, arXiv:1612.05560.
\newblock \doarXiv{1612.05560}

\bibitem[{{Charalampopoulos} {et~al.}(2023){Charalampopoulos}, {Pursiainen},
  {Leloudas}, {Arcavi}, {Newsome}, {Schulze}, {Burke}, \&
  {Nicholl}}]{Charalampopoulos2023}
{Charalampopoulos}, P., {Pursiainen}, M., {Leloudas}, G., {et~al.} 2023, \aap,
  673, A95, \dodoi{10.1051/0004-6361/202245065}

\bibitem[{{Charalampopoulos} {et~al.}(2022){Charalampopoulos}, {Leloudas},
  {Malesani}, {Wevers}, {Arcavi}, {Nicholl}, {Pursiainen}, {Lawrence},
  {Anderson}, {Benetti}, {Cannizzaro}, {Chen}, {Galbany}, {Gromadzki},
  {Guti{\'e}rrez}, {Inserra}, {Jonker}, {M{\"u}ller-Bravo}, {Onori}, {Short},
  {Sollerman}, \& {Young}}]{Charalampopoulos2022_spec}
{Charalampopoulos}, P., {Leloudas}, G., {Malesani}, D.~B., {et~al.} 2022, \aap,
  659, A34, \dodoi{10.1051/0004-6361/202142122}

\bibitem[{{Chomiuk} {et~al.}(2011){Chomiuk}, {Chornock}, {Soderberg}, {Berger},
  {Chevalier}, {Foley}, {Huber}, {Narayan}, {Rest}, {Gezari}, {Kirshner},
  {Riess}, {Rodney}, {Smartt}, {Stubbs}, {Tonry}, {Wood-Vasey}, {Burgett},
  {Chambers}, {Czekala}, {Flewelling}, {Forster}, {Kaiser}, {Kudritzki},
  {Magnier}, {Martin}, {Morgan}, {Neill}, {Price}, {Roth}, {Sanders}, \&
  {Wainscoat}}]{Chomiuk2011}
{Chomiuk}, L., {Chornock}, R., {Soderberg}, A.~M., {et~al.} 2011, \apj, 743,
  114, \dodoi{10.1088/0004-637X/743/2/114}

\bibitem[{{Chornock} {et~al.}(2014){Chornock}, {Berger}, {Gezari}, {Zauderer},
  {Rest}, {Chomiuk}, {Kamble}, {Soderberg}, {Czekala}, {Dittmann}, {Drout},
  {Foley}, {Fong}, {Huber}, {Kirshner}, {Lawrence}, {Lunnan}, {Marion},
  {Narayan}, {Riess}, {Roth}, {Sanders}, {Scolnic}, {Smartt}, {Smith},
  {Stubbs}, {Tonry}, {Burgett}, {Chambers}, {Flewelling}, {Hodapp}, {Kaiser},
  {Magnier}, {Martin}, {Neill}, {Price}, \& {Wainscoat}}]{Chornock2014}
{Chornock}, R., {Berger}, E., {Gezari}, S., {et~al.} 2014, \apj, 780, 44,
  \dodoi{10.1088/0004-637X/780/1/44}

\bibitem[{{Chu} {et~al.}(2021{\natexlab{a}}){Chu}, {Dahiwale}, \&
  {Fremling}}]{Chu2021_ZTF21abzciqh}
{Chu}, M., {Dahiwale}, A., \& {Fremling}, C. 2021{\natexlab{a}}, Transient Name
  Server Classification Report, 2021-3438, 1

\bibitem[{{Chu} {et~al.}(2021{\natexlab{b}}){Chu}, {Dahiwale}, \&
  {Fremling}}]{Chu2021_21mhg_CR}
---. 2021{\natexlab{b}}, Transient Name Server Classification Report,
  2021-2672, 1

\bibitem[{{Chu} {et~al.}(2021{\natexlab{c}}){Chu}, {Dahiwale}, \&
  {Fremling}}]{Chu2021_21sdu_CR}
---. 2021{\natexlab{c}}, Transient Name Server Classification Report,
  2021-2712, 1

\bibitem[{{Chu} {et~al.}(2022){Chu}, {Dahiwale}, \&
  {Fremling}}]{Chu2022_21yzv_CR}
---. 2022, Transient Name Server Classification Report, 2022-363, 1

\bibitem[{{Conroy} {et~al.}(2009){Conroy}, {Gunn}, \& {White}}]{Conroy2009a}
{Conroy}, C., {Gunn}, J.~E., \& {White}, M. 2009, \apj, 699, 486,
  \dodoi{10.1088/0004-637X/699/1/486}

\bibitem[{{Coughlin} \& {Begelman}(2014)}]{coughlin14_ZEBRA}
{Coughlin}, E.~R., \& {Begelman}, M.~C. 2014, \apj, 781, 82,
  \dodoi{10.1088/0004-637X/781/2/82}

\bibitem[{{Dahiwale} \&
  {Fremling}(2020{\natexlab{a}})}]{Dahiwale2020_ZTF20abisysx}
{Dahiwale}, A., \& {Fremling}, C. 2020{\natexlab{a}}, Transient Name Server
  Classification Report, 2020-3800, 1

\bibitem[{{Dahiwale} \&
  {Fremling}(2020{\natexlab{b}})}]{Dahiwale2020_ZTF20aaivego}
---. 2020{\natexlab{b}}, Transient Name Server Classification Report, 2020-534,
  1

\bibitem[{{Dahiwale} \&
  {Fremling}(2020{\natexlab{c}})}]{Dahiwale2020_ZTF20ackdkva}
---. 2020{\natexlab{c}}, Transient Name Server Classification Report,
  2020-3308, 1

\bibitem[{{Dahiwale} \&
  {Fremling}(2020{\natexlab{d}})}]{Dahiwale2020_ZTF20aayxdse}
---. 2020{\natexlab{d}}, Transient Name Server Classification Report,
  2020-1756, 1

\bibitem[{{Dahiwale} \&
  {Fremling}(2020{\natexlab{e}})}]{Dahiwale2020_ZTF19abulzhy}
---. 2020{\natexlab{e}}, Transient Name Server Classification Report, 2020-601,
  1

\bibitem[{{Dahiwale} \& {Fremling}(2021)}]{Dahiwale2021_ZTF21aaglrzc}
---. 2021, Transient Name Server Classification Report, 2021-1378, 1

\bibitem[{{Dai} {et~al.}(2015){Dai}, {McKinney}, \& {Miller}}]{Dai2015}
{Dai}, L., {McKinney}, J.~C., \& {Miller}, M.~C. 2015, \apjl, 812, L39,
  \dodoi{10.1088/2041-8205/812/2/L39}

\bibitem[{{Dai} {et~al.}(2018){Dai}, {McKinney}, {Roth}, {Ramirez-Ruiz}, \&
  {Miller}}]{Dai2018}
{Dai}, L., {McKinney}, J.~C., {Roth}, N., {Ramirez-Ruiz}, E., \& {Miller},
  M.~C. 2018, \apjl, 859, L20, \dodoi{10.3847/2041-8213/aab429}

\bibitem[{{Dekany} {et~al.}(2020){Dekany}, {Smith}, {Riddle}, {Feeney},
  {Porter}, {Hale}, {Zolkower}, {Belicki}, {Kaye}, {Henning}, {Walters},
  {Cromer}, {Delacroix}, {Rodriguez}, {Reiley}, {Mao}, {Hover}, {Murphy},
  {Burruss}, {Baker}, {Kowalski}, {Reif}, {Mueller}, {Bellm}, {Graham}, \&
  {Kulkarni}}]{Dekany2020}
{Dekany}, R., {Smith}, R.~M., {Riddle}, R., {et~al.} 2020, \pasp, 132, 038001,
  \dodoi{10.1088/1538-3873/ab4ca2}

\bibitem[{{Demircan} \& {Kahraman}(1991)}]{Demircan1991}
{Demircan}, O., \& {Kahraman}, G. 1991, \apss, 181, 313,
  \dodoi{10.1007/BF00639097}

\bibitem[{{Du} {et~al.}(2022){Du}, {Ega{\~n}a-Ugrinovic}, {Essig}, {Fragione},
  \& {Perna}}]{Du2022}
{Du}, P., {Ega{\~n}a-Ugrinovic}, D., {Essig}, R., {Fragione}, G., \& {Perna},
  R. 2022, Nature Communications, 13, 4626, \dodoi{10.1038/s41467-022-32301-4}

\bibitem[{{Duev} {et~al.}(2019){Duev}, {Mahabal}, {Masci}, {Graham},
  {Rusholme}, {Walters}, {Karmarkar}, {Frederick}, {Kasliwal}, {Rebbapragada},
  \& {Ward}}]{Duev2019}
{Duev}, D.~A., {Mahabal}, A., {Masci}, F.~J., {et~al.} 2019, \mnras, 489, 3582,
  \dodoi{10.1093/mnras/stz2357}

\bibitem[{{Flesch}(2019)}]{Flesch2019}
{Flesch}, E.~W. 2019, arXiv e-prints, arXiv:1912.05614,
  \dodoi{10.48550/arXiv.1912.05614}

\bibitem[{{Flewelling} {et~al.}(2020){Flewelling}, {Magnier}, {Chambers},
  {Heasley}, {Holmberg}, {Huber}, {Sweeney}, {Waters}, {Calamida}, {Casertano},
  {Chen}, {Farrow}, {Hasinger}, {Henderson}, {Long}, {Metcalfe}, {Narayan},
  {Nieto-Santisteban}, {Norberg}, {Rest}, {Saglia}, {Szalay}, {Thakar},
  {Tonry}, {Valenti}, {Werner}, {White}, {Denneau}, {Draper}, {Hodapp},
  {Jedicke}, {Kaiser}, {Kudritzki}, {Price}, {Wainscoat}, {Chastel}, {McLean},
  {Postman}, \& {Shiao}}]{Flewelling2020}
{Flewelling}, H.~A., {Magnier}, E.~A., {Chambers}, K.~C., {et~al.} 2020, \apjs,
  251, 7, \dodoi{10.3847/1538-4365/abb82d}

\bibitem[{{Foreman-Mackey} {et~al.}(2013){Foreman-Mackey}, {Hogg}, {Lang}, \&
  {Goodman}}]{Foreman-Mackey2013}
{Foreman-Mackey}, D., {Hogg}, D.~W., {Lang}, D., \& {Goodman}, J. 2013,
  Publications of the Astronomical Society of the Pacific, 125, 306,
  \dodoi{10.1086/670067}

\bibitem[{{Foreman-Mackey} {et~al.}(2014){Foreman-Mackey}, {Sick}, \&
  {Johnson}}]{ForemanMackey2014a}
{Foreman-Mackey}, D., {Sick}, J., \& {Johnson}, B. 2014, {Python-Fsps: Python
  Bindings To Fsps (V0.1.1)}, v0.1.1,  Zenodo, \dodoi{10.5281/zenodo.12157}

\bibitem[{{Frederick} {et~al.}(2021){Frederick}, {Gezari}, {Graham},
  {Sollerman}, {van Velzen}, {Perley}, {Stern}, {Ward}, {Hammerstein}, {Hung},
  {Yan}, {Andreoni}, {Bellm}, {Duev}, {Kowalski}, {Mahabal}, {Masci},
  {Medford}, {Rusholme}, {Smith}, \& {Walters}}]{Frederick2021}
{Frederick}, S., {Gezari}, S., {Graham}, M.~J., {et~al.} 2021, \apj, 920, 56,
  \dodoi{10.3847/1538-4357/ac110f}

\bibitem[{{Fremling} {et~al.}(2020){Fremling}, {Miller}, {Sharma}, {Dugas},
  {Perley}, {Taggart}, {Sollerman}, {Goobar}, {Graham}, {Neill}, {Nordin},
  {Rigault}, {Walters}, {Andreoni}, {Bagdasaryan}, {Belicki}, {Cannella},
  {Bellm}, {Cenko}, {De}, {Dekany}, {Frederick}, {Golkhou}, {Graham}, {Helou},
  {Ho}, {Kasliwal}, {Kupfer}, {Laher}, {Mahabal}, {Masci}, {Riddle},
  {Rusholme}, {Schulze}, {Shupe}, {Smith}, {van Velzen}, {Yan}, {Yao},
  {Zhuang}, \& {Kulkarni}}]{Fremling2020}
{Fremling}, C., {Miller}, A.~A., {Sharma}, Y., {et~al.} 2020, \apj, 895, 32,
  \dodoi{10.3847/1538-4357/ab8943}

\bibitem[{{French} {et~al.}(2016){French}, {Arcavi}, \&
  {Zabludoff}}]{French2016}
{French}, K.~D., {Arcavi}, I., \& {Zabludoff}, A. 2016, \apjl, 818, L21,
  \dodoi{10.3847/2041-8205/818/1/L21}

\bibitem[{{French} {et~al.}(2020){French}, {Wevers}, {Law-Smith}, {Graur}, \&
  {Zabludoff}}]{French2020}
{French}, K.~D., {Wevers}, T., {Law-Smith}, J., {Graur}, O., \& {Zabludoff},
  A.~I. 2020, \ssr, 216, 32, \dodoi{10.1007/s11214-020-00657-y}

\bibitem[{{Gallo} \& {Sesana}(2019)}]{Gallo2019}
{Gallo}, E., \& {Sesana}, A. 2019, \apjl, 883, L18,
  \dodoi{10.3847/2041-8213/ab40c6}

\bibitem[{{Gallo} {et~al.}(2008){Gallo}, {Treu}, {Jacob}, {Woo}, {Marshall}, \&
  {Antonucci}}]{Gallo2008}
{Gallo}, E., {Treu}, T., {Jacob}, J., {et~al.} 2008, \apj, 680, 154,
  \dodoi{10.1086/58801210.48550/arXiv.0711.2073}

\bibitem[{{Gallo} {et~al.}(2010){Gallo}, {Treu}, {Marshall}, {Woo}, {Leipski},
  \& {Antonucci}}]{Gallo2010}
{Gallo}, E., {Treu}, T., {Marshall}, P.~J., {et~al.} 2010, \apj, 714, 25,
  \dodoi{10.1088/0004-637X/714/1/2510.48550/arXiv.1002.3619}

\bibitem[{{Gallo} {et~al.}(2019){Gallo}, {Hodges-Kluck}, {Treu}, {Greene},
  {Wilkes}, {Seth}, {Reines}, {Baldassare}, {Plotkin}, \&
  {Chandar}}]{Gallo2019_wp}
{Gallo}, E., {Hodges-Kluck}, E., {Treu}, T., {et~al.} 2019, \baas, 51, 35,
  \dodoi{10.48550/arXiv.1903.06629}

\bibitem[{{Gebhardt} {et~al.}(2001){Gebhardt}, {Lauer}, {Kormendy}, {Pinkney},
  {Bower}, {Green}, {Gull}, {Hutchings}, {Kaiser}, {Nelson}, {Richstone}, \&
  {Weistrop}}]{Gebhardt2001}
{Gebhardt}, K., {Lauer}, T.~R., {Kormendy}, J., {et~al.} 2001, \aj, 122, 2469,
  \dodoi{10.1086/323481}

\bibitem[{{Gehrels}(1986)}]{Gehrels1986}
{Gehrels}, N. 1986, \apj, 303, 336, \dodoi{10.1086/164079}

\bibitem[{{Gehrels} {et~al.}(2004){Gehrels}, {Chincarini}, {Giommi}, {Mason},
  {Nousek}, {Wells}, {White}, {Barthelmy}, {Burrows}, \&
  {Cominsky}}]{Gehrels2004}
{Gehrels}, N., {Chincarini}, G., {Giommi}, P., {et~al.} 2004, \apj, 611, 1005,
  \dodoi{10.1086/422091}

\bibitem[{{Gezari}(2021)}]{Gezari2021}
{Gezari}, S. 2021, \araa, 59, 21, \dodoi{10.1146/annurev-astro-111720-030029}

\bibitem[{{Gezari} {et~al.}(2006){Gezari}, {Martin}, {Milliard}, {Basa},
  {Halpern}, {Forster}, {Friedman}, {Morrissey}, {Neff}, {Schiminovich},
  {Seibert}, {Small}, \& {Wyder}}]{Gezari2006}
{Gezari}, S., {Martin}, D.~C., {Milliard}, B., {et~al.} 2006, \apjl, 653, L25,
  \dodoi{10.1086/509918}

\bibitem[{{Gezari} {et~al.}(2008){Gezari}, {Basa}, {Martin}, {Bazin},
  {Forster}, {Milliard}, {Halpern}, {Friedman}, {Morrissey}, {Neff},
  {Schiminovich}, {Seibert}, {Small}, \& {Wyder}}]{Gezari2008}
{Gezari}, S., {Basa}, S., {Martin}, D.~C., {et~al.} 2008, \apj, 676, 944,
  \dodoi{10.1086/529008}

\bibitem[{{Gezari} {et~al.}(2009){Gezari}, {Heckman}, {Cenko}, {Eracleous},
  {Forster}, {Gon{\c{c}}alves}, {Martin}, {Morrissey}, {Neff}, {Seibert},
  {Schiminovich}, \& {Wyder}}]{Gezari2009}
{Gezari}, S., {Heckman}, T., {Cenko}, S.~B., {et~al.} 2009, \apj, 698, 1367,
  \dodoi{10.1088/0004-637X/698/2/1367}

\bibitem[{{Gezari} {et~al.}(2012){Gezari}, {Chornock}, {Rest}, {Huber},
  {Forster}, {Berger}, {Challis}, {Neill}, {Martin}, {Heckman}, {Lawrence},
  {Norman}, {Narayan}, {Foley}, {Marion}, {Scolnic}, {Chomiuk}, {Soderberg},
  {Smith}, {Kirshner}, {Riess}, {Smartt}, {Stubbs}, {Tonry}, {Wood-Vasey},
  {Burgett}, {Chambers}, {Grav}, {Heasley}, {Kaiser}, {Kudritzki}, {Magnier},
  {Morgan}, \& {Price}}]{Gezari2012}
{Gezari}, S., {Chornock}, R., {Rest}, A., {et~al.} 2012, \nat, 485, 217,
  \dodoi{10.1038/nature10990}

\bibitem[{{Gomez} {et~al.}(2023){Gomez}, {Villar}, {Berger}, {Gezari}, {van
  Velzen}, {Nicholl}, {Blanchard}, \& {Alexander}}]{Gomez2023}
{Gomez}, S., {Villar}, V.~A., {Berger}, E., {et~al.} 2023, \apj, 949, 113,
  \dodoi{10.3847/1538-4357/acc535}

\bibitem[{{Graham} {et~al.}(2019){Graham}, {Kulkarni}, {Bellm}, {Adams},
  {Barbarino}, {Blagorodnova}, {Bodewits}, {Bolin}, {Brady}, {Cenko}, {Chang},
  {Coughlin}, {De}, {Eadie}, {Farnham}, {Feindt}, {Franckowiak}, {Fremling},
  {Gezari}, {Ghosh}, {Goldstein}, {Golkhou}, {Goobar}, {Ho}, {Huppenkothen},
  {Ivezi{\'c}}, {Jones}, {Juric}, {Kaplan}, {Kasliwal}, {Kelley}, {Kupfer},
  {Lee}, {Lin}, {Lunnan}, {Mahabal}, {Miller}, {Ngeow}, {Nugent}, {Ofek},
  {Prince}, {Rauch}, {van Roestel}, {Schulze}, {Singer}, {Sollerman}, {Taddia},
  {Yan}, {Ye}, {Yu}, {Barlow}, {Bauer}, {Beck}, {Belicki}, {Biswas}, {Brinnel},
  {Brooke}, {Bue}, {Bulla}, {Burruss}, {Connolly}, {Cromer}, {Cunningham},
  {Dekany}, {Delacroix}, {Desai}, {Duev}, {Feeney}, {Flynn}, {Frederick},
  {Gal-Yam}, {Giomi}, {Groom}, {Hacopians}, {Hale}, {Helou}, {Henning},
  {Hover}, {Hillenbrand}, {Howell}, {Hung}, {Imel}, {Ip}, {Jackson}, {Kaspi},
  {Kaye}, {Kowalski}, {Kramer}, {Kuhn}, {Landry}, {Laher}, {Mao}, {Masci},
  {Monkewitz}, {Murphy}, {Nordin}, {Patterson}, {Penprase}, {Porter},
  {Rebbapragada}, {Reiley}, {Riddle}, {Rigault}, {Rodriguez}, {Rusholme}, {van
  Santen}, {Shupe}, {Smith}, {Soumagnac}, {Stein}, {Surace}, {Szkody}, {Terek},
  {Van Sistine}, {van Velzen}, {Vestrand}, {Walters}, {Ward}, {Zhang}, \&
  {Zolkower}}]{Graham2019}
{Graham}, M.~J., {Kulkarni}, S.~R., {Bellm}, E.~C., {et~al.} 2019, \pasp, 131,
  078001, \dodoi{10.1088/1538-3873/ab006c}

\bibitem[{{Greene} {et~al.}(2020){Greene}, {Strader}, \& {Ho}}]{Greene2020}
{Greene}, J.~E., {Strader}, J., \& {Ho}, L.~C. 2020, \araa, 58, 257,
  \dodoi{10.1146/annurev-astro-032620-021835}

\bibitem[{{Grupe} {et~al.}(1999){Grupe}, {Thomas}, \& {Leighly}}]{Grupe1999}
{Grupe}, D., {Thomas}, H.~C., \& {Leighly}, K.~M. 1999, \aap, 350, L31,
  \dodoi{10.48550/arXiv.astro-ph/9909101}

\bibitem[{{G{\"u}ltekin} {et~al.}(2009){G{\"u}ltekin}, {Richstone}, {Gebhardt},
  {Lauer}, {Tremaine}, {Aller}, {Bender}, {Dressler}, {Faber}, {Filippenko},
  {Green}, {Ho}, {Kormendy}, {Magorrian}, {Pinkney}, \&
  {Siopis}}]{Gultekin2009}
{G{\"u}ltekin}, K., {Richstone}, D.~O., {Gebhardt}, K., {et~al.} 2009, \apj,
  698, 198, \dodoi{10.1088/0004-637X/698/1/198}

\bibitem[{{Hammerstein} {et~al.}(2021{\natexlab{a}}){Hammerstein}, {Gezari},
  {Velzen}, {Kulkarni}, {Cenko}, {Graham}, {Ravi}, {Lu}, {Duev}, {Stern},
  {Somalwar}, \& {Yao}}]{Hammerstein2021_20vwl_CR}
{Hammerstein}, E., {Gezari}, S., {Velzen}, S.~V., {et~al.} 2021{\natexlab{a}},
  Transient Name Server Classification Report, 2021-159, 1

\bibitem[{{Hammerstein} {et~al.}(2021{\natexlab{b}}){Hammerstein}, {Gezari},
  {Velzen}, {Kulkarni}, {Cenko}, {Graham}, {Ravi}, {Lu}, {Duev}, {Stern},
  {Somalwar}, \& {Yao}}]{Hammerstein2021_20acka_CR}
---. 2021{\natexlab{b}}, Transient Name Server Classification Report, 2021-262,
  1

\bibitem[{{Hammerstein} {et~al.}(2021{\natexlab{c}}){Hammerstein}, {Gezari},
  {Velzen}, {Yao}, {Somalwar}, {Cenko}, {Kulkarni}, {Graham}, \&
  {Ravi}}]{Hammerstein2021_21axu_CR}
---. 2021{\natexlab{c}}, Transient Name Server Classification Report, 2021-955,
  1

\bibitem[{{Hammerstein} {et~al.}(2021{\natexlab{d}}){Hammerstein}, {Gezari},
  {van Velzen}, {Cenko}, {Roth}, {Ward}, {Frederick}, {Hung}, {Graham},
  {Foley}, {Bellm}, {Cannella}, {Drake}, {Kupfer}, {Laher}, {Mahabal}, {Masci},
  {Riddle}, {Rojas-Bravo}, \& {Smith}}]{Hammerstein2021}
{Hammerstein}, E., {Gezari}, S., {van Velzen}, S., {et~al.} 2021{\natexlab{d}},
  \apjl, 908, L20, \dodoi{10.3847/2041-8213/abdcb4}

\bibitem[{{Hammerstein} {et~al.}(2023){Hammerstein}, {van Velzen}, {Gezari},
  {Cenko}, {Yao}, {Ward}, {Frederick}, {Villanueva}, {Somalwar}, {Graham},
  {Kulkarni}, {Stern}, {Andreoni}, {Bellm}, {Dekany}, {Dhawan}, {Drake},
  {Fremling}, {Gatkine}, {Groom}, {Ho}, {Kasliwal}, {Karambelkar}, {Kool},
  {Masci}, {Medford}, {Perley}, {Purdum}, {Roestel}, {Sharma}, {Sollerman},
  {Taggart}, \& {Yan}}]{Hammerstein2023}
{Hammerstein}, E., {van Velzen}, S., {Gezari}, S., {et~al.} 2023, \apj, 942, 9,
  \dodoi{10.3847/1538-4357/aca283}

\bibitem[{{Hills}(1975)}]{Hills1975}
{Hills}, J.~G. 1975, \nat, 254, 295, \dodoi{10.1038/254295a0}

\bibitem[{{Hinkle} {et~al.}(2021){Hinkle}, {Holoien}, {Auchettl}, {Shappee},
  {Neustadt}, {Payne}, {Brown}, {Kochanek}, {Stanek}, {Graham}, {Tucker}, {Do},
  {Anderson}, {Bose}, {Chen}, {Coulter}, {Dimitriadis}, {Dong}, {Foley},
  {Huber}, {Hung}, {Kilpatrick}, {Pignata}, {Piro}, {Rojas-Bravo}, {Siebert},
  {Stalder}, {Thompson}, {Tonry}, {Vallely}, \& {Wisniewski}}]{Hinkle2021}
{Hinkle}, J.~T., {Holoien}, T.~W.~S., {Auchettl}, K., {et~al.} 2021, \mnras,
  500, 1673, \dodoi{10.1093/mnras/staa3170}

\bibitem[{{Ho} {et~al.}(2023){Ho}, {Perley}, {Gal-Yam}, {Lunnan}, {Sollerman},
  {Schulze}, {Das}, {Dobie}, {Yao}, {Fremling}, {Adams}, {Anand}, {Andreoni},
  {Bellm}, {Bruch}, {Burdge}, {Castro-Tirado}, {Dahiwale}, {De}, {Dekany},
  {Drake}, {Duev}, {Graham}, {Helou}, {Kaplan}, {Karambelkar}, {Kasliwal},
  {Kool}, {Kulkarni}, {Mahabal}, {Medford}, {Miller}, {Nordin}, {Ofek},
  {Petitpas}, {Riddle}, {Sharma}, {Smith}, {Stewart}, {Taggart}, {Tartaglia},
  {Tzanidakis}, \& {Winters}}]{Ho2023_fbot_sample}
{Ho}, A. Y.~Q., {Perley}, D.~A., {Gal-Yam}, A., {et~al.} 2023, \apj, 949, 120,
  \dodoi{10.3847/1538-4357/acc533}

\bibitem[{{Holoien} {et~al.}(2014){Holoien}, {Prieto}, {Bersier}, {Kochanek},
  {Stanek}, {Shappee}, {Grupe}, {Basu}, {Beacom}, {Brimacombe}, {Brown},
  {Davis}, {Jencson}, {Pojmanski}, \& {Szczygie{\l}}}]{Holoien2014}
{Holoien}, T.~W.~S., {Prieto}, J.~L., {Bersier}, D., {et~al.} 2014, \mnras,
  445, 3263, \dodoi{10.1093/mnras/stu1922}

\bibitem[{{Huang} \& {Lu}(2022)}]{Huang2023}
{Huang}, H.-T., \& {Lu}, W. 2022, arXiv e-prints, arXiv:2301.00259.
\newblock \doarXiv{2301.00259}

\bibitem[{{Hung} {et~al.}(2017){Hung}, {Gezari}, {Blagorodnova}, {Roth},
  {Cenko}, {Kulkarni}, {Horesh}, {Arcavi}, {McCully}, {Yan}, {Lunnan},
  {Fremling}, {Cao}, {Nugent}, \& {Wozniak}}]{Hung2017}
{Hung}, T., {Gezari}, S., {Blagorodnova}, N., {et~al.} 2017, \apj, 842, 29,
  \dodoi{10.3847/1538-4357/aa7337}

\bibitem[{{Ivezi{\'c}} {et~al.}(2019){Ivezi{\'c}}, {Kahn}, {Tyson}, {Abel},
  {Acosta}, {Allsman}, {Alonso}, {AlSayyad}, {Anderson}, {Andrew}, {Angel},
  {Angeli}, {Ansari}, {Antilogus}, {Araujo}, {Armstrong}, {Arndt}, {Astier},
  {Aubourg}, {Auza}, {Axelrod}, {Bard}, {Barr}, {Barrau}, {Bartlett}, {Bauer},
  {Bauman}, {Baumont}, {Bechtol}, {Bechtol}, {Becker}, {Becla}, {Beldica},
  {Bellavia}, {Bianco}, {Biswas}, {Blanc}, {Blazek}, {Blandford}, {Bloom},
  {Bogart}, {Bond}, {Booth}, {Borgland}, {Borne}, {Bosch}, {Boutigny},
  {Brackett}, {Bradshaw}, {Brandt}, {Brown}, {Bullock}, {Burchat}, {Burke},
  {Cagnoli}, {Calabrese}, {Callahan}, {Callen}, {Carlin}, {Carlson},
  {Chandrasekharan}, {Charles-Emerson}, {Chesley}, {Cheu}, {Chiang}, {Chiang},
  {Chirino}, {Chow}, {Ciardi}, {Claver}, {Cohen-Tanugi}, {Cockrum}, {Coles},
  {Connolly}, {Cook}, {Cooray}, {Covey}, {Cribbs}, {Cui}, {Cutri}, {Daly},
  {Daniel}, {Daruich}, {Daubard}, {Daues}, {Dawson}, {Delgado}, {Dellapenna},
  {de Peyster}, {de Val-Borro}, {Digel}, {Doherty}, {Dubois},
  {Dubois-Felsmann}, {Durech}, {Economou}, {Eifler}, {Eracleous}, {Emmons},
  {Fausti Neto}, {Ferguson}, {Figueroa}, {Fisher-Levine}, {Focke}, {Foss},
  {Frank}, {Freemon}, {Gangler}, {Gawiser}, {Geary}, {Gee}, {Geha}, {Gessner},
  {Gibson}, {Gilmore}, {Glanzman}, {Glick}, {Goldina}, {Goldstein}, {Goodenow},
  {Graham}, {Gressler}, {Gris}, {Guy}, {Guyonnet}, {Haller}, {Harris},
  {Hascall}, {Haupt}, {Hernandez}, {Herrmann}, {Hileman}, {Hoblitt}, {Hodgson},
  {Hogan}, {Howard}, {Huang}, {Huffer}, {Ingraham}, {Innes}, {Jacoby}, {Jain},
  {Jammes}, {Jee}, {Jenness}, {Jernigan}, {Jevremovi{\'c}}, {Johns}, {Johnson},
  {Johnson}, {Jones}, {Juramy-Gilles}, {Juri{\'c}}, {Kalirai}, {Kallivayalil},
  {Kalmbach}, {Kantor}, {Karst}, {Kasliwal}, {Kelly}, {Kessler}, {Kinnison},
  {Kirkby}, {Knox}, {Kotov}, {Krabbendam}, {Krughoff}, {Kub{\'a}nek},
  {Kuczewski}, {Kulkarni}, {Ku}, {Kurita}, {Lage}, {Lambert}, {Lange},
  {Langton}, {Le Guillou}, {Levine}, {Liang}, {Lim}, {Lintott}, {Long},
  {Lopez}, {Lotz}, {Lupton}, {Lust}, {MacArthur}, {Mahabal}, {Mandelbaum},
  {Markiewicz}, {Marsh}, {Marshall}, {Marshall}, {May}, {McKercher}, {McQueen},
  {Meyers}, {Migliore}, {Miller}, {Mills}, {Miraval}, {Moeyens}, {Moolekamp},
  {Monet}, {Moniez}, {Monkewitz}, {Montgomery}, {Morrison}, {Mueller},
  {Muller}, {Mu{\~n}oz Arancibia}, {Neill}, {Newbry}, {Nief}, {Nomerotski},
  {Nordby}, {O'Connor}, {Oliver}, {Olivier}, {Olsen}, {O'Mullane}, {Ortiz},
  {Osier}, {Owen}, {Pain}, {Palecek}, {Parejko}, {Parsons}, {Pease},
  {Peterson}, {Peterson}, {Petravick}, {Libby Petrick}, {Petry},
  {Pierfederici}, {Pietrowicz}, {Pike}, {Pinto}, {Plante}, {Plate}, {Plutchak},
  {Price}, {Prouza}, {Radeka}, {Rajagopal}, {Rasmussen}, {Regnault}, {Reil},
  {Reiss}, {Reuter}, {Ridgway}, {Riot}, {Ritz}, {Robinson}, {Roby}, {Roodman},
  {Rosing}, {Roucelle}, {Rumore}, {Russo}, {Saha}, {Sassolas}, {Schalk},
  {Schellart}, {Schindler}, {Schmidt}, {Schneider}, {Schneider}, {Schoening},
  {Schumacher}, {Schwamb}, {Sebag}, {Selvy}, {Sembroski}, {Seppala}, {Serio},
  {Serrano}, {Shaw}, {Shipsey}, {Sick}, {Silvestri}, {Slater}, {Smith},
  {Smith}, {Sobhani}, {Soldahl}, {Storrie-Lombardi}, {Stover}, {Strauss},
  {Street}, {Stubbs}, {Sullivan}, {Sweeney}, {Swinbank}, {Szalay}, {Takacs},
  {Tether}, {Thaler}, {Thayer}, {Thomas}, {Thornton}, {Thukral}, {Tice},
  {Trilling}, {Turri}, {Van Berg}, {Vanden Berk}, {Vetter}, {Virieux},
  {Vucina}, {Wahl}, {Walkowicz}, {Walsh}, {Walter}, {Wang}, {Wang}, {Warner},
  {Wiecha}, {Willman}, {Winters}, {Wittman}, {Wolff}, {Wood-Vasey}, {Wu},
  {Xin}, {Yoachim}, \& {Zhan}}]{Ivezic2019}
{Ivezi{\'c}}, {\v{Z}}., {Kahn}, S.~M., {Tyson}, J.~A., {et~al.} 2019, \apj,
  873, 111, \dodoi{10.3847/1538-4357/ab042c}

\bibitem[{{Jani} {et~al.}(2020){Jani}, {Shoemaker}, \& {Cutler}}]{Jani2020}
{Jani}, K., {Shoemaker}, D., \& {Cutler}, C. 2020, Nature Astronomy, 4, 260,
  \dodoi{10.1038/s41550-019-0932-7}

\bibitem[{{Jiang} {et~al.}(2016){Jiang}, {Guillochon}, \&
  {Loeb}}]{Jiang2016_self_crossing_shock}
{Jiang}, Y.-F., {Guillochon}, J., \& {Loeb}, A. 2016, \apj, 830, 125,
  \dodoi{10.3847/0004-637X/830/2/125}

\bibitem[{{Johnson} {et~al.}(2021){Johnson}, {Leja}, {Conroy}, \&
  {Speagle}}]{Johnson2021a}
{Johnson}, B.~D., {Leja}, J., {Conroy}, C., \& {Speagle}, J.~S. 2021, \apjs,
  254, 22, \dodoi{10.3847/1538-4365/abef67}

\bibitem[{{Kangas} {et~al.}(2022){Kangas}, {Yan}, {Schulze}, {Fransson},
  {Sollerman}, {Lunnan}, {Omand}, {Andreoni}, {Burruss}, {Chen}, {Drake},
  {Fremling}, {Gal-Yam}, {Graham}, {Groom}, {Lezmy}, {Mahabal}, {Masci},
  {Perley}, {Riddle}, {Tartaglia}, \& {Yao}}]{Kangas2022}
{Kangas}, T., {Yan}, L., {Schulze}, S., {et~al.} 2022, \mnras, 516, 1193,
  \dodoi{10.1093/mnras/stac2218}

\bibitem[{{Kelly} \& {Merloni}(2012)}]{Kelly2012}
{Kelly}, B.~C., \& {Merloni}, A. 2012, Advances in Astronomy, 2012, 970858,
  \dodoi{10.1155/2012/970858}

\bibitem[{{Kelly} \& {Shen}(2013)}]{Kelly&Shen2013}
{Kelly}, B.~C., \& {Shen}, Y. 2013, \apj, 764, 45,
  \dodoi{10.1088/0004-637X/764/1/45}

\bibitem[{{Kesden}(2012)}]{Kesden2012}
{Kesden}, M. 2012, \prd, 85, 024037, \dodoi{10.1103/PhysRevD.85.024037}

\bibitem[{{Kewley} {et~al.}(2006){Kewley}, {Groves}, {Kauffmann}, \&
  {Heckman}}]{Kewley2006}
{Kewley}, L.~J., {Groves}, B., {Kauffmann}, G., \& {Heckman}, T. 2006, \mnras,
  372, 961, \dodoi{10.1111/j.1365-2966.2006.10859.x}

\bibitem[{{Kim} {et~al.}(2022){Kim}, {Rigault}, {Neill}, {Briday}, {Copin},
  {Lezmy}, {Nicolas}, {Riddle}, {Sharma}, {Smith}, {Sollerman}, \&
  {Walters}}]{Kim2022}
{Kim}, Y.~L., {Rigault}, M., {Neill}, J.~D., {et~al.} 2022, \pasp, 134, 024505,
  \dodoi{10.1088/1538-3873/ac50a0}

\bibitem[{{Kochanek}(2016)}]{Kochanek2016}
{Kochanek}, C.~S. 2016, \mnras, 461, 371, \dodoi{10.1093/mnras/stw1290}

\bibitem[{{Komossa} {et~al.}(2008){Komossa}, {Zhou}, {Wang}, {Ajello}, {Ge},
  {Greiner}, {Lu}, {Salvato}, {Saxton}, {Shan}, {Xu}, \& {Yuan}}]{Komossa2008}
{Komossa}, S., {Zhou}, H., {Wang}, T., {et~al.} 2008, \apjl, 678, L13,
  \dodoi{10.1086/588281}

\bibitem[{{Kormendy} \& {Ho}(2013)}]{Kormendy2013}
{Kormendy}, J., \& {Ho}, L.~C. 2013, \araa, 51, 511,
  \dodoi{10.1146/annurev-astro-082708-101811}

\bibitem[{{Lacy} {et~al.}(2020){Lacy}, {Baum}, {Chandler}, {Chatterjee},
  {Clarke}, {Deustua}, {English}, {Farnes}, {Gaensler}, {Gugliucci},
  {Hallinan}, {Kent}, {Kimball}, {Law}, {Lazio}, {Marvil}, {Mao}, {Medlin},
  {Mooley}, {Murphy}, {Myers}, {Osten}, {Richards}, {Rosolowsky}, {Rudnick},
  {Schinzel}, {Sivakoff}, {Sjouwerman}, {Taylor}, {White}, {Wrobel},
  {Andernach}, {Beasley}, {Berger}, {Bhatnager}, {Birkinshaw}, {Bower},
  {Brandt}, {Brown}, {Burke-Spolaor}, {Butler}, {Comerford}, {Demorest}, {Fu},
  {Giacintucci}, {Golap}, {G{\"u}th}, {Hales}, {Hiriart}, {Hodge}, {Horesh},
  {Ivezi{\'c}}, {Jarvis}, {Kamble}, {Kassim}, {Liu}, {Loinard}, {Lyons},
  {Masters}, {Mezcua}, {Moellenbrock}, {Mroczkowski}, {Nyland},
  {O{\textquoteright}Dea}, {O{\textquoteright}Sullivan}, {Peters}, {Radford},
  {Rao}, {Robnett}, {Salcido}, {Shen}, {Sobotka}, {Witz}, {Vaccari}, {van
  Weeren}, {Vargas}, {Williams}, \& {Yoon}}]{Lacy2020}
{Lacy}, M., {Baum}, S.~A., {Chandler}, C.~J., {et~al.} 2020, \pasp, 132,
  035001, \dodoi{10.1088/1538-3873/ab63eb}

\bibitem[{{Lang}(2014)}]{Lang2014}
{Lang}, D. 2014, \aj, 147, 108, \dodoi{10.1088/0004-6256/147/5/108}

\bibitem[{{Latimer} {et~al.}(2021{\natexlab{a}}){Latimer}, {Reines}, {Bogdan},
  \& {Kraft}}]{Latimer2021_AGN_fraction}
{Latimer}, L.~J., {Reines}, A.~E., {Bogdan}, A., \& {Kraft}, R.
  2021{\natexlab{a}}, \apjl, 922, L40, \dodoi{10.3847/2041-8213/ac3af6}

\bibitem[{{Latimer} {et~al.}(2021{\natexlab{b}}){Latimer}, {Reines},
  {Hainline}, {Greene}, \& {Stern}}]{Latimer2021_color}
{Latimer}, L.~J., {Reines}, A.~E., {Hainline}, K.~N., {Greene}, J.~E., \&
  {Stern}, D. 2021{\natexlab{b}}, \apj, 914, 133,
  \dodoi{10.3847/1538-4357/abfe0c}

\bibitem[{{Law-Smith} {et~al.}(2017){Law-Smith}, {Ramirez-Ruiz}, {Ellison}, \&
  {Foley}}]{Law-Smith2017}
{Law-Smith}, J., {Ramirez-Ruiz}, E., {Ellison}, S.~L., \& {Foley}, R.~J. 2017,
  \apj, 850, 22, \dodoi{10.3847/1538-4357/aa94c7}

\bibitem[{{Lezhnin} \& {Vasiliev}(2015)}]{Lezhnin2015}
{Lezhnin}, K., \& {Vasiliev}, E. 2015, \apjl, 808, L5,
  \dodoi{10.1088/2041-8205/808/1/L5}

\bibitem[{{Lin} {et~al.}(2022){Lin}, {Jiang}, {Kong}, {Huang}, {Lin}, {Zhu}, \&
  {Wang}}]{Lin2022}
{Lin}, Z., {Jiang}, N., {Kong}, X., {et~al.} 2022, \apjl, 939, L33,
  \dodoi{10.3847/2041-8213/ac9c63}

\bibitem[{{Loeb} \& {Ulmer}(1997)}]{loeb97_spherical_envelope}
{Loeb}, A., \& {Ulmer}, A. 1997, \apj, 489, 573, \dodoi{10.1086/304814}

\bibitem[{{Lu} \& {Bonnerot}(2020)}]{Lu2020}
{Lu}, W., \& {Bonnerot}, C. 2020, \mnras, 492, 686,
  \dodoi{10.1093/mnras/stz3405}

\bibitem[{{MacLeod} {et~al.}(2012){MacLeod}, {Guillochon}, \&
  {Ramirez-Ruiz}}]{MacLeod2012}
{MacLeod}, M., {Guillochon}, J., \& {Ramirez-Ruiz}, E. 2012, \apj, 757, 134,
  \dodoi{10.1088/0004-637X/757/2/134}

\bibitem[{{MacLeod} {et~al.}(2013){MacLeod}, {Ramirez-Ruiz}, {Grady}, \&
  {Guillochon}}]{MacLeod2013}
{MacLeod}, M., {Ramirez-Ruiz}, E., {Grady}, S., \& {Guillochon}, J. 2013, \apj,
  777, 133, \dodoi{10.1088/0004-637X/777/2/133}

\bibitem[{{Magorrian} \& {Tremaine}(1999)}]{Magorrian1999}
{Magorrian}, J., \& {Tremaine}, S. 1999, \mnras, 309, 447,
  \dodoi{10.1046/j.1365-8711.1999.02853.x}

\bibitem[{{Mahabal} {et~al.}(2019){Mahabal}, {Rebbapragada}, {Walters},
  {Masci}, {Blagorodnova}, {van Roestel}, {Ye}, {Biswas}, {Burdge}, {Chang},
  {Duev}, {Golkhou}, {Miller}, {Nordin}, {Ward}, {Adams}, {Bellm}, {Branton},
  {Bue}, {Cannella}, {Connolly}, {Dekany}, {Feindt}, {Hung}, {Fortson},
  {Frederick}, {Fremling}, {Gezari}, {Graham}, {Groom}, {Kasliwal}, {Kulkarni},
  {Kupfer}, {Lin}, {Lintott}, {Lunnan}, {Parejko}, {Prince}, {Riddle},
  {Rusholme}, {Saunders}, {Sedaghat}, {Shupe}, {Singer}, {Soumagnac}, {Szkody},
  {Tachibana}, {Tirumala}, {van Velzen}, \& {Wright}}]{Mahabal2019}
{Mahabal}, A., {Rebbapragada}, U., {Walters}, R., {et~al.} 2019, \pasp, 131,
  038002, \dodoi{10.1088/1538-3873/aaf3fa}

\bibitem[{{Mainzer} {et~al.}(2011){Mainzer}, {Bauer}, {Grav}, {Masiero},
  {Cutri}, {Dailey}, {Eisenhardt}, {McMillan}, {Wright}, {Walker}, {Jedicke},
  {Spahr}, {Tholen}, {Alles}, {Beck}, {Brandenburg}, {Conrow}, {Evans},
  {Fowler}, {Jarrett}, {Marsh}, {Masci}, {McCallon}, {Wheelock}, {Wittman},
  {Wyatt}, {DeBaun}, {Elliott}, {Elsbury}, {Gautier}, {Gomillion}, {Leisawitz},
  {Maleszewski}, {Micheli}, \& {Wilkins}}]{Mainzer2011}
{Mainzer}, A., {Bauer}, J., {Grav}, T., {et~al.} 2011, \apj, 731, 53,
  \dodoi{10.1088/0004-637X/731/1/53}

\bibitem[{{Marconi} {et~al.}(2004){Marconi}, {Risaliti}, {Gilli}, {Hunt},
  {Maiolino}, \& {Salvati}}]{Marconi2004}
{Marconi}, A., {Risaliti}, G., {Gilli}, R., {et~al.} 2004, \mnras, 351, 169,
  \dodoi{10.1111/j.1365-2966.2004.07765.x}

\bibitem[{{Margalit} {et~al.}(2022){Margalit}, {Quataert}, \&
  {Ho}}]{Margalit2022}
{Margalit}, B., {Quataert}, E., \& {Ho}, A. Y.~Q. 2022, \apj, 928, 122,
  \dodoi{10.3847/1538-4357/ac53b0}

\bibitem[{{Martin} {et~al.}(2005){Martin}, {Fanson}, {Schiminovich},
  {Morrissey}, {Friedman}, {Barlow}, {Conrow}, {Grange}, {Jelinsky},
  {Milliard}, {Siegmund}, {Bianchi}, {Byun}, {Donas}, {Forster}, {Heckman},
  {Lee}, {Madore}, {Malina}, {Neff}, {Rich}, {Small}, {Surber}, {Szalay},
  {Welsh}, \& {Wyder}}]{Martin2005a}
{Martin}, D.~C., {Fanson}, J., {Schiminovich}, D., {et~al.} 2005, \apj, 619,
  L1, \dodoi{10.1086/426387}

\bibitem[{{Masci} {et~al.}(2019){Masci}, {Laher}, {Rusholme}, {Shupe}, {Groom},
  {Surace}, {Jackson}, {Monkewitz}, {Beck}, {Flynn}, {Terek}, {Landry},
  {Hacopians}, {Desai}, {Howell}, {Brooke}, {Imel}, {Wachter}, {Ye}, {Lin},
  {Cenko}, {Cunningham}, {Rebbapragada}, {Bue}, {Miller}, {Mahabal}, {Bellm},
  {Patterson}, {Juri{\'c}}, {Golkhou}, {Ofek}, {Walters}, {Graham}, {Kasliwal},
  {Dekany}, {Kupfer}, {Burdge}, {Cannella}, {Barlow}, {Van Sistine}, {Giomi},
  {Fremling}, {Blagorodnova}, {Levitan}, {Riddle}, {Smith}, {Helou}, {Prince},
  \& {Kulkarni}}]{Masci2019}
{Masci}, F.~J., {Laher}, R.~R., {Rusholme}, B., {et~al.} 2019, \pasp, 131,
  018003, \dodoi{10.1088/1538-3873/aae8ac}

\bibitem[{{Mendel} {et~al.}(2014){Mendel}, {Simard}, {Palmer}, {Ellison}, \&
  {Patton}}]{Mendel2014}
{Mendel}, J.~T., {Simard}, L., {Palmer}, M., {Ellison}, S.~L., \& {Patton},
  D.~R. 2014, \apjs, 210, 3, \dodoi{10.1088/0067-0049/210/1/3}

\bibitem[{{Merloni} \& {Heinz}(2008)}]{Merloni2008}
{Merloni}, A., \& {Heinz}, S. 2008, \mnras, 388, 1011,
  \dodoi{10.1111/j.1365-2966.2008.13472.x}

\bibitem[{{Merritt} \& {Ferrarese}(2001)}]{Merritt2001}
{Merritt}, D., \& {Ferrarese}, L. 2001, \apj, 547, 140, \dodoi{10.1086/318372}

\bibitem[{{Metzger}(2022)}]{metzger22_colling_envelope}
{Metzger}, B.~D. 2022, \apjl, 937, L12, \dodoi{10.3847/2041-8213/ac90ba}

\bibitem[{{Metzger} \& {Stone}(2016)}]{metzger16_reprocessing}
{Metzger}, B.~D., \& {Stone}, N.~C. 2016, \mnras, 461, 948,
  \dodoi{10.1093/mnras/stw1394}

\bibitem[{{Miller} \& {Hall}(2021)}]{Miller2021}
{Miller}, A.~A., \& {Hall}, X.~J. 2021, \pasp, 133, 054502,
  \dodoi{10.1088/1538-3873/abf038}

\bibitem[{{Miller} {et~al.}(2012){Miller}, {Gallo}, {Treu}, \&
  {Woo}}]{Miller2012}
{Miller}, B., {Gallo}, E., {Treu}, T., \& {Woo}, J.-H. 2012, \apj, 747, 57,
  \dodoi{10.1088/0004-637X/747/1/57}

\bibitem[{{Miller} {et~al.}(2015){Miller}, {Gallo}, {Greene}, {Kelly}, {Treu},
  {Woo}, \& {Baldassare}}]{Miller2015}
{Miller}, B.~P., {Gallo}, E., {Greene}, J.~E., {et~al.} 2015, \apj, 799, 98,
  \dodoi{10.1088/0004-637X/799/1/98}

\bibitem[{{Miller}(2015)}]{Miller2015_disk_wind}
{Miller}, M.~C. 2015, \apj, 805, 83, \dodoi{10.1088/0004-637X/805/1/83}

\bibitem[{{Million} {et~al.}(2016){Million}, {Fleming}, {Shiao}, {Seibert},
  {Loyd}, {Tucker}, {Smith}, {Thompson}, \& {White}}]{Million2016}
{Million}, C., {Fleming}, S.~W., {Shiao}, B., {et~al.} 2016, \apj, 833, 292,
  \dodoi{10.3847/1538-4357/833/2/292}

\bibitem[{{Nicholl} {et~al.}(2019{\natexlab{a}}){Nicholl}, {Berger},
  {Blanchard}, {Gomez}, \& {Chornock}}]{Nicholl2019_SLSN}
{Nicholl}, M., {Berger}, E., {Blanchard}, P.~K., {Gomez}, S., \& {Chornock}, R.
  2019{\natexlab{a}}, \apj, 871, 102, \dodoi{10.3847/1538-4357/aaf470}

\bibitem[{{Nicholl} {et~al.}(2022){Nicholl}, {Lanning}, {Ramsden}, {Mockler},
  {Lawrence}, {Short}, \& {Ridley}}]{Nicholl2022}
{Nicholl}, M., {Lanning}, D., {Ramsden}, P., {et~al.} 2022, \mnras, 515, 5604,
  \dodoi{10.1093/mnras/stac2206}

\bibitem[{{Nicholl} {et~al.}(2019{\natexlab{b}}){Nicholl}, {Short}, {Lawrence},
  {Ross}, {Smartt}, \& {Oates}}]{Nicholl2019_ZTF19acfwynw}
{Nicholl}, M., {Short}, P., {Lawrence}, A., {et~al.} 2019{\natexlab{b}},
  Transient Name Server Classification Report, 2019-2271, 1

\bibitem[{{Nicholl} {et~al.}(2019{\natexlab{c}}){Nicholl}, {Blanchard},
  {Berger}, {Gomez}, {Margutti}, {Alexander}, {Guillochon}, {Leja}, {Chornock},
  {Snios}, {Auchettl}, {Bruce}, {Challis}, {D'Orazio}, {Drout}, {Eftekhari},
  {Foley}, {Graur}, {Kilpatrick}, {Lawrence}, {Piro}, {Rojas-Bravo}, {Ross},
  {Short}, {Smartt}, {Smith}, \& {Stalder}}]{Nicholl2019}
{Nicholl}, M., {Blanchard}, P.~K., {Berger}, E., {et~al.} 2019{\natexlab{c}},
  \mnras, 488, 1878, \dodoi{10.1093/mnras/stz1837}

\bibitem[{{Nicholl} {et~al.}(2020){Nicholl}, {Wevers}, {Oates}, {Alexander},
  {Leloudas}, {Onori}, {Jerkstrand}, {Gomez}, {Campana}, {Arcavi},
  {Charalampopoulos}, {Gromadzki}, {Ihanec}, {Jonker}, {Lawrence}, {Mandel},
  {Schulze}, {Short}, {Burke}, {McCully}, {Hiramatsu}, {Howell}, {Pellegrino},
  {Abbot}, {Anderson}, {Berger}, {Blanchard}, {Cannizzaro}, {Chen},
  {Dennefeld}, {Galbany}, {Gonz{\'a}lez-Gait{\'a}n}, {Hosseinzadeh}, {Inserra},
  {Irani}, {Kuin}, {M{\"u}ller-Bravo}, {Pineda}, {Ross}, {Roy}, {Smartt},
  {Smith}, {Tucker}, {Wyrzykowski}, \& {Young}}]{Nicholl2020}
{Nicholl}, M., {Wevers}, T., {Oates}, S.~R., {et~al.} 2020, \mnras, 499, 482,
  \dodoi{10.1093/mnras/staa2824}

\bibitem[{{Nordin} {et~al.}(2019){Nordin}, {Brinnel}, {van Santen}, {Bulla},
  {Feindt}, {Franckowiak}, {Fremling}, {Gal-Yam}, {Giomi}, {Kowalski},
  {Mahabal}, {Miranda}, {Rauch}, {Reusch}, {Rigault}, {Schulze}, {Sollerman},
  {Stein}, {Yaron}, {van Velzen}, \& {Ward}}]{Nordin2019}
{Nordin}, J., {Brinnel}, V., {van Santen}, J., {et~al.} 2019, \aap, 631, A147,
  \dodoi{10.1051/0004-6361/201935634}

\bibitem[{{Oke} \& {Gunn}(1982)}]{Oke1982}
{Oke}, J.~B., \& {Gunn}, J.~E. 1982, \pasp, 94, 586, \dodoi{10.1086/131027}

\bibitem[{{Oke} {et~al.}(1995){Oke}, {Cohen}, {Carr}, {Cromer}, {Dingizian},
  {Harris}, {Labrecque}, {Lucinio}, {Schaal}, {Epps}, \& {Miller}}]{Oke1995}
{Oke}, J.~B., {Cohen}, J.~G., {Carr}, M., {et~al.} 1995, \pasp, 107, 375,
  \dodoi{10.1086/133562}

\bibitem[{{Onori} {et~al.}(2019){Onori}, {Cannizzaro}, {Jonker}, {Fraser},
  {Kostrzewa-Rutkowska}, {Martin-Carrillo}, {Benetti}, {Elias-Rosa},
  {Gromadzki}, {Harmanen}, {Mattila}, {Strizinger}, {Terreran}, \&
  {Wevers}}]{Onori2019}
{Onori}, F., {Cannizzaro}, G., {Jonker}, P.~G., {et~al.} 2019, \mnras, 489,
  1463, \dodoi{10.1093/mnras/stz2053}

\bibitem[{{Palaversa} {et~al.}(2016){Palaversa}, {Gezari}, {Sesar}, {Stuart},
  {Wozniak}, {Holl}, \& {Ivezi{\'c}}}]{Palaversa2016}
{Palaversa}, L., {Gezari}, S., {Sesar}, B., {et~al.} 2016, \apj, 819, 151,
  \dodoi{10.3847/0004-637X/819/2/151}

\bibitem[{{Patterson} {et~al.}(2019){Patterson}, {Bellm}, {Rusholme}, {Masci},
  {Juric}, {Krughoff}, {Golkhou}, {Graham}, {Kulkarni}, {Helou}, \& {Zwicky
  Transient Facility Collaboration}}]{Patterson2019}
{Patterson}, M.~T., {Bellm}, E.~C., {Rusholme}, B., {et~al.} 2019, \pasp, 131,
  018001, \dodoi{10.1088/1538-3873/aae904}

\bibitem[{{Perez-Fournon} {et~al.}(2020){Perez-Fournon}, {Poidevin}, {Angel},
  {Shirley}, {Marques-Chaves}, {Geier}, {Shu}, {Rodney}, {Roberts-Pierel},
  {Bolton}, {Chakrabarti}, {Craig}, \&
  {Alamiri}}]{Perez-Fournon2020_ZTF20abobpcb}
{Perez-Fournon}, I., {Poidevin}, F., {Angel}, C.~J., {et~al.} 2020, Transient
  Name Server Classification Report, 2020-2456, 1

\bibitem[{{Perley} {et~al.}(2020{\natexlab{a}}){Perley}, {Taggart}, {Dahiwale},
  \& {Fremling}}]{Perley2020_ZTF20abgoocl}
{Perley}, D.~A., {Taggart}, K., {Dahiwale}, A., \& {Fremling}, C.
  2020{\natexlab{a}}, Transient Name Server Classification Report, 2020-2086, 1

\bibitem[{{Perley} {et~al.}(2020{\natexlab{b}}){Perley}, {Fremling},
  {Sollerman}, {Miller}, {Dahiwale}, {Sharma}, {Bellm}, {Biswas}, {Brink},
  {Bruch}, {De}, {Dekany}, {Drake}, {Duev}, {Filippenko}, {Gal-Yam}, {Goobar},
  {Graham}, {Graham}, {Ho}, {Irani}, {Kasliwal}, {Kim}, {Kulkarni}, {Mahabal},
  {Masci}, {Modak}, {Neill}, {Nordin}, {Riddle}, {Soumagnac}, {Strotjohann},
  {Schulze}, {Taggart}, {Tzanidakis}, {Walters}, \& {Yan}}]{Perley2020}
{Perley}, D.~A., {Fremling}, C., {Sollerman}, J., {et~al.} 2020{\natexlab{b}},
  \apj, 904, 35, \dodoi{10.3847/1538-4357/abbd98}

\bibitem[{{Pessi} {et~al.}(2020){Pessi}, {Anderson}, {Galbany}, \&
  {Yaron}}]{Pessi2020_ZTF20acbcfaa}
{Pessi}, P.~J., {Anderson}, J., {Galbany}, L., \& {Yaron}, O. 2020, Transient
  Name Server Classification Report, 2020-3712, 1

\bibitem[{{Pinkney} {et~al.}(2003){Pinkney}, {Gebhardt}, {Bender}, {Bower},
  {Dressler}, {Faber}, {Filippenko}, {Green}, {Ho}, {Kormendy}, {Lauer},
  {Magorrian}, {Richstone}, \& {Tremaine}}]{Pinkney2003}
{Pinkney}, J., {Gebhardt}, K., {Bender}, R., {et~al.} 2003, \apj, 596, 903,
  \dodoi{10.1086/378118}

\bibitem[{{Piran} {et~al.}(2015){Piran}, {Svirski}, {Krolik}, {Cheng}, \&
  {Shiokawa}}]{piran15_shock_model}
{Piran}, T., {Svirski}, G., {Krolik}, J., {Cheng}, R.~M., \& {Shiokawa}, H.
  2015, \apj, 806, 164, \dodoi{10.1088/0004-637X/806/2/164}

\bibitem[{{Predehl} {et~al.}(2021){Predehl}, {Andritschke}, {Arefiev},
  {Babyshkin}, {Batanov}, {Becker}, {B{\"o}hringer}, {Bogomolov}, {Boller},
  {Borm}, {Bornemann}, {Br{\"a}uninger}, {Br{\"u}ggen}, {Brunner}, {Brusa},
  {Bulbul}, {Buntov}, {Burwitz}, {Burkert}, {Clerc}, {Churazov}, {Coutinho},
  {Dauser}, {Dennerl}, {Doroshenko}, {Eder}, {Emberger}, {Eraerds},
  {Finoguenov}, {Freyberg}, {Friedrich}, {Friedrich}, {F{\"u}rmetz},
  {Georgakakis}, {Gilfanov}, {Granato}, {Grossberger}, {Gueguen}, {Gureev},
  {Haberl}, {H{\"a}lker}, {Hartner}, {Hasinger}, {Huber}, {Ji}, {Kienlin},
  {Kink}, {Korotkov}, {Kreykenbohm}, {Lamer}, {Lomakin}, {Lapshov}, {Liu},
  {Maitra}, {Meidinger}, {Menz}, {Merloni}, {Mernik}, {Mican}, {Mohr},
  {M{\"u}ller}, {Nandra}, {Nazarov}, {Pacaud}, {Pavlinsky}, {Perinati},
  {Pfeffermann}, {Pietschner}, {Ramos-Ceja}, {Rau}, {Reiffers}, {Reiprich},
  {Robrade}, {Salvato}, {Sanders}, {Santangelo}, {Sasaki}, {Scheuerle},
  {Schmid}, {Schmitt}, {Schwope}, {Shirshakov}, {Steinmetz}, {Stewart},
  {Str{\"u}der}, {Sunyaev}, {Tenzer}, {Tiedemann}, {Tr{\"u}mper}, {Voron},
  {Weber}, {Wilms}, \& {Yaroshenko}}]{Predehl2021}
{Predehl}, P., {Andritschke}, R., {Arefiev}, V., {et~al.} 2021, \aap, 647, A1,
  \dodoi{10.1051/0004-6361/202039313}

\bibitem[{{Prugniel} \& {Soubiran}(2001)}]{Prugniel2001}
{Prugniel}, P., \& {Soubiran}, C. 2001, \aap, 369, 1048,
  \dodoi{10.1051/0004-6361:20010163}

\bibitem[{{Prugniel} {et~al.}(2007){Prugniel}, {Soubiran}, {Koleva}, \& {Le
  Borgne}}]{Prugniel2007}
{Prugniel}, P., {Soubiran}, C., {Koleva}, M., \& {Le Borgne}, D. 2007, arXiv
  e-prints, astro.
\newblock \doarXiv{astro-ph/0703658}

\bibitem[{{Quimby} {et~al.}(2011){Quimby}, {Kulkarni}, {Kasliwal}, {Gal-Yam},
  {Arcavi}, {Sullivan}, {Nugent}, {Thomas}, {Howell}, {Nakar}, {Bildsten},
  {Theissen}, {Law}, {Dekany}, {Rahmer}, {Hale}, {Smith}, {Ofek}, {Zolkower},
  {Velur}, {Walters}, {Henning}, {Bui}, {McKenna}, {Poznanski}, {Cenko}, \&
  {Levitan}}]{Quimby2011}
{Quimby}, R.~M., {Kulkarni}, S.~R., {Kasliwal}, M.~M., {et~al.} 2011, \nat,
  474, 487, \dodoi{10.1038/nature10095}

\bibitem[{Raftery(1995)}]{Raftery1995}
Raftery, A.~E. 1995, Sociological Methodology, 25, 111.
\newblock \url{http://www.jstor.org/stable/271063}

\bibitem[{{Ramsden} {et~al.}(2022){Ramsden}, {Lanning}, {Nicholl}, \&
  {McGee}}]{Ramsden2022}
{Ramsden}, P., {Lanning}, D., {Nicholl}, M., \& {McGee}, S.~L. 2022, \mnras,
  515, 1146, \dodoi{10.1093/mnras/stac1810}

\bibitem[{{Rees}(1988)}]{Rees1988}
{Rees}, M.~J. 1988, \nat, 333, 523, \dodoi{10.1038/333523a0}

\bibitem[{{Reines} \& {Volonteri}(2015)}]{Reines2015}
{Reines}, A.~E., \& {Volonteri}, M. 2015, \apj, 813, 82,
  \dodoi{10.1088/0004-637X/813/2/82}

\bibitem[{{Reynolds}(2021)}]{Reynolds2021}
{Reynolds}, C.~S. 2021, \araa, 59, 117,
  \dodoi{10.1146/annurev-astro-112420-035022}

\bibitem[{{Ricarte} \& {Natarajan}(2018{\natexlab{a}})}]{Ricarte2018b}
{Ricarte}, A., \& {Natarajan}, P. 2018{\natexlab{a}}, \mnras, 481, 3278,
  \dodoi{10.1093/mnras/sty2448}

\bibitem[{{Ricarte} \& {Natarajan}(2018{\natexlab{b}})}]{Ricarte2018a}
---. 2018{\natexlab{b}}, \mnras, 474, 1995, \dodoi{10.1093/mnras/stx2851}

\bibitem[{{Ricarte} {et~al.}(2019){Ricarte}, {Pacucci}, {Cappelluti}, \&
  {Natarajan}}]{Ricarte2019}
{Ricarte}, A., {Pacucci}, F., {Cappelluti}, N., \& {Natarajan}, P. 2019,
  \mnras, 489, 1006, \dodoi{10.1093/mnras/stz1891}

\bibitem[{{Rigault} {et~al.}(2019){Rigault}, {Neill}, {Blagorodnova}, {Dugas},
  {Feeney}, {Walters}, {Brinnel}, {Copin}, {Fremling}, {Nordin}, \&
  {Sollerman}}]{Rigault2019}
{Rigault}, M., {Neill}, J.~D., {Blagorodnova}, N., {et~al.} 2019, \aap, 627,
  A115, \dodoi{10.1051/0004-6361/201935344}

\bibitem[{{Roming} {et~al.}(2005){Roming}, {Kennedy}, {Mason}, {Nousek}, {Ahr},
  {Bingham}, {Broos}, {Carter}, {Hancock}, \& {Huckle}}]{Roming2005}
{Roming}, P. W.~A., {Kennedy}, T.~E., {Mason}, K.~O., {et~al.} 2005, \ssr, 120,
  95, \dodoi{10.1007/s11214-005-5095-4}

\bibitem[{{Roth} \& {Kasen}(2018)}]{Roth2018}
{Roth}, N., \& {Kasen}, D. 2018, \apj, 855, 54,
  \dodoi{10.3847/1538-4357/aaaec6}

\bibitem[{{Roth} {et~al.}(2016){Roth}, {Kasen}, {Guillochon}, \&
  {Ramirez-Ruiz}}]{roth16_reprocessing}
{Roth}, N., {Kasen}, D., {Guillochon}, J., \& {Ramirez-Ruiz}, E. 2016, \apj,
  827, 3, \dodoi{10.3847/0004-637X/827/1/3}

\bibitem[{{Roth} {et~al.}(2021){Roth}, {van Velzen}, {Cenko}, \&
  {Mushotzky}}]{Roth2021}
{Roth}, N., {van Velzen}, S., {Cenko}, S.~B., \& {Mushotzky}, R.~F. 2021, \apj,
  910, 93, \dodoi{10.3847/1538-4357/abdf50}

\bibitem[{{S{\'a}nchez-Janssen} {et~al.}(2019){S{\'a}nchez-Janssen},
  {C{\^o}t{\'e}}, {Ferrarese}, {Peng}, {Roediger}, {Blakeslee}, {Emsellem},
  {Puzia}, {Spengler}, {Taylor}, {{\'A}lamo-Mart{\'\i}nez}, {Boselli},
  {Cantiello}, {Cuillandre}, {Duc}, {Durrell}, {Gwyn}, {MacArthur},
  {Lan{\c{c}}on}, {Lim}, {Liu}, {Mei}, {Miller}, {Mu{\~n}oz}, {Mihos},
  {Paudel}, {Powalka}, \& {Toloba}}]{Sanchez-Janssen2019}
{S{\'a}nchez-Janssen}, R., {C{\^o}t{\'e}}, P., {Ferrarese}, L., {et~al.} 2019,
  \apj, 878, 18, \dodoi{10.3847/1538-4357/aaf4fd}

\bibitem[{{Saxton} {et~al.}(2020){Saxton}, {Komossa}, {Auchettl}, \&
  {Jonker}}]{Saxton2020}
{Saxton}, R., {Komossa}, S., {Auchettl}, K., \& {Jonker}, P.~G. 2020, \ssr,
  216, 85, \dodoi{10.1007/s11214-020-00708-4}

\bibitem[{{Sazonov} {et~al.}(2021){Sazonov}, {Gilfanov}, {Medvedev}, {Yao},
  {Khorunzhev}, {Semena}, {Sunyaev}, {Burenin}, {Lyapin}, {Meshcheryakov},
  {Uskov}, {Zaznobin}, {Postnov}, {Dodin}, {Belinski}, {Cherepashchuk},
  {Eselevich}, {Dodonov}, {Grokhovskaya}, {Kotov}, {Bikmaev}, {Zhuchkov},
  {Gumerov}, {van Velzen}, \& {Kulkarni}}]{Sazonov2021}
{Sazonov}, S., {Gilfanov}, M., {Medvedev}, P., {et~al.} 2021, \mnras, 508,
  3820, \dodoi{10.1093/mnras/stab2843}

\bibitem[{{Schawinski} {et~al.}(2014){Schawinski}, {Urry}, {Simmons},
  {Fortson}, {Kaviraj}, {Keel}, {Lintott}, {Masters}, {Nichol}, {Sarzi},
  {Skibba}, {Treister}, {Willett}, {Wong}, \& {Yi}}]{Schawinski2014}
{Schawinski}, K., {Urry}, C.~M., {Simmons}, B.~D., {et~al.} 2014, \mnras, 440,
  889, \dodoi{10.1093/mnras/stu327}

\bibitem[{{Schlafly} \& {Finkbeiner}(2011)}]{Schlafly2011}
{Schlafly}, E.~F., \& {Finkbeiner}, D.~P. 2011, \apj, 737, 103,
  \dodoi{10.1088/0004-637X/737/2/103}

\bibitem[{{Schmidt}(1968)}]{Schmidt1968}
{Schmidt}, M. 1968, \apj, 151, 393, \dodoi{10.1086/149446}

\bibitem[{{Schulze} {et~al.}(2021){Schulze}, {Yaron}, {Sollerman}, {Leloudas},
  {Gal}, {Wright}, {Lunnan}, {Gal-Yam}, {Ofek}, {Perley}, {Filippenko},
  {Kasliwal}, {Kulkarni}, {Neill}, {Nugent}, {Quimby}, {Sullivan},
  {Strotjohann}, {Arcavi}, {Ben-Ami}, {Bianco}, {Bloom}, {De}, {Fraser},
  {Fremling}, {Horesh}, {Johansson}, {Kelly}, {Kne{\v{z}}evi{\'c}},
  {Kne{\v{z}}evi{\'c}}, {Maguire}, {Nyholm}, {Papadogiannakis}, {Petrushevska},
  {Rubin}, {Yan}, {Yang}, {Adams}, {Bufano}, {Clubb}, {Foley}, {Green},
  {Harmanen}, {Ho}, {Hook}, {Hosseinzadeh}, {Howell}, {Kong}, {Kotak},
  {Matheson}, {McCully}, {Milisavljevic}, {Pan}, {Poznanski}, {Shivvers}, {van
  Velzen}, \& {Verbeek}}]{Schulze2021a}
{Schulze}, S., {Yaron}, O., {Sollerman}, J., {et~al.} 2021, \apjs, 255, 29,
  \dodoi{10.3847/1538-4365/abff5e}

\bibitem[{{Shankar}(2013)}]{Shankar2013}
{Shankar}, F. 2013, Classical and Quantum Gravity, 30, 244001,
  \dodoi{10.1088/0264-9381/30/24/244001}

\bibitem[{{Shankar} {et~al.}(2009){Shankar}, {Weinberg}, \&
  {Miralda-Escud{\'e}}}]{Shankar2009}
{Shankar}, F., {Weinberg}, D.~H., \& {Miralda-Escud{\'e}}, J. 2009, \apj, 690,
  20, \dodoi{10.1088/0004-637X/690/1/20}

\bibitem[{{Shankar} {et~al.}(2016){Shankar}, {Bernardi}, {Sheth}, {Ferrarese},
  {Graham}, {Savorgnan}, {Allevato}, {Marconi}, {L{\"a}sker}, \&
  {Lapi}}]{Shankar2016}
{Shankar}, F., {Bernardi}, M., {Sheth}, R.~K., {et~al.} 2016, \mnras, 460,
  3119, \dodoi{10.1093/mnras/stw678}

\bibitem[{{Sheinis} {et~al.}(2002){Sheinis}, {Bolte}, {Epps}, {Kibrick},
  {Miller}, {Radovan}, {Bigelow}, \& {Sutin}}]{Sheinis2002}
{Sheinis}, A.~I., {Bolte}, M., {Epps}, H.~W., {et~al.} 2002, \pasp, 114, 851,
  \dodoi{10.1086/341706}

\bibitem[{{Shingles} {et~al.}(2021){Shingles}, {Smith}, {Young}, {Smartt},
  {Tonry}, {Denneau}, {Heinze}, {Weiland}, {Flewelling}, {Stalder},
  {Clocchiatti}, {F{\"o}rster}, {Pignata}, {Rest}, {Anderson}, {Stubbs}, \&
  {Erasmus}}]{Shingles2021}
{Shingles}, L., {Smith}, K.~W., {Young}, D.~R., {et~al.} 2021, Transient Name
  Server AstroNote, 7, 1

\bibitem[{{Siebert}(2020)}]{Siebert2020_ZTF20aaurjbj}
{Siebert}, M. 2020, Transient Name Server Classification Report, 2020-1469, 1

\bibitem[{{Skrutskie} {et~al.}(2006){Skrutskie}, {Cutri}, {Stiening},
  {Weinberg}, {Schneider}, {Carpenter}, {Beichman}, {Capps}, {Chester},
  {Elias}, {Huchra}, {Liebert}, {Lonsdale}, {Monet}, {Price}, {Seitzer},
  {Jarrett}, {Kirkpatrick}, {Gizis}, {Howard}, {Evans}, {Fowler}, {Fullmer},
  {Hurt}, {Light}, {Kopan}, {Marsh}, {McCallon}, {Tam}, {Van Dyk}, \&
  {Wheelock}}]{Skrutskie2006}
{Skrutskie}, M.~F., {Cutri}, R.~M., {Stiening}, R., {et~al.} 2006, \aj, 131,
  1163, \dodoi{10.1086/498708}

\bibitem[{{Smith} {et~al.}(2020){Smith}, {Smartt}, {Young}, {Tonry}, {Denneau},
  {Flewelling}, {Heinze}, {Weiland}, {Stalder}, {Rest}, {Stubbs}, {Anderson},
  {Chen}, {Clark}, {Do}, {F{\"o}rster}, {Fulton}, {Gillanders}, {McBrien},
  {O'Neill}, {Srivastav}, \& {Wright}}]{Smith2020}
{Smith}, K.~W., {Smartt}, S.~J., {Young}, D.~R., {et~al.} 2020, \pasp, 132,
  085002, \dodoi{10.1088/1538-3873/ab936e}

\bibitem[{{SNIascore}(2021)}]{SNIascore2021_ZTF21abcmepi}
{SNIascore}. 2021, Transient Name Server Classification Report, 2021-1939, 1

\bibitem[{{Somalwar} {et~al.}(2022){Somalwar}, {Ravi}, {Dong}, {Graham},
  {Hallinan}, {Law}, {Lu}, \& {Myers}}]{Somalwar2022}
{Somalwar}, J.~J., {Ravi}, V., {Dong}, D., {et~al.} 2022, \apj, 929, 184,
  \dodoi{10.3847/1538-4357/ac5e29}

\bibitem[{{Soumagnac} \& {Ofek}(2018)}]{Soumagnac2018}
{Soumagnac}, M.~T., \& {Ofek}, E.~O. 2018, \pasp, 130, 075002,
  \dodoi{10.1088/1538-3873/aac410}

\bibitem[{{Stein} {et~al.}(2021){Stein}, {Velzen}, {Kowalski}, {Franckowiak},
  {Gezari}, {Miller-Jones}, {Frederick}, {Sfaradi}, {Bietenholz}, {Horesh},
  {Fender}, {Garrappa}, {Ahumada}, {Andreoni}, {Belicki}, {Bellm},
  {B{\"o}ttcher}, {Brinnel}, {Burruss}, {Cenko}, {Coughlin}, {Cunningham},
  {Drake}, {Farrar}, {Feeney}, {Foley}, {Gal-Yam}, {Golkhou}, {Goobar},
  {Graham}, {Hammerstein}, {Helou}, {Hung}, {Kasliwal}, {Kilpatrick}, {Kong},
  {Kupfer}, {Laher}, {Mahabal}, {Masci}, {Necker}, {Nordin}, {Perley},
  {Rigault}, {Reusch}, {Rodriguez}, {Rojas-Bravo}, {Rusholme}, {Shupe},
  {Singer}, {Sollerman}, {Soumagnac}, {Stern}, {Taggart}, {van Santen}, {Ward},
  {Woudt}, \& {Yao}}]{Stein2021}
{Stein}, R., {Velzen}, S.~v., {Kowalski}, M., {et~al.} 2021, Nature Astronomy,
  5, 510, \dodoi{10.1038/s41550-020-01295-8}

\bibitem[{{Stone} {et~al.}(2019){Stone}, {Kesden}, {Cheng}, \& {van
  Velzen}}]{Stone2019}
{Stone}, N.~C., {Kesden}, M., {Cheng}, R.~M., \& {van Velzen}, S. 2019, General
  Relativity and Gravitation, 51, 30, \dodoi{10.1007/s10714-019-2510-9}

\bibitem[{{Stone} \& {Metzger}(2016)}]{Stone2016}
{Stone}, N.~C., \& {Metzger}, B.~D. 2016, \mnras, 455, 859,
  \dodoi{10.1093/mnras/stv2281}

\bibitem[{{Stone} {et~al.}(2020){Stone}, {Vasiliev}, {Kesden}, {Rossi},
  {Perets}, \& {Amaro-Seoane}}]{Stone2020}
{Stone}, N.~C., {Vasiliev}, E., {Kesden}, M., {et~al.} 2020, \ssr, 216, 35,
  \dodoi{10.1007/s11214-020-00651-4}

\bibitem[{{Strauss} {et~al.}(2002){Strauss}, {Weinberg}, {Lupton}, {Narayanan},
  {Annis}, {Bernardi}, {Blanton}, {Burles}, {Connolly}, {Dalcanton}, {Doi},
  {Eisenstein}, {Frieman}, {Fukugita}, {Gunn}, {Ivezi{\'c}}, {Kent}, {Kim},
  {Knapp}, {Kron}, {Munn}, {Newberg}, {Nichol}, {Okamura}, {Quinn}, {Richmond},
  {Schlegel}, {Shimasaku}, {SubbaRao}, {Szalay}, {Vanden Berk}, {Vogeley},
  {Yanny}, {Yasuda}, {York}, \& {Zehavi}}]{Strauss2002}
{Strauss}, M.~A., {Weinberg}, D.~H., {Lupton}, R.~H., {et~al.} 2002, \aj, 124,
  1810, \dodoi{10.1086/342343}

\bibitem[{{Sunyaev} {et~al.}(2021){Sunyaev}, {Arefiev}, {Babyshkin},
  {Bogomolov}, {Borisov}, {Buntov}, {Brunner}, {Burenin}, {Churazov},
  {Coutinho}, {Eder}, {Eismont}, {Freyberg}, {Gilfanov}, {Gureyev}, {Hasinger},
  {Khabibullin}, {Kolmykov}, {Komovkin}, {Krivonos}, {Lapshov}, {Levin},
  {Lomakin}, {Lutovinov}, {Medvedev}, {Merloni}, {Mernik}, {Mikhailov},
  {Molodtsov}, {Mzhelsky}, {M{\"u}ller}, {Nandra}, {Nazarov}, {Pavlinsky},
  {Poghodin}, {Predehl}, {Robrade}, {Sazonov}, {Scheuerle}, {Shirshakov},
  {Tkachenko}, \& {Voron}}]{Sunyaev2021}
{Sunyaev}, R., {Arefiev}, V., {Babyshkin}, V., {et~al.} 2021, \aap, 656, A132,
  \dodoi{10.1051/0004-6361/202141179}

\bibitem[{{Tachibana} \& {Miller}(2018)}]{Tachibana2018}
{Tachibana}, Y., \& {Miller}, A.~A. 2018, \pasp, 130, 128001,
  \dodoi{10.1088/1538-3873/aae3d9}

\bibitem[{{Teboul} {et~al.}(2022){Teboul}, {Stone}, \& {Ostriker}}]{Teboul2022}
{Teboul}, O., {Stone}, N.~C., \& {Ostriker}, J.~P. 2022, arXiv e-prints,
  arXiv:2211.05858.
\newblock \doarXiv{2211.05858}

\bibitem[{{Thomsen} {et~al.}(2022){Thomsen}, {Kwan}, {Dai}, {Wu}, {Roth}, \&
  {Ramirez-Ruiz}}]{Thomsen2022_disk_wind}
{Thomsen}, L.~L., {Kwan}, T.~M., {Dai}, L., {et~al.} 2022, \apjl, 937, L28,
  \dodoi{10.3847/2041-8213/ac911f}

\bibitem[{{Tonry} {et~al.}(2018){Tonry}, {Denneau}, {Heinze}, {Stalder},
  {Smith}, {Smartt}, {Stubbs}, {Weiland }, \& {Rest}}]{Tonry2018}
{Tonry}, J.~L., {Denneau}, L., {Heinze}, A.~N., {et~al.} 2018, \pasp, 130,
  064505, \dodoi{10.1088/1538-3873/aabadf}

\bibitem[{{Tucci} \& {Volonteri}(2017)}]{Tucci&Volonteri2017}
{Tucci}, M., \& {Volonteri}, M. 2017, \aap, 600, A64,
  \dodoi{10.1051/0004-6361/201628419}

\bibitem[{{Tucker}(2021)}]{Tucker2021_ZTF20aasuiks}
{Tucker}, M.~A. 2021, Transient Name Server Classification Report, 2021-433, 1

\bibitem[{{Valluri} {et~al.}(2005){Valluri}, {Ferrarese}, {Merritt}, \&
  {Joseph}}]{Valluri2005}
{Valluri}, M., {Ferrarese}, L., {Merritt}, D., \& {Joseph}, C.~L. 2005, \apj,
  628, 137, \dodoi{10.1086/430752}

\bibitem[{{van Velzen}(2018)}]{vanVelzen2018}
{van Velzen}, S. 2018, \apj, 852, 72, \dodoi{10.3847/1538-4357/aa998e}

\bibitem[{{van Velzen} {et~al.}(2020){van Velzen}, {Holoien}, {Onori}, {Hung},
  \& {Arcavi}}]{vanVelzen2020}
{van Velzen}, S., {Holoien}, T. W.~S., {Onori}, F., {Hung}, T., \& {Arcavi}, I.
  2020, \ssr, 216, 124, \dodoi{10.1007/s11214-020-00753-z}

\bibitem[{{van Velzen} {et~al.}(2011){van Velzen}, {Farrar}, {Gezari},
  {Morrell}, {Zaritsky}, {{\"O}stman}, {Smith}, {Gelfand}, \&
  {Drake}}]{vanVelzen2011}
{van Velzen}, S., {Farrar}, G.~R., {Gezari}, S., {et~al.} 2011, \apj, 741, 73,
  \dodoi{10.1088/0004-637X/741/2/73}

\bibitem[{{van Velzen} {et~al.}(2019){van Velzen}, {Gezari}, {Cenko}, {Kara},
  {Miller-Jones}, {Hung}, {Bright}, {Roth}, {Blagorodnova}, {Huppenkothen},
  {Yan}, {Ofek}, {Sollerman}, {Frederick}, {Ward}, {Graham}, {Fender},
  {Kasliwal}, {Canella}, {Stein}, {Giomi}, {Brinnel}, {van Santen}, {Nordin},
  {Bellm}, {Dekany}, {Fremling}, {Golkhou}, {Kupfer}, {Kulkarni}, {Laher},
  {Mahabal}, {Masci}, {Miller}, {Neill}, {Riddle}, {Rigault}, {Rusholme},
  {Soumagnac}, \& {Tachibana}}]{vanVelzen2019}
{van Velzen}, S., {Gezari}, S., {Cenko}, S.~B., {et~al.} 2019, \apj, 872, 198,
  \dodoi{10.3847/1538-4357/aafe0c}

\bibitem[{{van Velzen} {et~al.}(2021){van Velzen}, {Gezari}, {Hammerstein},
  {Roth}, {Frederick}, {Ward}, {Hung}, {Cenko}, {Stein}, {Perley}, {Taggart},
  {Foley}, {Sollerman}, {Blagorodnova}, {Andreoni}, {Bellm}, {Brinnel}, {De},
  {Dekany}, {Feeney}, {Fremling}, {Giomi}, {Golkhou}, {Graham}, {Ho},
  {Kasliwal}, {Kilpatrick}, {Kulkarni}, {Kupfer}, {Laher}, {Mahabal}, {Masci},
  {Miller}, {Nordin}, {Riddle}, {Rusholme}, {van Santen}, {Sharma}, {Shupe}, \&
  {Soumagnac}}]{vanVelzen2021}
{van Velzen}, S., {Gezari}, S., {Hammerstein}, E., {et~al.} 2021, \apj, 908, 4,
  \dodoi{10.3847/1538-4357/abc258}

\bibitem[{{Vika} {et~al.}(2009){Vika}, {Driver}, {Graham}, \&
  {Liske}}]{Vika2009}
{Vika}, M., {Driver}, S.~P., {Graham}, A.~W., \& {Liske}, J. 2009, \mnras, 400,
  1451, \dodoi{10.1111/j.1365-2966.2009.15544.x}

\bibitem[{{Wang} \& {Merritt}(2004)}]{Wang2004}
{Wang}, J., \& {Merritt}, D. 2004, \apj, 600, 149, \dodoi{10.1086/379767}

\bibitem[{{Wevers} {et~al.}(2017){Wevers}, {van Velzen}, {Jonker}, {Stone},
  {Hung}, {Onori}, {Gezari}, \& {Blagorodnova}}]{Wevers2017}
{Wevers}, T., {van Velzen}, S., {Jonker}, P.~G., {et~al.} 2017, \mnras, 471,
  1694, \dodoi{10.1093/mnras/stx1703}

\bibitem[{{Wevers} {et~al.}(2019){Wevers}, {Stone}, {van Velzen}, {Jonker},
  {Hung}, {Auchettl}, {Gezari}, {Onori}, {Mata S{\'a}nchez},
  {Kostrzewa-Rutkowska}, \& {Casares}}]{Wevers2019_Mbh}
{Wevers}, T., {Stone}, N.~C., {van Velzen}, S., {et~al.} 2019, \mnras, 487,
  4136, \dodoi{10.1093/mnras/stz1602}

\bibitem[{{Woods} {et~al.}(2019){Woods}, {Agarwal}, {Bromm}, {Bunker}, {Chen},
  {Chon}, {Ferrara}, {Glover}, {Haemmerl{\'e}}, {Haiman}, {Hartwig}, {Heger},
  {Hirano}, {Hosokawa}, {Inayoshi}, {Klessen}, {Kobayashi}, {Koliopanos},
  {Latif}, {Li}, {Mayer}, {Mezcua}, {Natarajan}, {Pacucci}, {Rees}, {Regan},
  {Sakurai}, {Salvadori}, {Schneider}, {Surace}, {Tanaka}, {Whalen}, \&
  {Yoshida}}]{Woods2019}
{Woods}, T.~E., {Agarwal}, B., {Bromm}, V., {et~al.} 2019, \pasa, 36, e027,
  \dodoi{10.1017/pasa.2019.14}

\bibitem[{{Wright} {et~al.}(2016){Wright}, {Robotham}, {Bourne}, {Driver},
  {Dunne}, {Maddox}, {Alpaslan}, {Andrews}, {Bauer}, {Bland-Hawthorn},
  {Brough}, {Brown}, {Clarke}, {Cluver}, {Davies}, {Grootes}, {Holwerda},
  {Hopkins}, {Jarrett}, {Kafle}, {Lange}, {Liske}, {Loveday}, {Moffett},
  {Norberg}, {Popescu}, {Smith}, {Taylor}, {Tuffs}, {Wang}, \&
  {Wilkins}}]{Wright2016a}
{Wright}, A.~H., {Robotham}, A.~S.~G., {Bourne}, N., {et~al.} 2016, \mnras,
  460, 765, \dodoi{10.1093/mnras/stw832}

\bibitem[{{Wright} {et~al.}(2017){Wright}, {Robotham}, {Driver}, {Alpaslan},
  {Andrews}, {Baldry}, {Bland-Hawthorn}, {Brough}, {Brown}, {Colless}, {da
  Cunha}, {Davies}, {Graham}, {Holwerda}, {Hopkins}, {Kafle}, {Kelvin},
  {Loveday}, {Maddox}, {Meyer}, {Moffett}, {Norberg}, {Phillipps}, {Rowlands},
  {Taylor}, {Wang}, \& {Wilkins}}]{Wright2017}
{Wright}, A.~H., {Robotham}, A.~S.~G., {Driver}, S.~P., {et~al.} 2017, \mnras,
  470, 283, \dodoi{10.1093/mnras/stx1149}

\bibitem[{{Yan} {et~al.}(2020){Yan}, {Lunnan}, {Perley}, {Schulze}, \&
  {Chen}}]{Yan2020_ZTF20achuhlt}
{Yan}, L., {Lunnan}, R., {Perley}, D., {Schulze}, S., \& {Chen}, T.~W. 2020,
  Transient Name Server Classification Report, 2020-3640, 1

\bibitem[{{Yao}(2021)}]{Yao2021_21uqv_CR}
{Yao}, Y. 2021, Transient Name Server Classification Report, 2021-3411, 1

\bibitem[{{Yao}(2022)}]{Yao2022_ZTF21abwjibi}
---. 2022, Transient Name Server Classification Report, 2022-2915, 1

\bibitem[{{Yao} {et~al.}(2021{\natexlab{a}}){Yao}, {Chu}, {Das}, {Kulkarni},
  {Somalwar}, {Gezari}, {Velzen}, \& {Hammerstein}}]{Yao2021_21yte_CR}
{Yao}, Y., {Chu}, M., {Das}, K.~K., {et~al.} 2021{\natexlab{a}}, Transient Name
  Server Classification Report, 2021-3611, 1

\bibitem[{{Yao} {et~al.}(2021{\natexlab{b}}){Yao}, {Gezari}, {Velzen},
  {Hammerstein}, \& {Somalwar}}]{Yao2021_21nwa_CR}
{Yao}, Y., {Gezari}, S., {Velzen}, S.~V., {Hammerstein}, E., \& {Somalwar}, J.
  2021{\natexlab{b}}, Transient Name Server Classification Report, 2021-2155, 1

\bibitem[{{Yao} {et~al.}(2021{\natexlab{c}}){Yao}, {Hammerstein}, {Gezari},
  {Velzen}, {Somalwar}, \& {Kulkarni}}]{Yao2021_ZTF21aavdqgf}
{Yao}, Y., {Hammerstein}, E., {Gezari}, S., {et~al.} 2021{\natexlab{c}},
  Transient Name Server Classification Report, 2021-2535, 1

\bibitem[{{Yao} {et~al.}(2021{\natexlab{d}}){Yao}, {Velzen}, {Perley},
  {Gezari}, {Hammerstein}, {Somalwar}, {Sharma}, \&
  {Kulkarni}}]{Yao2021_21jjm_CR}
{Yao}, Y., {Velzen}, S.~V., {Perley}, D., {et~al.} 2021{\natexlab{d}},
  Transient Name Server Classification Report, 2021-1632, 1

\bibitem[{{Yao} {et~al.}(2019){Yao}, {Miller}, {Kulkarni}, {Bulla}, {Masci},
  {Goldstein}, {Goobar}, {Nugent}, {Dugas}, {Blagorodnova}, {Neill}, {Rigault},
  {Sollerman}, {Nordin}, {Bellm}, {Cenko}, {De}, {Dhawan}, {Feindt},
  {Fremling}, {Gatkine}, {Graham}, {Graham}, {Ho}, {Hung}, {Kasliwal},
  {Kupfer}, {Laher}, {Perley}, {Rusholme}, {Shupe}, {Soumagnac}, {Taggart},
  {Walters}, \& {Yan}}]{Yao2019}
{Yao}, Y., {Miller}, A.~A., {Kulkarni}, S.~R., {et~al.} 2019, \apj, 886, 152,
  \dodoi{10.3847/1538-4357/ab4cf5}

\bibitem[{{Yao} {et~al.}(2022{\natexlab{a}}){Yao}, {Lu}, {Guolo}, {Pasham},
  {Gezari}, {Gilfanov}, {Gendreau}, {Harrison}, {Cenko}, {Kulkarni}, {Miller},
  {Walton}, {Garc{\'\i}a}, {van Velzen}, {Alexander}, {Miller-Jones},
  {Nicholl}, {Hammerstein}, {Medvedev}, {Stern}, {Ravi}, {Sunyaev}, {Bloom},
  {Graham}, {Kool}, {Mahabal}, {Masci}, {Purdum}, {Rusholme}, {Sharma},
  {Smith}, \& {Sollerman}}]{Yao2022}
{Yao}, Y., {Lu}, W., {Guolo}, M., {et~al.} 2022{\natexlab{a}}, \apj, 937, 8,
  \dodoi{10.3847/1538-4357/ac898a}

\bibitem[{{Yao} {et~al.}(2022{\natexlab{b}}){Yao}, {Ho}, {Medvedev}, {Nayana},
  {Perley}, {Kulkarni}, {Chandra}, {Sazonov}, {Gilfanov}, {Khorunzhev},
  {Khatami}, \& {Sunyaev}}]{Yao2022_20mrf}
{Yao}, Y., {Ho}, A. Y.~Q., {Medvedev}, P., {et~al.} 2022{\natexlab{b}}, \apj,
  934, 104, \dodoi{10.3847/1538-4357/ac7a41}

\bibitem[{{Yaron} \& {Gal-Yam}(2012)}]{Yaron2012}
{Yaron}, O., \& {Gal-Yam}, A. 2012, \pasp, 124, 668, \dodoi{10.1086/666656}

\bibitem[{{Yu} \& {Lu}(2008)}]{Yu2008}
{Yu}, Q., \& {Lu}, Y. 2008, \apj, 689, 732, \dodoi{10.1086/592770}

\bibitem[{{Yu} {et~al.}(2022){Yu}, {Kochanek}, {Mathur}, {Auchettl}, {Grupe},
  \& {Holoien}}]{Yu2022}
{Yu}, Z., {Kochanek}, C.~S., {Mathur}, S., {et~al.} 2022, \mnras, 515, 5198,
  \dodoi{10.1093/mnras/stac2073}

\end{thebibliography}
\bibliographystyle{aasjournal}

\end{document}